\documentclass[twocolumn,floatfix,showpacs,prd,aps,tightenlines,letterpaper]{revtex4-1}
\usepackage{amsmath}
\usepackage{longtable}
\usepackage{dcolumn}
\usepackage{graphicx}
\usepackage{psfrag}
\usepackage{color}
\usepackage{dcolumn}
\usepackage{bm}
\usepackage[mathscr]{eucal}
\usepackage{mathrsfs}
\usepackage{xspace}

\def\msun{\rm M_{\odot}}
\def\kms{\rm km \, s^{-1}}
\def\etal{{\it et al.}\xspace}
\def\simlt{\mathrel{\rlap{\lower 3pt\hbox{$\sim$}}\raise 2.0pt\hbox{$<$}}}
\def\simgt{\mathrel{\rlap{\lower 3pt\hbox{$\sim$}} \raise 2.0pt\hbox{$>$}}}
\def\lsim{\mathrel{\rlap{\lower 3pt\hbox{$\sim$}}\raise 2.0pt\hbox{$<$}}}
\def\gsim{\mathrel{\rlap{\lower 3pt\hbox{$\sim$}} \raise 2.0pt\hbox{$>$}}}

\def\msunpc3{\msun~{\rm {pc^{-3}}}}
\newcommand{\be}{\begin{equation}}
\newcommand{\ee}{\end{equation}}
\def\kms{{\rm\,km\,s^{-1}}}

\newcommand{\hangup}{{\it hangup kick}\xspace}

\newcommand{\cross}{{\it cross kick}\xspace}

\newcommand{\bea}{\begin{eqnarray}}
\newcommand{\eea}{\end{eqnarray}}
\newcommand{\beq}{\begin{equation}}
\newcommand{\eeq}{\end{equation}}
\newcommand{\KMS}{\rm km\ s^{-1}}

\newcommand{\Spar}{\tilde S_\|}
\newcommand{\Dpar}{\tilde \Delta_\|} 
\newcommand{\dmt}{\delta{m}}

\begin{document}

\def\fun#1#2{\lower3.6pt\vbox{\baselineskip0pt\lineskip.9pt
  \ialign{$\mathsurround=0pt#1\hfil##\hfil$\crcr#2\crcr\sim\crcr}}}
\def\lap{\mathrel{\mathpalette\fun <}}
\def\gap{\mathrel{\mathpalette\fun >}}
\def\kms{{\rm km\ s}^{-1}}
\def\vk{V_{\rm recoil}}

\title{Modeling the remnant mass, spin, and recoil from unequal-mass,
precessing black-hole binaries:\\
 The
Intermediate Mass Ratio Regime}

\author{
Yosef Zlochower}
\author{
Carlos O. Lousto
}
\affiliation{Center for Computational Relativity and Gravitation,\\
and School of Mathematical Sciences, Rochester Institute of
Technology, 85 Lomb Memorial Drive, Rochester, New York 14623}

\begin{abstract}
We revisit the modeling of the properties of the remnant black hole 
resulting from the merger of a black-hole binary as a function of the
parameters of the binary.
We provide a set of empirical formulas for the final mass, spin, and
recoil velocity of the final black hole as a function of the
mass ratio and individual spins of the progenitor.
In order to determine the fitting coefficients for these formulas,
we perform a set of 128 new numerical evolutions of precessing, unequal-mass
black-hole binaries, and fit to the resulting remnant mass, spin, and
recoil. In order to reduce the complexity of the
analysis, we chose  configurations that
have one of the black holes spinning, with dimensionless spin
$\alpha=0.8$, at
different angles with respect to the orbital angular momentum, and
the other nonspinning. In addition to evolving families of binaries
with different spin-inclination angles, we also evolved binaries with
mass ratios as small as $q=M_1/M_2=1/6$.
We use the resulting empirical formulas to predict the probabilities
of black hole mergers leading to
a given recoil velocity, total radiated gravitational energy,
and final black hole spin.
\end{abstract}

 \maketitle

\section{Introduction}\label{sec:Introduction} Black holes and
black-hole binaries (BHBs) are thought to be ubiquitous in nature.
Supermassive BHs, which have masses from $\sim 10^6M_\odot$ to $\sim
10^{10}M_\odot$ ($M_\odot$ is the mass of the sun) are thought to be at the centers of most galaxies with
a bulge, while stellar-mass BHs generated in the collapse of massive
stars, have masses from $\sim10 M_\odot$ to $\sim 100 M_\odot$. There
is strong observational evidence for both binaries and
solitary black holes from
these two populations. More speculative is the
intermediate mass BH population, which would have masses from
$100M_\odot$ to $\sim 10^6 M_\odot$ (see, e.g.,~\cite{Miller:2003sc}).

In 2005, there was a series of remarkable breakthroughs in numerical
relativity (NR)~\cite{Pretorius:2005gq, Campanelli:2005dd,
Baker:2005vv}, that allowed, for the first time, simulations of
merging BHBs.  One of the most remarkable results that came from these
simulations is that the merger remnant can recoil at thousands of
kilometers per second (see~\cite{Campanelli:2004zw,Baker:2006vn, Baker:2007gi,
Baker:2008md, Brugmann:2007zj, Campanelli:2007cga, Campanelli:2007ew,
Choi:2007eu, Dain:2008ck, Gonzalez:2006md, Gonzalez:2007hi,
Healy:2008js, Herrmann:2006cd, Lousto:2007db, Herrmann:2007ac,
Herrmann:2007ex, Herrmann:2007zz, HolleyBockelmann:2007eh,
Jaramillo:2011re, Koppitz:2007ev, Laguna:2009zz, Lousto:2008dn,
Lousto:2010xk, Nakano:2010kv, Lousto:2011kp, Lousto:2012gt, Lousto:2012su,
Miller:2008en, Pollney:2007ss, Rezzolla:2010df, Schnittman:2007ij,
Sopuerta:2006et, Sopuerta:2006wj, Tichy:2007hk, vanMeter:2010md,
Zlochower:2010sn}).

The first in-depth modeling of the recoil from the merger of
nonspinning asymmetric BHBs was done in Ref.~\cite{Gonzalez:2006md},
where it was shown that the maximum recoil is limited to $\approx 175\
\KMS$. Soon after, other groups showed that the maximum recoil for
spinning binaries, where the spins are aligned and antialigned
with the angular momentum, is much larger. In
Ref.~\cite{Herrmann:2007ac}~and~\cite{Koppitz:2007ev}, it was shown
that the
maximum recoil for an equal mass,
spinning binary with one BH spin aligned with the
orbital angular momentum and other antialigned is $\sim475\ \KMS$.
However in Ref.~\cite{Healy:2014yta} we find that for a mass ratio of
$q\approx0.62$ there is a maximum recoil of $V_{max}\sim525\ \KMS$.

The recoils induced by unequal masses and aligned/antialigned spins
is always in the orbital plane of the binary (which, by symmetry,
does not precess). In~\cite{Campanelli:2007ew}, our group performed a
set of simulations that showed that the out-of-plane recoil, which is
induced by spins lying in the orbital plane, can be much larger.
These {\it superkicks}~\cite{Campanelli:2007ew, Gonzalez:2007hi,
Campanelli:2007cga, Dain:2008ck, Lousto:2010xk} were found to be up to
$4000\ \KMS$ when the spins were exactly in the orbital plane.
Originally, it was thought that these in-plane spins maximized the
recoil, however, as our group found out
in~\cite{Lousto:2011kp, Lousto:2012su, Lousto:2012gt},
due to the hangup and other nonlinear-in-spin effects \cite{Campanelli:2006uy},
having partially miss-aligned spins actually leads to a substantially larger
recoil (up to $5000\ \KMS$).

An open question remained, however, of how the recoil behaves as a
function of the binary's mass ratio. This problem was first examined
in detail in~\cite{Baker:2008md}, where minimally precessing
configurations were examined,  and later in~\cite{Lousto:2008dn}.

The next major challenge was to distill the results from large
numbers of numerical simulations into convenient empirical
formulas that map the initial conditions of the binary (individual masses and
spins) to the final state of the merged black
hole~\cite{Boyle:2007sz, Boyle:2007ru, Buonanno:2007sv, Tichy:2008du,
Kesden:2008ga, Barausse:2009uz, Rezzolla:2008sd, Lousto:2009mf,
Healy:2014yta, Lousto:2013wta}.

Here we report on an 
effort to create both a bank of
a large number of unequal-mass, precessing BHB simulations and the
subsequent modeling of the recoil as a function of the binary's
initial configuration. Our goal in this paper is the produce an
interpolative formula that is accurate within the mass ratio range
$1/6\lesssim q\leq1$ and provides a reasonable extrapolative formula
down to mass ratios as small as $q=M_1/M_2=1/10$, as well as for intrinsic
spins $\alpha_i=S_i/M_i^2$ as large as $0.95 - 0.97$ (here $S_i$ is the
spin angular momentum of BH $i$).

In constructing the new formula, we will enforce the particle
limit behavior $v_{\rm rec} \propto {\cal O}(q^2)$, which is the expected
behavior provided that the central BH is not spinning too fast (see
Refs.~\cite{Hirata:2010xn, vandeMeent:2014raa} for a
discussion on resonance recoil which scale as $q^{1.5}$, see also
\cite{Baker:2008md, Lousto:2008dn} for a discussion on whether or not
the recoil should scale generically as ${\cal O} (q^2)$).

A note of caution. We will be basing our formulas on runs performed for
moderate to high spins $\alpha\lesssim 0.8$. The dynamics of particles
in the vicinity of a Kerr BH vary in a non-differentiable way at
$\alpha=1$. Therefore, for extremely high spins, there are likely
interesting effects that cannot be elucidated using lower spin
simulations. Fortunately, these effects occur at spins higher than
what is expected astrophysically. See~\cite{Schnittman:2014zsa,
Berti:2014lva, Yang:2014tla} for discussions about these effects.

In addition to modeling the recoil, we also provide new interpolative
formulas for the total radiated mass and final remnant spin.

This paper is organized as follows. In Sec.~\ref{sec:numerical_kicks}
we summarize the numerical techniques used and describe
the configurations we evolve.
In Sec.~\ref{sec:symmetries} we review how symmetry arguments can be
used to limit the number of terms in an expansion of the recoil and
remnant mass and spin, and then
explicitly give the form of each expansion term up through
fourth-order.
In Sec.~\ref{sec:Fits},
we provide the procedure used to fit the remnant properties to the
parameters of the binary and provide the resulting fitting formulas.
In Sec.~\ref{sec:analysis}, we use these fitting formulas to
calculate the statistical probabilities for a given recoil and remnant
mass and spin given several plausible
distributions for the possible parameters of the binary.
Finally, in Sec.~\ref{sec:discussion},
we discuss the relevance of our results
in the context of galactic and supermassive black-hole evolutions.
We also provide an appendix with an extensive list of simulation
results that can be used for further modeling.

\section{Simulations and Results}
\label{sec:numerical_kicks}
We evolve the following BHB data sets using the {\sc
LazEv}~\cite{Zlochower:2005bj} implementation of the moving puncture
approach~\cite{Campanelli:2005dd,Baker:2005vv} with the conformal
function $W=\sqrt{\chi}=\exp(-2\phi)$ suggested by
Ref.~\cite{Marronetti:2007wz}.  For the runs presented here, we use
centered, eighth-order finite differencing in
space~\cite{Lousto:2007rj} and a fourth-order Runge Kutta time
integrator. (Note that we do not upwind the advection terms.)

Our code uses the {\sc EinsteinToolkit}~\cite{Loffler:2011ay,
einsteintoolkit} / {\sc Cactus}~\cite{cactus_web} /
{\sc Carpet}~\cite{Schnetter-etal-03b}
infrastructure.  The {\sc
Carpet} mesh refinement driver provides a
``moving boxes'' style of mesh refinement. In this approach, refined
grids of fixed size are arranged about the coordinate centers of both
holes.  The {\sc Carpet} code then moves these fine grids about the
computational domain by following the trajectories of the two BHs.

We use {\sc AHFinderDirect}~\cite{Thornburg2003:AH-finding} to locate
apparent horizons.  We measure the magnitude of the horizon spin using
the {\it isolated horizon} (IH) algorithm detailed in
Ref.~\cite{Dreyer02a}.
Note that once we have the horizon spin, we can calculate the horizon
mass via the Christodoulou formula
\begin{equation}
{M_H} = \sqrt{M_{\rm irr}^2 + S_H^2/(4 M_{\rm irr}^2)} \,,
\end{equation}
where $M_{\rm irr} = \sqrt{A/(16 \pi)}$, $A$ is the surface area of
the horizon, and $S_H$ is the spin angular momentum of the BH (in
units of $M^2$).  In the tables below, we use the variation in the
measured horizon irreducible mass and spin during the simulation as a
measure of the error in these quantities.  We measure radiated energy,
linear momentum, and angular momentum, in terms of the radiative Weyl
scalar $\psi_4$, using the formulas provided in
Refs.~\cite{Campanelli:1998jv,Lousto:2007mh}. However, rather than
using the full $\psi_4$, we decompose it into $\ell$ and $m$ modes and
solve for the radiated linear momentum, dropping terms with $\ell \geq
5$ \footnote{At least in the similar mass regime, we found that
the $\ell\geq5$ modes 
contributes a smaller amount to the recoil than
the extrapolation error to infinite radius, while, at the same time, these modes are
poorly resolved using our choice of gridspacing. We therefore suppress
these modes in our calculation.}.  The formulas in Refs.~\cite{Campanelli:1998jv,Lousto:2007mh} are
valid at $r=\infty$.  We extract the radiated energy-momentum at
finite radius and extrapolate to $r=\infty$ using both linear and
quadratic extrapolations. We use the difference of these two
extrapolations as a measure of the error.

Both the variation (with time) of the remnant parameters (as measured
using the isolated horizons formalism), and the variation in the
extrapolation of the radiation to infinity (as a function of different
extraction radii) underestimate the actual
errors in the quantity of interest. However, because quantities like
the total radiated energy can be obtained from either extrapolations
of $\psi_4$ or, quite independently, from the remnant BHs mass, the
difference between these two is a reasonable estimate for the actual
error. Furthermore, in~\cite{Healy:2014yta}, the errors associated
with finite resolution, finite extraction radii, and using low
$\ell$ modes only were examined
in detail. There it was found that for the recoil
the errors associated with dropping
$\ell\geq5$, the errors associated with finite extraction radii, and
the truncation error were all of a similar size (roughly $5-10\
\KMS$).

We use the {\sc
TwoPunctures}
thorn~\cite{Ansorg:2004ds} to generate initial puncture
data~\cite{Brandt97b} for the BHB simulations described below. These
data are characterized by mass parameters $m_{p1/2}$, momenta $\vec
p_{1/2}$, spins $\vec S_{1/2}$, and coordinate locations $\vec
x_{1/2}$ of each hole. We obtain parameters for the location,
momentum, and spin of each BH using the 2.5 PN
quasicircular parameters.
We normalize our data such that the total Arnowitt-Deser-Misner (ADM)
energy is $1M$ and the mass ratio, as measured by the horizons masses
on the initial slice, has a given value. Because the BHs absorb energy
during the first few $M$ of evolution, the actual mass ratio will be
altered. In the fits below, we always use the mass ratio calculated
when the BHs have equilibrated.

\begin{figure}
\includegraphics[width=\columnwidth]{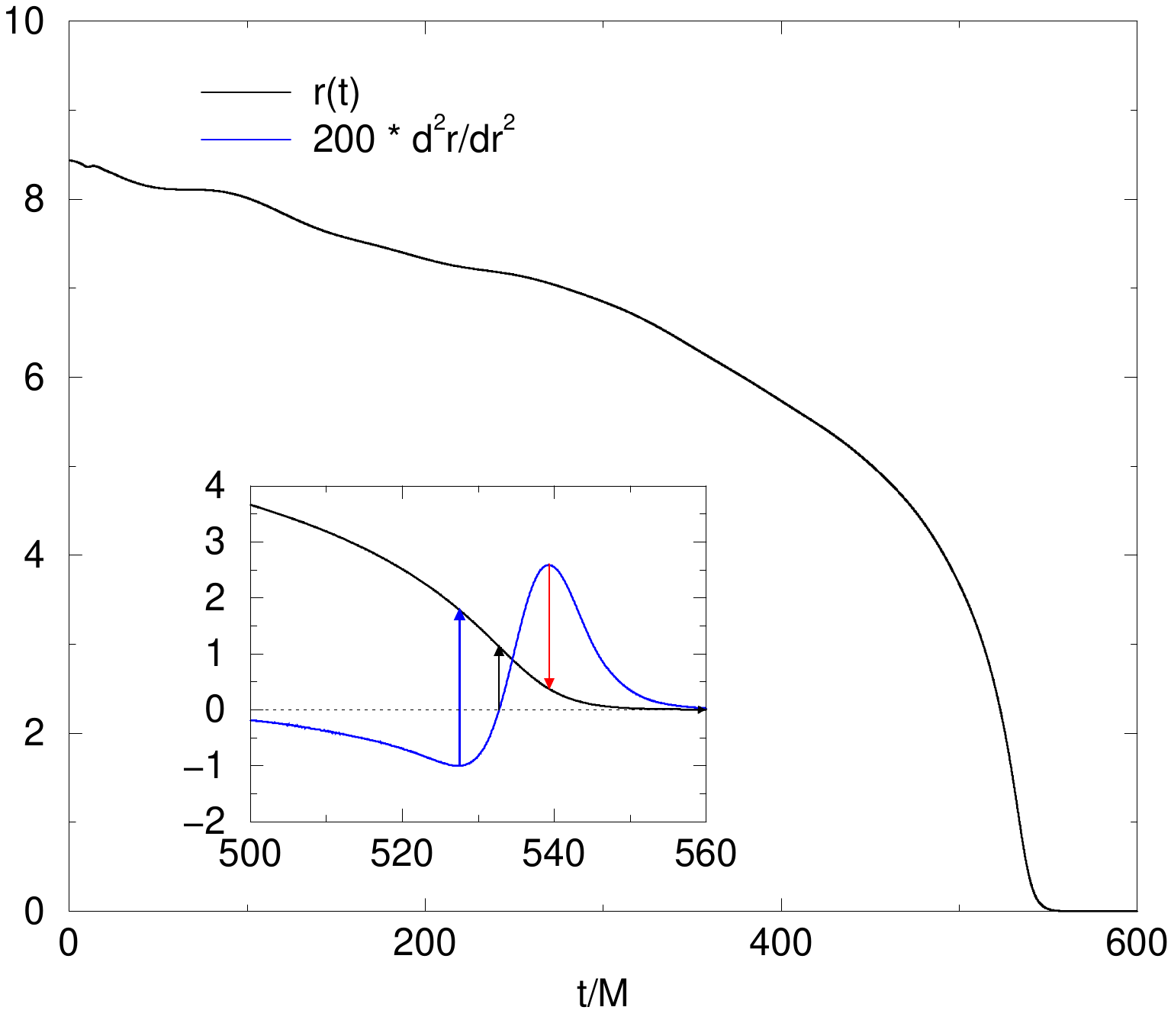}\\
\includegraphics[width=\columnwidth]{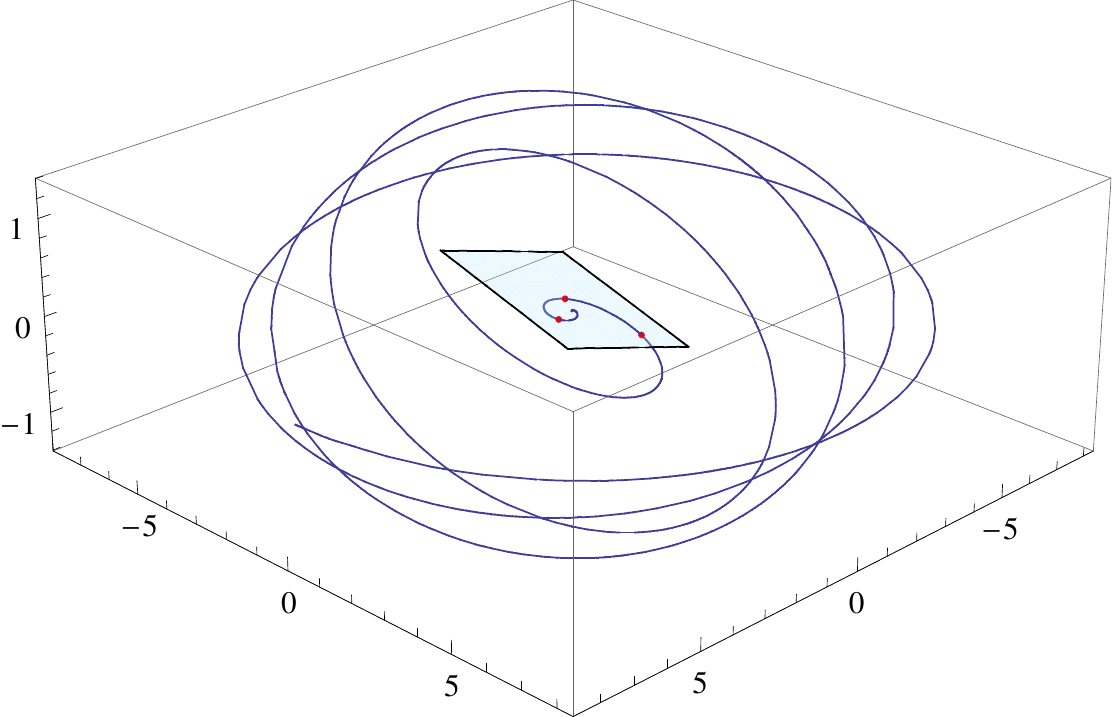}

\caption{Finding the orbital plane near merger. The upper plot
shows the orbital separation $r(t)$ versus time. The inset
shows $r(t)$ near merger and $\ddot r(t)$ (rescaled by 200 for
clarity). The points $\vec r_+$, $\vec r_0$, and $\vec r_-$ correspond
to the times where $\ddot r$ is maximized, zero, and minimized,
respectively
(denoted with arrows here). The plot below
shows the trajectory,  the points $\vec r_+, \vec r_0,
\vec r_-$  (large red dots) and the ``merger'' plane. }
\label{fig:find_plane}
\end{figure}

Our empirical formula will depend on the spins measured with respect
to the orbital plane at merger. In Ref~\cite{Lousto:2008dn} we
described a procedure for determining an approximate plane. This is
based on locating three fiducial points on the BHBs trajectory $\vec
r_+$, $\vec r_0$, and $\vec r_-$, where $\vec r_+$ is the point where
$\ddot r(t)$ [$r(t)$ is the orbital separation] reaches its maximum,
$\vec r_-$ is the point where $\ddot r(t)$ reaches its minimum, and
$\vec r_0$ is the point between the two where $\ddot r(t)=0$.  These
three points can then be used to define an approximate merger plane
(see Fig.~\ref{fig:find_plane}).
We then need to rotate each trajectory such that the infall directions
all align (as much as possible).
This is accomplished by rotating the system, keeping the merger
plane's orientation fixed, such that the
vector $\vec r_+ - \vec r_0$  is aligned with the $y$
axis. The azimuthal angle $\varphi$, described below, is measured in
this rotated frame.

\subsection{Configurations}

\begin{figure}
\includegraphics[width=0.9\columnwidth]{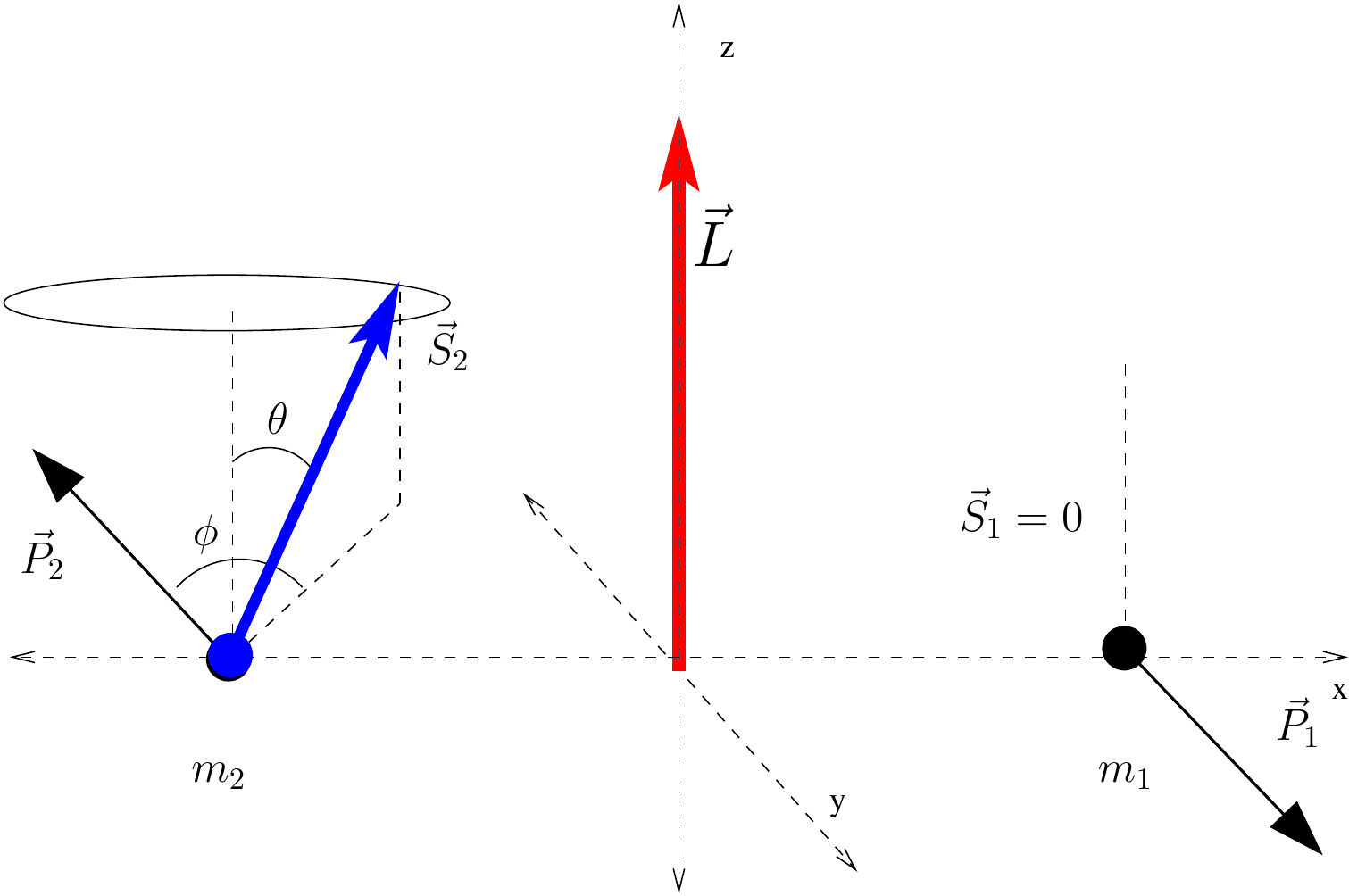}
\caption{The NQ configuration. Here one BH is spinning (typically the
larger one) and one is nonspinning.
 Numerical evolutions preserve the NQ configurations
approximately.}
\label{fig:NConfig}
\end{figure}

For this exploration of the dependence of the recoil, total radiated
energy, and remnant spin on the mass ratio we
will use an extension of the basic N configuration
of~\cite{Lousto:2012gt}, which we will denote by NQ here. The
difference between the N configurations and the new NQ configurations
is that the NQ configurations will have non-unit mass ratios.
For the N/NQ configuration (see Fig.~\ref{fig:NConfig}) one BH is spinning and
the other nonspinning. 
By convention, we choose BH2 to
be spinning and define the mass ratio $q$ by  $q=M_1/M_2$. So $q<1$ implies
that the larger BH is spinning, while $q>1$ implies that the smaller
BH is spinning.
 The polar orientation of the N/NQ
configurations will in general change over the course of the
evolution. However, a family of fixed starting polar angle $\theta$
and different azimuthal angles $\phi$ will evolve to a family
of configurations at merger with very similar polar orientations. This
will be critical to our fitting as we will be examining the maximum
recoil over $\phi$ for a given (ending) polar angle $\theta$ and
mass ratio $q$.

\begin{figure}
\includegraphics[width=0.9\columnwidth]{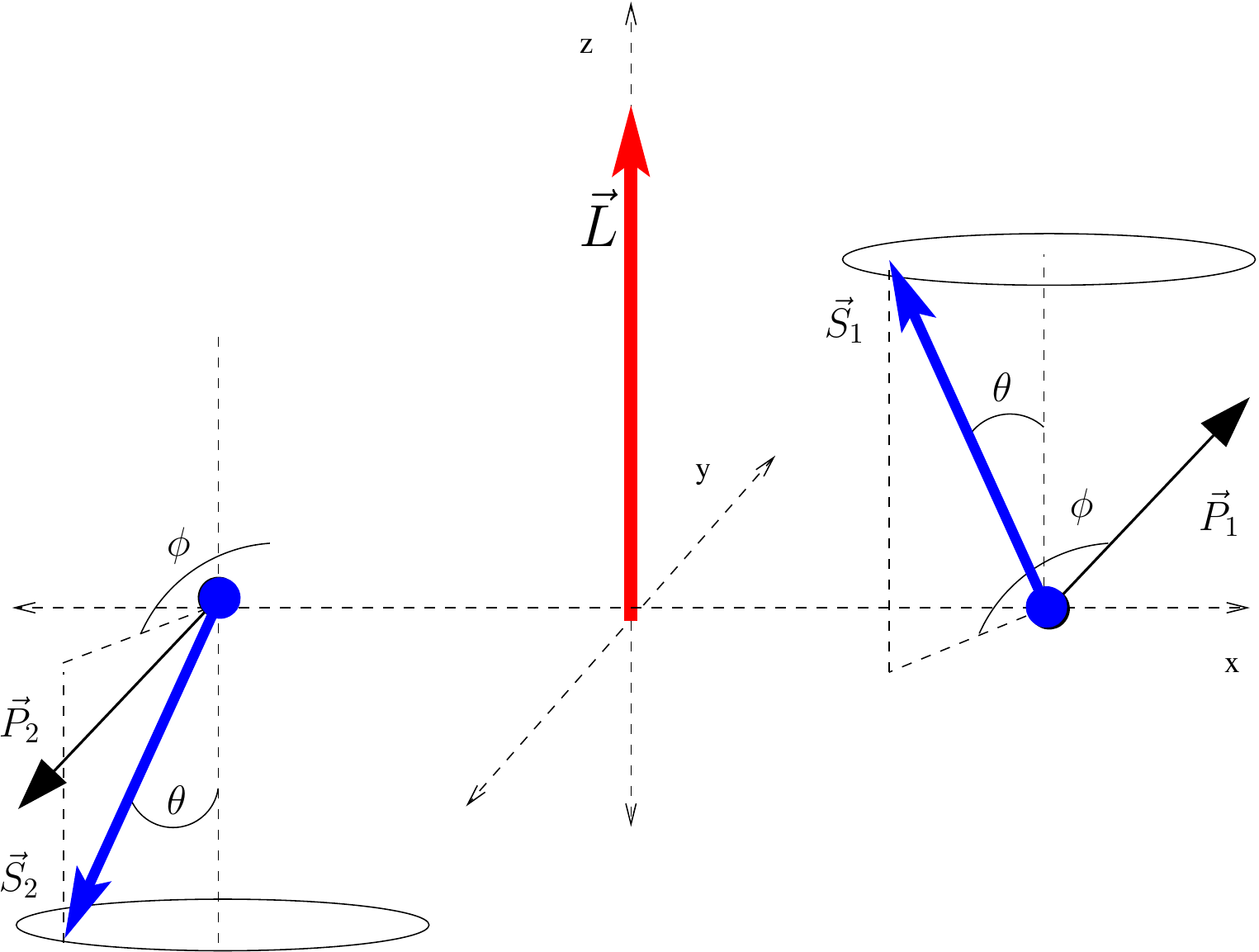}
\caption{The K configuration. $S_{1z}=-S_{2z}$, while
$S_{1x}=S_{2x}$ and $S_{1y}=S_{2y}$, initially.
}
\label{fig:KConfig}
\end{figure}

We will denote these configurations by NQxxxTHyyyPHzzz,
where xxx indicated the mass ratio, yyy indicates the initial
polar angle of the spin, and zzz indicated the initial azimuthal orientation
of the spin. We will also reexamine the fitting of the equal-mass N
and K configurations of
Ref.~\cite{Lousto:2012gt}. Note that while the K configurations start
with the two in-plane components of the spins aligned (see
Fig.~\ref{fig:KConfig}), the in-plane components of the two spins
rotate with respect to each other.

 A detailed list of initial data parameters for the new
NQxxxTHyyyPHzzz
configurations is given in Table~\ref{tab:ID_part1}.
The measured radiated mass, angular momentum, and
recoil is given in Table~\ref{tab:rad_part1}. As
we saw previously~\cite{Lousto:2012gt, Lousto:2013wta, Healy:2014yta}, the isolated horizon
quantities are more accurate than their radiative counterparts. The two
are shown for various configurations in
Table~\ref{tab:rem_rad_cmp_part1}. The
differences between the radiative and isolated horizon measures is a
reasonable measure of the true error in the radiative quantities.

Finally, in Table~\ref{tab:rot_part1} we give the spins near merger and the
recoils in a frame adapted to the averaged orbital plane at merger.
Results from the K configurations are also reported in these tables.

\section{Expansions for unequal mass binaries}
\label{sec:symmetries}
In the sections below we use the following conventions. We denote the
horizon
mass of the two BHs in a binary by $M_1$ and $M_2$ and the
total mass by $m$, where $m=M_1+M_2$.
The symbol $M$ will always
denote the unit of mass. We will use $\vec S_1$ and $\vec S_2$ to
denote the spins (in units of $M^2$) of the two BHs.
For our expansion formulas, we will use the variables,
\begin{eqnarray}
\delta{m}&=&(M_1-M_2)/m,\\
\vec S&=&\vec S_1 + \vec S_2,\\
\vec \Delta &=&m (\vec S_2/M_2 - \vec S_1/M_1),\\
 \vec S_0 &=& \vec S + (1/2) \delta m \vec \Delta,
\end{eqnarray}
as well as the dimensionless equivalent
variables
\begin{eqnarray}
 \vec{\tilde{S}} &=& \vec S/m^2,\\
 \vec{\tilde{\Delta}} &=&\vec \Delta/m^2,\\
 \vec {\tilde S}_0 &=& \vec S_0/m^2.
\end{eqnarray}
Note that for generic BHBs, the component of
$\vec{S}_0=m(\vec{S}_1/M_1+\vec{S}_2/M_2)$
in the direction of the orbital angular momentum is conserved
at low PN order \cite{Racine:2008qv} and
approximately conserved in full numerical simulations
\cite{Lousto:2013vpa}.

The set of variables \{$\vec S$, $\vec \Delta$, $\vec
S_0$\} is
linearly dependent. We will only use the pair of spin variables $(\vec S, \vec
\Delta)$ or the pair $(\vec S_0, \vec \Delta)$ in any one fit.
Finally, we will decompose vectors in terms of components parallel to
the orbital angular momentum, which we will denote with the subscript
$\|$, and components in the orbital plane, which we will denote with the
subscript $\perp$.

We have adopted Taylor-like expansion formulas \cite{Boyle:2007ru}
to model the remnant black holes
mass and spin~\cite{Lousto:2013wta} and recoil~\cite{Lousto:2012gt}.
In the approach above, we considered polynomial formulas
in powers of the spin parameters only.

In this paper we generalize the
fitting formulas in \cite{Lousto:2012gt,Lousto:2013wta}
for unequal, but comparable,
mass binaries. To do this, we  consider the expansion variable
$\delta{m}$ to be  on the same footing as the spin variables.

A Taylor expansion of a function with $v$ independent variables
of a given order of expansion $o$ has
$n$ terms, where $n$ is given by
\cite{2005mmp..book.....A}
\begin{equation}\label{eq:terms}
n=\frac{(o+v-1)!}{o!\,(v-1)!}.
\end{equation}
However, only certain combinations of variables are allowed due to
symmetries of both the remnant quantity to be modeled and the binary
parameters entering the model. The two key symmetry operations are
parity ($x\to-x$, $y\to-y$, $z\to -z$) and exchange of labels
$1\leftrightarrow2$ for the two BHs. These symmetry properties are
summarized in Table~\ref{table:symmetries}.
\begin{table}
\caption{Symmetry properties of key quantities under parity (P) and
exchange of labels (X). Note that $\vec S_0$ has the same symmetries
as $\vec S$.}
\label{table:symmetries}
\begin{ruledtabular}
\begin{tabular}{lcc}
Quantity & P & X \\
\hline\hline
$S_\perp/m^2=(S_1+S_2)_\perp/m^2$ & -- & -- \\
$S_\|/m^2=(S_1+S_2)_\|/m^2$ & + & + \\
$\Delta_\perp/m^2=(S_2/M_2-S_1/M_1)_\perp/m$ & -- & + \\
$\Delta_\|/m^2=(S_2/M_2-S_1/M_1)_\|/m$ & + & -- \\
$\hat{n}=\hat{r}_1-\hat{r}_2$ & + & -- \\
$\delta m=(M_1-M_2)/m$ & + & -- \\
$V_\perp$ & + & -- \\
$V_\|$ & -- & + \\
$J_\perp/m^2$ & -- & -- \\
$J_\|/m^2$ & + & + \\
$M_{rem}/m$ & + & + \\
\end{tabular}
\end{ruledtabular}
\end{table}

Our particular expansion functions for the recoil are
summarized in Tables~\ref{table:Vz}~and~\ref{table:Vp}.
Note that each term in these tables is multiplied by a fitting
constant. The total number of terms for the expansion of
the recoil, and a comparison to a generic Taylor expansion,
is given in Table~\ref{table:numbertermsV}.

Despite the symmetries, which reduce the total number of terms in
the two components of the recoil by a factor of $\approx 4$ compared
to the generic Taylor expansion,
there still are many parameters to fit and aliasing
can lead to large statistical uncertainty in the values of the fitting
constants. To partially overcome this,
we  use a hierarchical procedure where we fit the full set of
coefficients and then reduce the number of fitting constants by
setting all constants with
large statistical errors in the original fit to zero.
The fit is repeated and again the constants with the largest
statistical
uncertainties are set to zero. This procedure is repeated until the
remaining constants have acceptable statistical uncertainties (in
practice we demand that the uncertainty in a constant is less than
half its absolute value).

\begin{table}
\caption{Parameter dependence at each order of expansion
 for the out-of-plane recoil.
}
\label{table:Vz}
\begin{ruledtabular}
\begin{tabular}{cl}
Order & Terms in $V_\|$\\
\hline
0th& 0\\
\\
1st& $\Delta_\perp$\\
\\
2nd&
$\Delta_\perp.S_\|+\Delta_\|.S_\perp$\\
 & $+\delta m (S_\perp) $\\
\\
3rd&
$\Delta_\|.S_\perp.S_\|+\Delta_\perp.S_\|^2+\Delta_\perp.\Delta_\|^2+\Delta_\perp^3+\Delta_\perp.S_\perp^2$\\
&  $+\delta m\left(\Delta_\perp.\Delta_\|+S_\perp.S_\|\right)$\\
& $+\delta m^2(\Delta_\perp)$\\
\\
4th&
 $S_\perp.\Delta_\|^3+\Delta_\perp.S_\|^3+\Delta_\perp.S_\|.\Delta_\|^2+S_\perp.\Delta_\|.S_\|^2$\\
& $+\Delta_\perp^3.S_\|+S_\perp^3.\Delta_\|+\Delta_\perp^2.S_\perp.\Delta_\|+\Delta_\perp.S_\perp^2.S_\|$\\
& $+\delta
m(S_\perp.\Delta_\|^2+S_\perp.S_\|^2+\Delta_\perp.\Delta_\|.S_\|
+S_\perp.\Delta_\perp^2+S_\perp^3)$\\
& $+\delta m^2(\Delta_\perp.S_\|+\Delta_\|.S_\perp)$\\
& $+\delta m^3(S_\perp)$\\
\end{tabular}
\end{ruledtabular}
\end{table}

\begin{table}
\caption{Parameter dependence at each order of expansion for the
in-plane recoil.}
\label{table:Vp}
\begin{ruledtabular}
\begin{tabular}{cl}
Order & Terms in $V_\perp$\\
\hline
0th & $0$ \\
\\
1st & $ \Delta_\|$\\
 & + $\delta m$ \\
\\
2nd & $\Delta_\|.S_\|+\Delta_\perp.S_\perp$\\
 & $+\delta m$ ($S_\|)$\\
\\
3rd &  $\Delta_\perp.S_\perp.S_\|+\Delta_\|.S_\|^2+\Delta_\|.\Delta_\perp^2+\Delta_\|^3+\Delta_\|.S_\perp^2$\\
& $+\delta m (\Delta_\|^2+S_\|^2+\Delta_\perp^2+S_\perp^2)$\\
& $+\delta m^2(\Delta_\|)$\\
& $+\delta m^3$ \\
\\
4th& $S_\perp.\Delta_\perp^3+\Delta_\|.S_\|^3+\Delta_\|.S_\|.\Delta_\perp^2+S_\perp.\Delta_\perp.S_\|^2$\\
& $+\Delta_\|^3.S_\|+S_\perp^3.\Delta_\perp+\Delta_\|^2.S_\perp.\Delta_\perp+\Delta_\|.S_\perp^2.S_\|$\\
& $+\delta m
(S_\|.\Delta_\|^2+S_\|.S_\perp^2+\Delta_\perp.S_\perp.\Delta_\|+S_\|.\Delta_\perp^2+S_\|^3)$\\
& $+\delta m^2(\Delta_\|.S_\|+\Delta_\perp.S_\perp)$\\
& $+\delta m^3(S_\|)$\\
\end{tabular}
\end{ruledtabular}
\end{table}

\begin{table}
\caption{Number of possible terms at a given order of expansion (with
respect to $\vec S$ or $\vec \Delta$ and  $\delta m$)}
\label{table:numbertermsV}
\begin{tabular}{cccccccc}
\hline\hline
Order & 0th & 1st & 2nd & 3rd & 4th & 5th & 6th \\
\hline
$V_\perp$ & 0 & 2 & 3 & 11 & 16 & 36 & 50\\
$V_\|$ & 0 & 1 & 3 & 8 & 16 & 30 & 50\\
Total & 0 & 3 & 6 & 19 & 32 & 66 & 100\\
Taylor & 1 & 5 & 15 & 35 & 70 & 126 & 210\\
Difference & -1 & -2 & -9 & -16 & -38 & -60 & -110\\
\hline\hline
\end{tabular}
\end{table}

Our particular expansion functions for the remnant spin are
summarized in Tables~\ref{table:Jzm}~and~\ref{table:Jp}.
Note that each term in these tables is multiplied by a fitting
constant. The total number of terms for the expansion of
the remnant spin, and a comparison to a generic Taylor expansion,
is given in Table~\ref{table:numbertermsJ}.

Note that the combined number of terms in the expansions of
the two components of $\vec V$ and $\vec J$ at any given order
matches the total number of terms in the Taylor expansion
for a scalar function with no symmetries.

\begin{table}
\caption{Parameter dependence at each order of expansion
 for the final spin component perpendicular to the reference
 $\vec{L}$ direction.
}
\label{table:Jp}
\begin{ruledtabular}
\begin{tabular}{cl}
Order & Terms in $J_\perp$\\
\hline
0th & $0$ \\
\\
1st & $S_\perp$\\
\\
2nd & $S_\perp.S_\|+\Delta_\|.\Delta_\perp$\\
& $+\delta m(\Delta_\perp)$\\
\\
3rd &  $\Delta_\|.\Delta_\perp.S_\|+S_\perp.S_\|^2+S_\perp.\Delta_\|^2+S_\perp^3+S_\perp.\Delta_\perp^2$\\
& $+\delta m (S_\perp.\Delta_\|+\Delta_\perp.S_\|)$\\
& $+\delta m^2(S_\perp)$\\
\\
4th & $\Delta_\perp.\Delta_\|^3+S_\perp.S_\|^3+S_\perp.S_\|.\Delta_\|^2+\Delta_\perp.\Delta_\|.S_\|^2$\\
&$+S_\perp^3.S_\|+\Delta_\perp^3.\Delta_\|+S_\perp^2.\Delta_\perp.\Delta_\|+S_\perp.\Delta_\perp^2.S_\|$\\
& $+\delta
m(\Delta_\perp.\Delta_\|^2+\Delta_\perp.S_\|^2+S_\perp.\Delta_\|.S_\|+\Delta_\perp.S_\perp^2+\Delta_\perp^3)$\\
& $+\delta m^2(S_\perp.S_\|+\Delta_\|.\Delta_\perp)$\\
& $+\delta m^3(\Delta_\perp)$\\
\end{tabular}
\end{ruledtabular}
\end{table}

\begin{table}
\caption{Parameter dependence at each order of expansion
for the final spin component along the reference
$\vec{L}$ direction and similarly for the remnant mass $M_{\rm rem}$
(or, equivalently, the mass loss of the binary $\delta {\cal M}$).
}
\label{table:Jzm}
\begin{ruledtabular}
\begin{tabular}{cl}
Order & Terms in $J_\|$ or $M_{\rm rem}$\\
\hline
0th & $L(S=0,\delta{m}=0)$ or $M(S=0,\delta{m}=0)$ \\
\\
1st & $S_\|$\\
    & $+\delta m$ \\
\\
2nd & $ \Delta_\|^2+S_\|^2+\Delta_\perp^2+S_\perp^2$\\
& +$\delta m (\Delta_\|)$\\
& $+\delta m^2$ \\
\\
3rd & $S_\|.\Delta_\|^2+S_\|.S_\perp^2+\Delta_\perp.S_\perp.\Delta_\|+S_\|.\Delta_\perp^2+S_\|^3$\\
& $+\delta m(\Delta_\|.S_\|+\Delta_\perp.S_\perp)$\\
& $+\delta m^2(S_\|)$\\
\\
4th &  $\Delta_\perp.\Delta_\|.S_\perp.S_\|+\Delta_\perp^4+\Delta_\|^4+S_\perp^4+S_\|^4+\Delta_\perp^2.\Delta_\|^2$\\
 & $+\Delta_\perp^2.S_\perp^2+\Delta_\perp^2.S_\|^2+\Delta_\|^2.S_\perp^2+\Delta_\|^2.S_\|^2+S_\perp^2.S_\|^2$\\
& $+\delta m
(\Delta_\perp.S_\perp.S_\|+\Delta_\|.S_\|^2+\Delta_\|.\Delta_\perp^2+\Delta_\|^3+\Delta_\|.S_\perp^2)$\\
& $+\delta m^2(\Delta_\|^2+S_\|^2+\Delta_\perp^2+S_\perp^2)$\\
&$+\delta m^3(\Delta_\|)$\\
& $+\delta m^4$\\
\end{tabular}
\end{ruledtabular}
\end{table}

\begin{table}
\caption{Number of possible terms at a given order of expansion (with
respect to $\vec S$ or $\vec \Delta$ and  $\delta m$}
\label{table:numbertermsJ}
\begin{tabular}{cccccccc}
\hline\hline
Order & 0th & 1st & 2nd & 3rd & 4th & 5th & 6th \\
\hline
$J_\perp$ & 0 & 1 & 3 & 8 & 16 & 30 & 50\\
$J_\|$ & 1 & 1 & 6 & 8 & 22 & 30 & 60\\
Total & 1 & 2 & 9 & 16 & 38 & 60 & 110\\
Taylor & 1 & 5 & 15 & 35 & 70 & 126 &210\\
Difference & 0 & -3 & -6 & -19 & -32 & -66 & -100 \\
\hline\hline
\end{tabular}
\end{table}

The expansion of the radiated mass will have an identical set
of terms to the expansion of $J_\|$ (see Table
\ref{table:numbertermsm}).
We have found that in practice this expansion (up through
fourth-order) provides an accurate
description (see~\cite{Healy:2014yta}) although alternative Pad\'e approximant expressions
are also possible as in Ref.~\cite{Hemberger:2013hsa}.

\begin{table}
\caption{Number of possible terms at a given order of expansion [with
respect to $\vec S$ or $\vec \Delta$ and  $\delta m$ for the final
mass ($M_{\rm rem}$)].}
\label{table:numbertermsm}
\begin{tabular}{cccccccc}
\hline\hline
Order & 0th & 1st & 2nd & 3rd & 4th & 5th & 6th \\
\hline
$M_{\rm rem}/m$ & 1 & 1 & 6 & 8 & 22 & 30 & 60\\
Total & 1 & 1 & 6 & 8 & 22 & 30 & 60\\
Taylor & 1 & 5 & 15 & 35 & 70 & 126 & 210\\
Difference & 0 & -4 & -9 & -27 & -48 & -96 & -150\\
\hline\hline
\end{tabular}
\end{table}

\section{Fits}
\label{sec:Fits}

In this section we fit for the total radiated mass and remnant spin and recoil
as a function of the spins of the binary at merger. Our expansion
variables, $ \vec {\tilde S}$ (or $ \vec {\tilde S}_0$), $ \vec
{\tilde \Delta}$, and $ \delta m$ are all measured with respect to
the {\it final} orbital plane (see Sec.~\ref{sec:numerical_kicks} and
Fig.~\ref{fig:find_plane}). For consistency with the particle limit,
we also include explicit dependence on $\eta=(1-\delta
m^2)/4$.

\subsection{Fitting the Recoil}
\label{sec:fit-recoil}

Before modeling the mass-ratio dependence of the recoil we will
reexamine the \cross of Ref.~\cite{Lousto:2012gt}.
As we noted there, the recoil should take the form
\begin{eqnarray}
V_\| &=& a_0 \vec{\tilde\Delta}\cdot\hat n_0 + a_1
  \vec{\tilde\Delta}\cdot\hat n_1 \left(2 S_\|\right)  + \cdots
\nonumber \\
&& + b_0 \left(2\vec{\tilde S}\right)\cdot\hat m_0 \tilde\Delta_\| + b_1
\left(2\vec{\tilde S}\right)\cdot\hat m_1 \tilde\Delta_\| \left(2
  S_\|\right) +
\cdots,
\end{eqnarray}
where the unit vectors $\hat n_i$ and $\hat m_i$ are all in the
orbital plane and need not be aligned
in any way. As a simplifying assumption, we fit the data assuming all
these unit vectors were aligned. While the fit for the N
configurations was quite good, we were not able to model the K
configurations with the same accuracy. The K configurations started
out with nontrivial $S_\perp$ and $\Delta_\|$ while having
$\Delta_\perp=0$ and $S_\|=0$, identically. However, these evolved to
configurations with nontrivial $\Delta_\perp$. The spin directions
and recoils for the K configurations are given in
Table~\ref{tab:rot_part1}.

Here we revisit the fitting of the equal-mass N and K configurations by assuming
that $\hat n_0 = \hat n_1 = \hat n_2 \cdots$, $\hat m_0 =
\hat m_1 = \hat m_2 \cdots$, and that $\hat n_0$ and $\hat m_0$ are not aligned.
Our procedure is as follows. We assume that the angle between $\hat
n_0$ and $\hat m_0$ is some given value, which we will denote by
$\zeta$. The expression for the maximum over azimuthal configurations
$\varphi$ of the recoil
for the N configurations (equal mass only, see Fig.~\ref{fig:NConfig}) then takes on the form
$V_{\|max}^2 = V_{hang}^2+ V_{cross}^2 + 2 V_{hang} V_{cross} \cos
\zeta$, where $V_{hang} = \tilde \Delta_\perp (h_1 + h_2 \left(2
\tilde S_\|\right) +
h_3 \left(2 \tilde S_\|\right)^2 + \cdots)$
and $V_{cross} = \left(2 \tilde S_\perp\right) \tilde \Delta_\|
(\sigma_1 + \sigma_2 \left(2\tilde
S_\|\right) +\sigma_3 \left( 2\tilde S_\|\right)^2 + \cdots)$.
Here $\tilde \Delta_\perp$ and $\tilde S_\perp$ are understood to be the magnitudes
of the projections of these two vectors in the plane. In practice,
we take the coefficients $(h_1, h_2, h_3, \cdots)$ from the expression
for the \hangup in~\cite{Lousto:2011kp}
and only fit to the coefficients $(\sigma_1,\sigma_2)$ (we take
$\sigma_3$ and higher
coefficients to be zero).

\begin{figure}
  \includegraphics[width=.9\columnwidth]{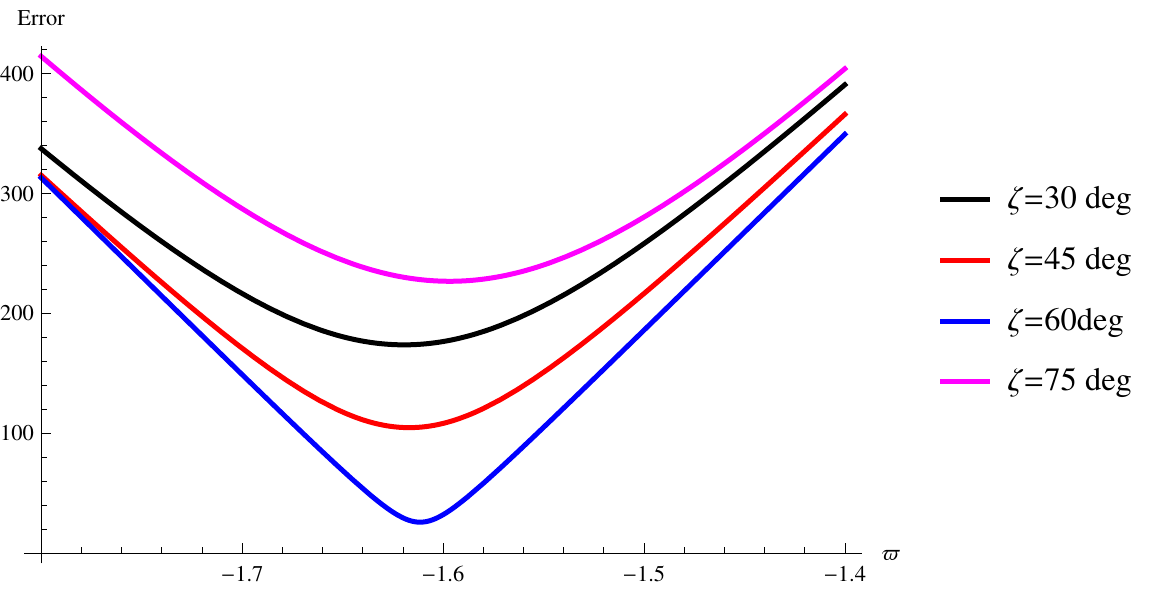}
  \caption{The RMS error in the prediction of the recoil for the
  K45 configurations as a function of $\varpi$ (the angle between the
  unit vector $\hat {n}_0$ and the $x$ axis of the rotated basis) for
several choices of $\zeta$. Note that $\varpi$ is measured in radians.  }
  \label{fig:Kfit}
\end{figure}

Once we have $\sigma_1$ and $\sigma_2$ for a given $\zeta$, we predict the
recoil for the K configurations. The prediction takes on the form
\begin{eqnarray}
  V_\| &=&  \vec{\tilde\Delta}\cdot \hat n_0 (h_1 + h_2 \left(2 \tilde
S_\|\right) +
h_3 \left(2 \tilde S_\|\right)^2 + \cdots) \nonumber\\
&& + \left[\left( 2\vec{\tilde S}\right)\cdot {\bf R}(\zeta)\hat
n_0\right] \tilde \Delta_\|
\left(\sigma_1 + \sigma_2 \left(2\tilde S_\|\right)\right),
\end{eqnarray}
where $\hat m_0 = {\bf R}(\zeta)\hat n_0$ is a unit vector in the
orbital plane rotated by
angle $\zeta$ from
$\hat n_0$. The remaining complication arises because we do not know the
direction of $\hat n_0$ with respect to the rotated frame where the
spins of the K configurations are given. To find this direction, we
take $\hat n_0 = ( \cos \varpi, \sin \varpi)$. The predicted recoil for
a given K configuration will then depend on the actual in-plane
components of the spins for that configuration and the angle $\varpi$.
We then find the value of $\varpi$ that minimizes the sum
$$
  \sum_{\varphi configs}\left(V_{\rm pred}(\hat m) - V_{\rm meas}\right)^2.
$$
The minimum over $\varpi$ of the sum is itself a function of
$\zeta$.
Finally, we adjust $\zeta$ until we find an absolute minimum.
This procedure is illustrated in Fig.~\ref{fig:Kfit}.

Interestingly, we
find that the error is minimized for both the K45 and K22.5
families by a single $\zeta$ value of about $-59^\circ$.
For example, the RMS error in the recoil
for the K45
configurations assuming that $\hat n_0 = \hat m_0$ (i.e., $\zeta=0$)
 is $275.3\ \KMS$,
while assuming the angle between $\hat n_0$ and $\hat m_0$ is  $-59^\circ$ gives an RMS error of
$25.2\ \KMS$ with a maximum recoil of $2234\pm12\ \KMS$.
Similarly, the RMS error in the prediction for the recoil of the K22.5
configurations is $48.9 \KMS$ with a maximum recoil of
$1731\pm25\ \KMS$ (the RMS errors is $253.4 \KMS$ if we assume
$\hat n_0 = \hat m_0$).

With this new fitting, the maximum recoil (over azimuthal
configurations) for a given polar configuration is given by
\begin{eqnarray}
V_\|^2& =& V_{\rm hang}^2 + V_{\rm cross}^2 + 2 V_{\rm hang}
  V_{\rm cross} \cos(59\pi/180), \label{eq:equal_mass_first}
\end{eqnarray}
where
\begin{eqnarray}
V_{\rm hang}&=&\tilde \Delta_\perp
 \left(3678+
2481 \left(2 \tilde S_\|\right)\right. \nonumber \\
&& \left.+ 1792 \left(2 \tilde S_\|\right)^2+  1507
\left(2 \tilde S_\|\right)^3\right),\\
V_{\rm cross} &=&
\left(2 \tilde S_\perp\right)
\tilde \Delta_\| \left(2160 + 3990 \left(2  \tilde S_\|\right)\right).
\end{eqnarray}
The errors in the last two coefficients are $2160\pm204$ and $3990\pm
680$.
If, however, we assume the angle $\zeta$ is zero, we get
\begin{eqnarray}
V_\|^2& =& V_{\rm hang}^2 + V_{\rm cross}^2 + 2 V_{\rm hang}
V_{\rm cross}
\end{eqnarray}
where
\begin{eqnarray}
V_{\rm hang}&=&\tilde \Delta_\perp
 \left(3678+
2481 \left(2 \tilde S_\|\right)\right. \nonumber \\
&& \left.+ 1792 \left(2 \tilde S_\|\right)^2+  1507
\left(2 \tilde S_\|\right)^3\right),\\
V_{\rm cross} &=&
\left(2 \tilde S_\perp\right)
\tilde \Delta_\| \left(1200 + 2550 \left(2  \tilde
S_\|\right)\right).\label{eq:equal_mass_last}
\end{eqnarray}
The errors in the last two coefficients are $1200\pm99$ and $2550\pm
340$.

We now move to the more general case of general mass ratios by
extending formulas(\ref{eq:equal_mass_first})-(\ref{eq:equal_mass_last})
to include terms proportional
to $\delta m$.

Simply adding all possible unequal mass corrections to the recoil
formula, even at low
order, is fraught with difficulty because of the sheer number of
terms (and hence the correspondingly large number of runs required).
Here we will settle on a compromise formula. One that is {\it accurate
enough} in a given mass ratio range (here $1/8 \lesssim q \leq 1$).

Our procedure is as follows. We fit each family of fixed mass ratio
and polar inclination angle to the form
\begin{equation}
V_{\rm kick} = V_{1} \cos(\varphi-\phi_1) +
   V_{3} \cos(3 \varphi-3 \phi_3),
\end{equation}
where $V_{1}$, $V_{3}$, $\phi_1$, and $\phi_3$ are fitting coefficients and
$\varphi$ is the angle (at merger) between $\vec \Delta_\perp$ for a
given PHzzz configuration and the corresponding PH0 configuration.
Our tests indicate that $V_1$ can be obtained accurately with six
choices of
the initial $\phi_i$ angles. These fitting parameters for each of the
NQ families are given in Table~\ref{tab:fit_NQ}.

\begin{table*}
\caption{Fitting parameters for the NQ families of configurations as a
function of $\varphi$ (see text) to the form
$V_\| = V_1 \cos((\varphi - \phi_1)\pi/180) + V_3 \cos(3(\varphi -
\phi_3)\pi/180)$. All angles are measured in degrees.}
\label{tab:fit_NQ}
\begin{ruledtabular}
\begin{tabular}{l|D{,}{\pm}{1}D{,}{\,\pm\,}{1}D{,}{\,\pm\,}{1}D{,}{\,\pm\,}{1}D{,}{\,\pm\,}{1}l}
Family & V_{1} & V_{3} & \phi_1 & \phi_3 & \mbox{RMS Err}\\
 \hline
NQ200TH30 & 390.27,0.59 & 11.85,0.57 &347.412,0.083 & 260.87,0.92 & 0.57\\
NQ200TH60 & 643.7,8.5 & 12.5,8.2 & 282.30,0.72 & 320,12 & 8.24\\
NQ200TH90 & 700.6,1.3 & 2.4,1.3 & 326.312,0.098 & 101,10 & 1.22\\
NQ200TH135 & 455.81,0.59 & 5.60,0.57 & 145.008,0.070 & 114.2,2.0 &  0.57 \\
NQ66TH60 & 1882,11 & 24,12 & 3.93,0.38 & 0.9,9.3 & 11.56\\
NQ50TH30 & 1313,18 & 65,16 & 309.29,0.70 & 241.8,5.0 & 16.34\\
NQ50TH60 & 1876,22 & 94,22 & 170.31,0.65 & 250.9,4.2 & 21.03\\
NQ50TH90 & 1720.3,7.9  & 89.3,9.2 & 29.32,0.30 & 353.4,1.7 & 7.13 \\
NQ50TH135 &  865.2,1.3 & 28.3,1.2 & 249.072,0.088 & 93.23,0.94 & 1.23 \\
NQ33TH45 &  1333.5,6.2 & 114.8,7.0 & 158.62,0.27 & 62.73,0.93 & 5.26 \\
NQ33TH75 & 1505.4,3.2 & 62.9,1.7 & 270.658,0.072 & 180.4,1.6 & 1.93 \\
NQ33TH100 & 1222.0,2.1 & 50.7,4.8 & 10.36,0.12 & 337.41,0.62 &  1.57 \\
NQ33TH135 & 632.88,0.21 & 9.33,0.24 & 221.306,0.024 & 72.97,0.47 & 0.21 \\
NQ25TH30 & 767.5,2.2 & 71.9,2.4 & 88.99,0.19 & 232.72,0.61 & 2.13 \\
NQ25TH60 & 1183.1,2.6 & 70.4,2.0 & 94.22,0.11 & 243.81,0.81 & 2.90 \\
NQ25TH90 & 1035.7,1.9 & 32.1,1.4 & 97.983,0.052 & 6.62,0.54 & 0.56 \\
NQ25TH135 & 454.58,0.54 & 5.56,0.75 & 195.54,0.09  & 404.4,1.9 & 0.54 \\
NQ25TH150 & 277.72,0.91 & 8.3,1.1 & 23.19,0.23 & 130.2,2.2 & 0.97 \\
NQ16TH45 & 627.0,5.2 & 65.6,6.8 & 124.85,0.27 & 35.2,1.1 & 2.37 \\
NQ16TH90 & 657,10 & 29.1,3.2 & 242.871,.093 & 149.4,3.7 & 1.18 \\
NQ16TH115 & 419.84,0.98 & 14.59,0.56 & 192.68,0.19 & 338.0,1.2 & 0.38 \\
NQ16TH135 & 253.4,1.4 & 6.9,3.1 & 277.89,0.38 & 27.0,4.0 & 0.90 \\
NQ16TH150 & 154.11,0.10 & 3.048,0.082 & 318.774,0.031 & 87.57,0.60 & 0.084 \\
\end{tabular}
\end{ruledtabular}
\end{table*}

We then model $V_{1}$ as a function of $S_\|$, $S_\perp$, $\Delta_\|$,
$\Delta_\perp$, and $\delta m$ using terms up
through fourth order in the expansion variables. However,
because we only consider contributions linear in $\cos\varphi$,
only those terms in Table~\ref{table:Vz} that are linear in the perpendicular
components of the spins enter the fit. A fit to this reduced form
still
leads to poor statistics for the fitting constants. We then
selectively remove the most poorly fit constants (i.e., set them to
zero) and refit. 
This process is repeated until a satisfactory fit is
obtained with the fewest number of free parameters. In particular, we remove only one parameter at a
time (always the one with the largest relative uncertainty). We stop removing
parameters when all the remaining coefficients have uncertainties
that are no larger in magnitude than 1/2 the value of the coefficient
itself.
Note that this
procedure does not lead to a unique minimal set of expansion terms.

\begin{figure*}
   \includegraphics[width=.48\textwidth]{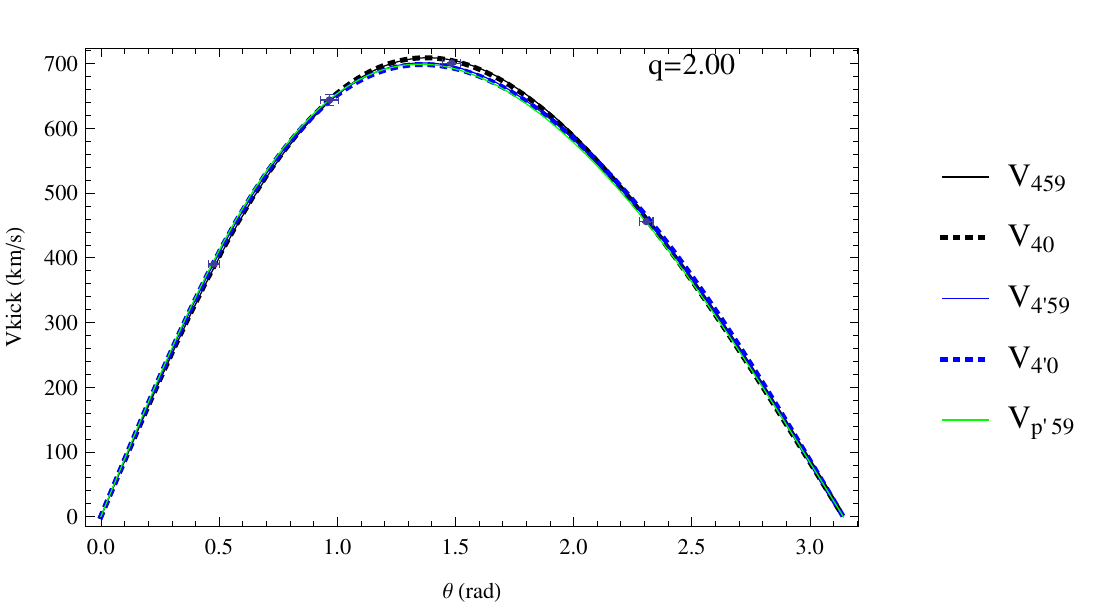}
   \includegraphics[width=.48\textwidth]{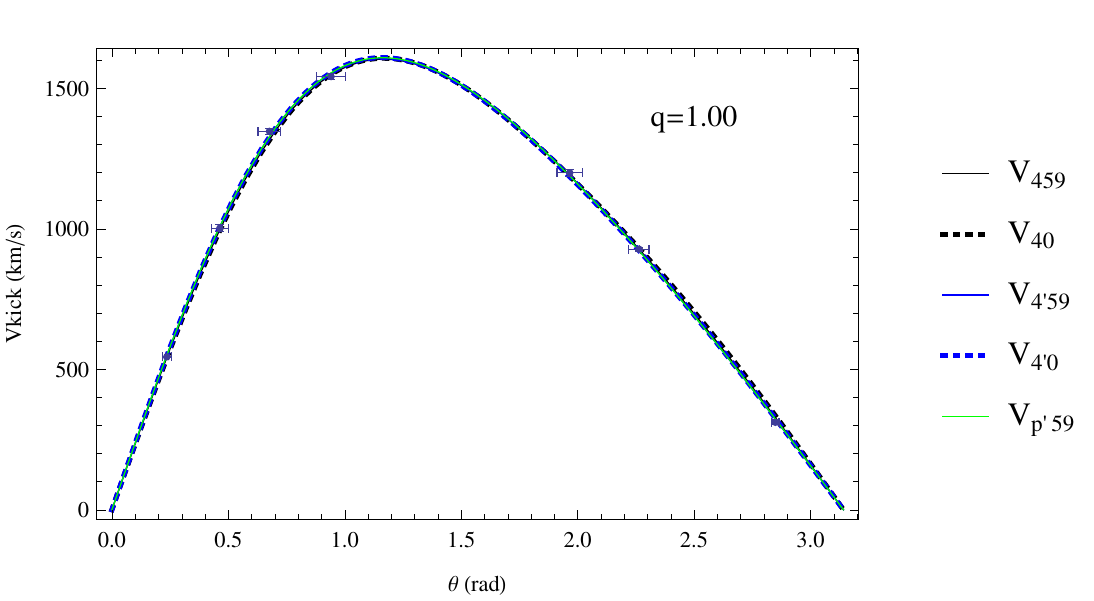}
   \includegraphics[width=.48\textwidth]{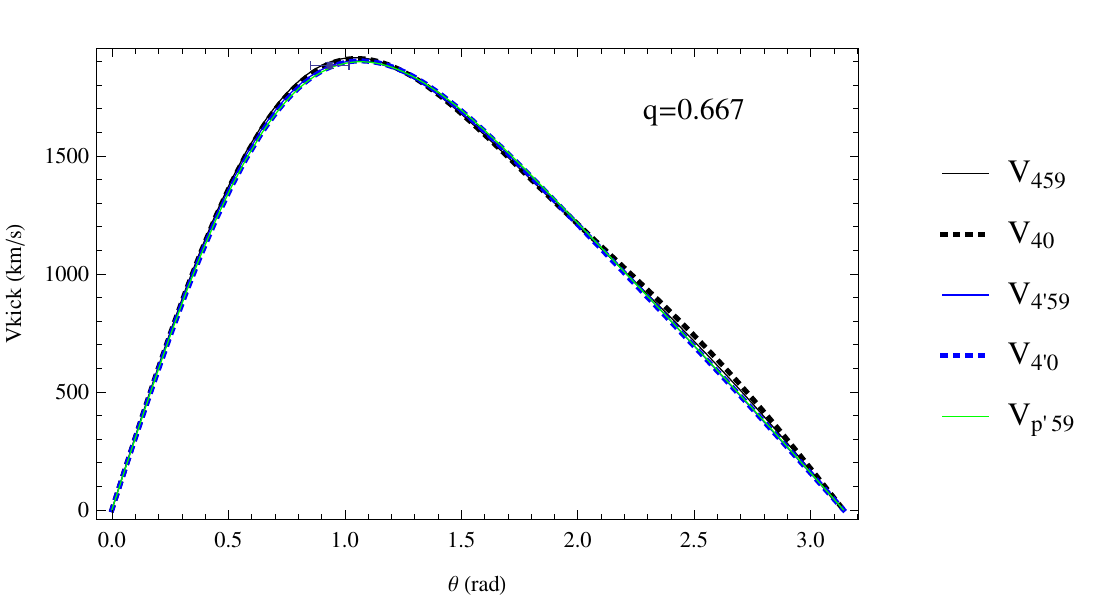}
   \includegraphics[width=.48\textwidth]{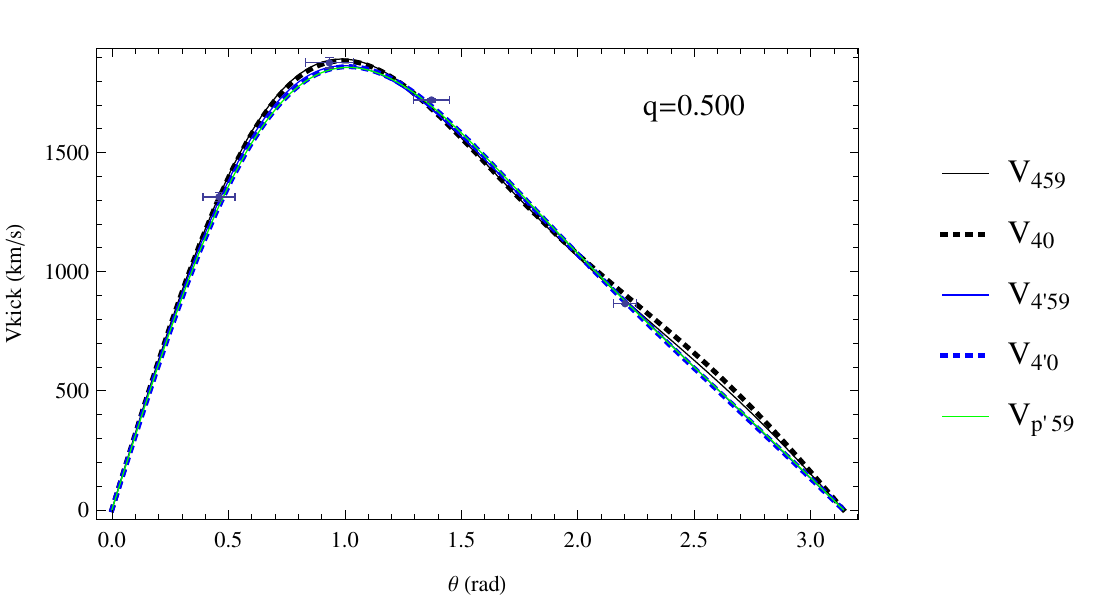}
   \includegraphics[width=.48\textwidth]{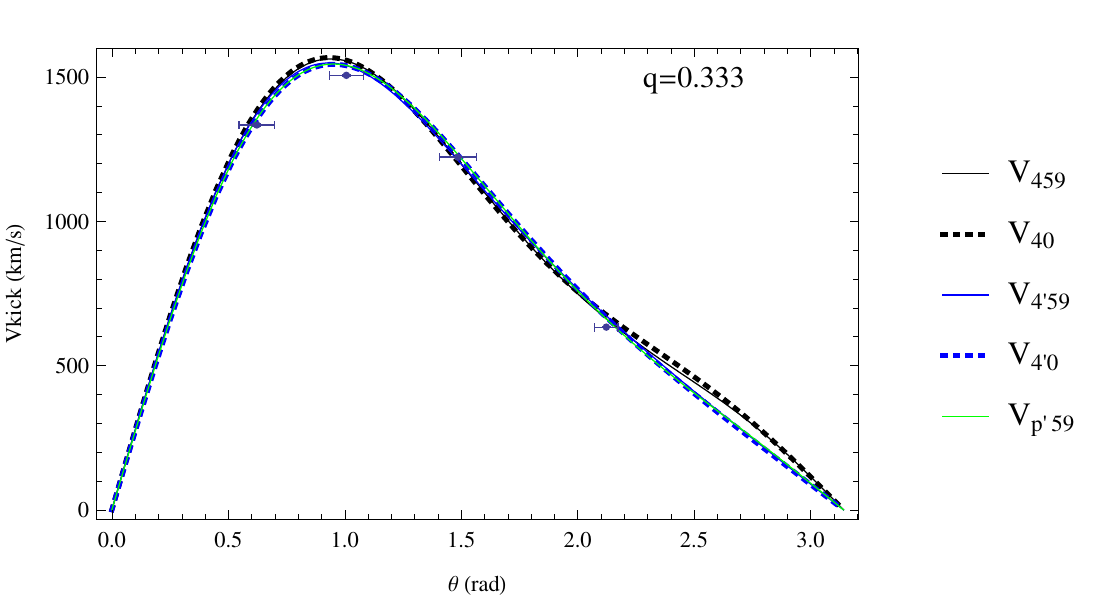}
   \includegraphics[width=.48\textwidth]{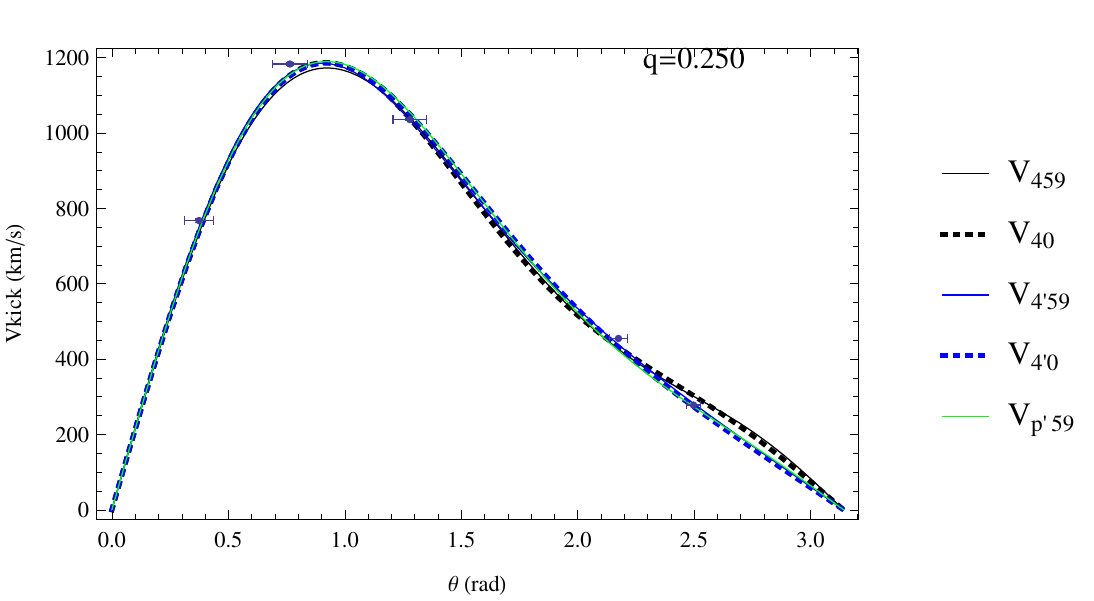}
   \includegraphics[width=.48\textwidth]{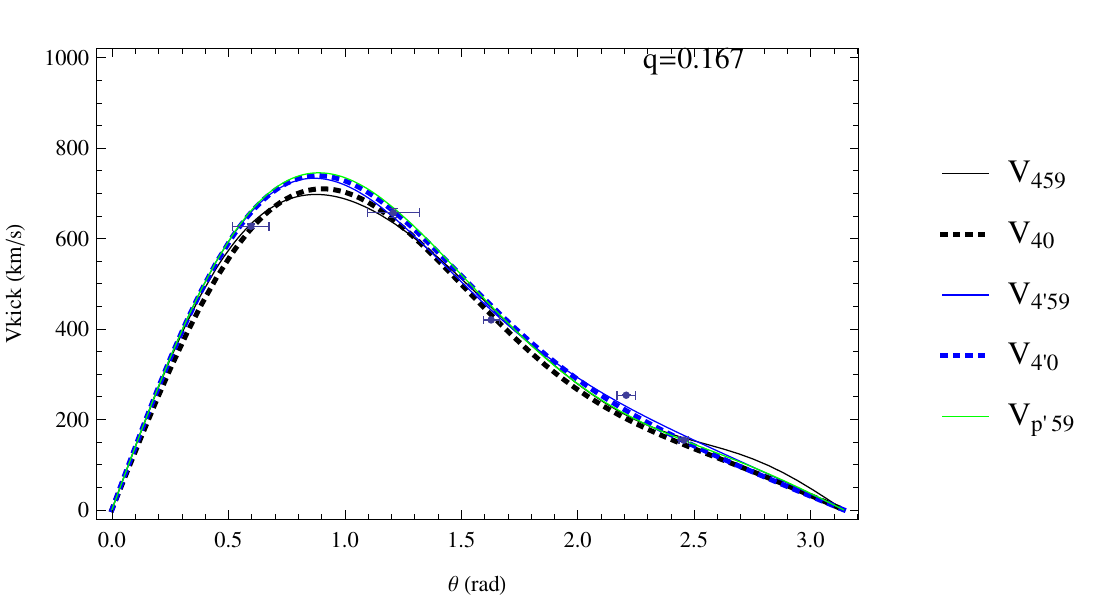}
   \includegraphics[width=.48\textwidth]{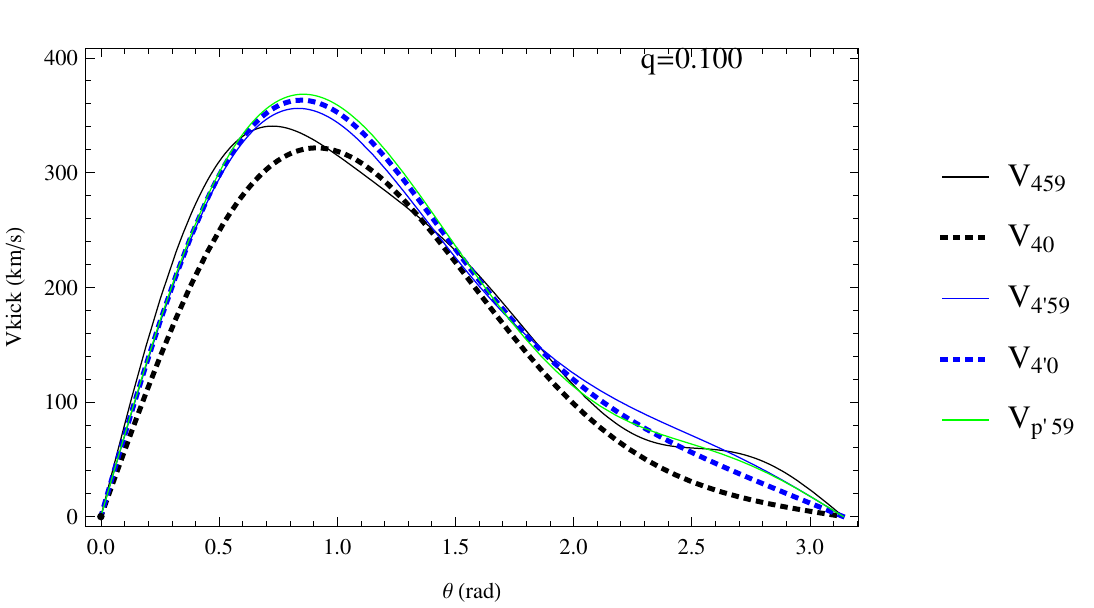}
  \caption{Plots of the fitted $V_1$ versus inclination angle $\theta$
    and $q$ for the NQ
configurations. Each data point represents the maximum of $V_1$ over a
family of azimuthal configurations with the same inclination angle and
mass ratio. The last plot shows an extrapolation to $q=1/10$.
}\label{fig:Vfits}
\end{figure*}

\begin{figure}
   \includegraphics[width=.9\columnwidth]{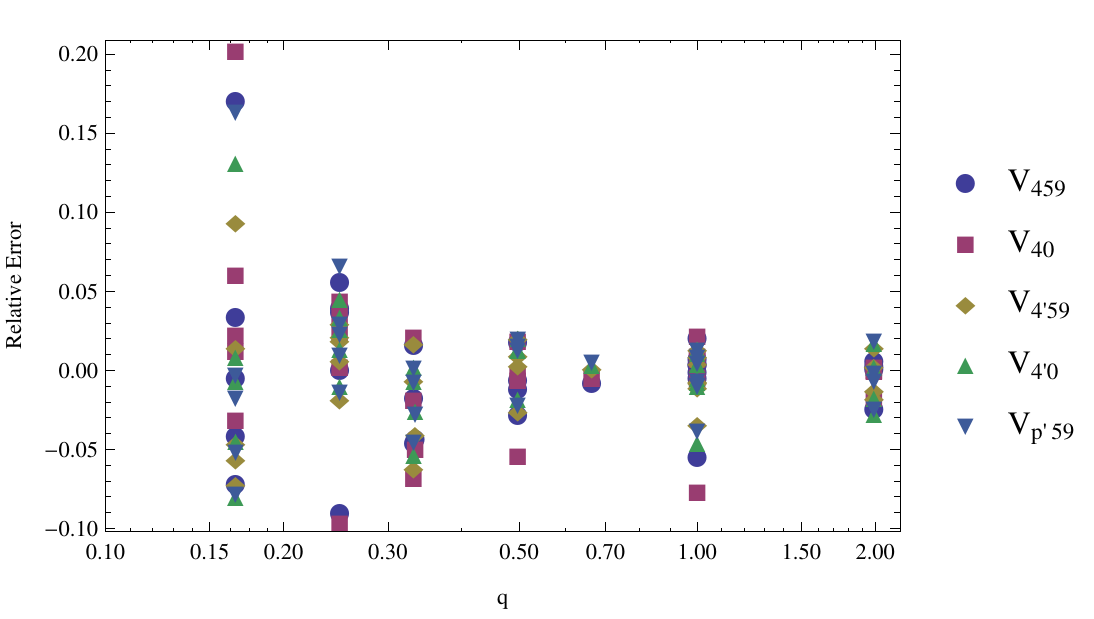}
   \includegraphics[width=.9\columnwidth]{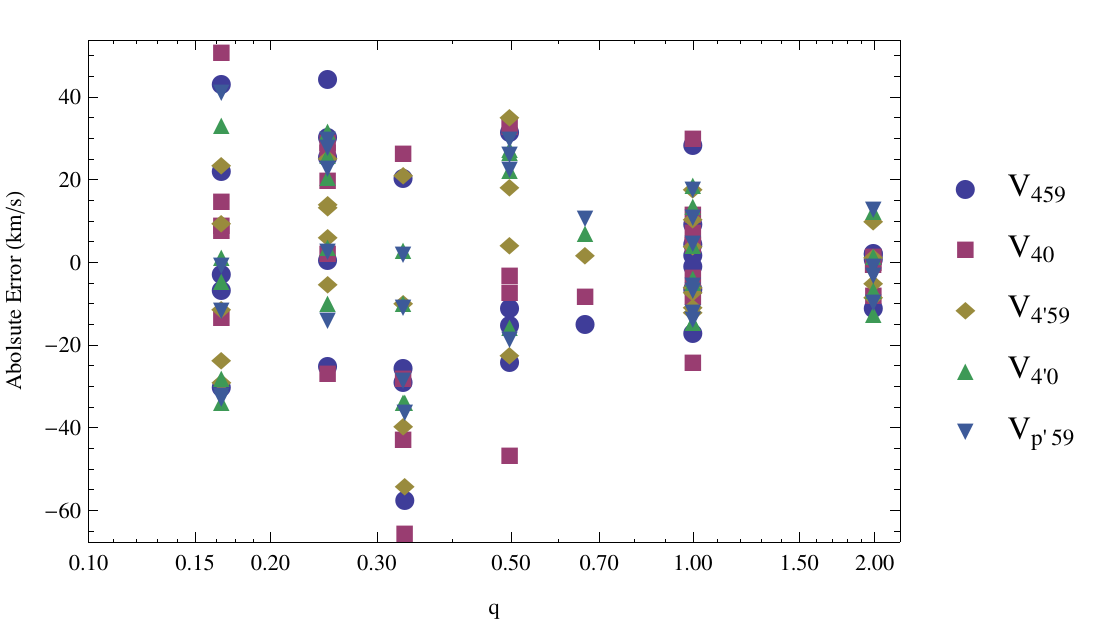}
  \caption{The relative errors (residuals) in the fit of $V_\phi$ versus $q$.
Note that there are multiple data points for each $q$.}\label{fig:Vresid}
\end{figure}

We fit the full set of unequal mass NQ configurations to the two forms
$V_{x0}$ and $V_{x59}$, where
\begin{eqnarray}
  V_{\rm h} &=& (4 \eta)^2\tilde \Delta_\perp (3678 (1 + c_1 \delta m^2) \nonumber \\
  && + 2481 (2 \tilde S_\|) (1 + c_2 \delta m^2)  + 1792 (2 \tilde S_\|)^2\nonumber \\
 &&  + 1507 (2 \tilde S_\|)^3 + c_{5} \tilde \Delta_\|^2 + \nonumber \\
 &&c_{7} \tilde \Delta_\|^2 \left(2 \tilde S_\|\right)+c_9 \delta m \tilde \Delta_\|),\label{eq:Vkick_final_1}\\
  V_{\rm c0 } &=&  (4\eta)^2 \left(2 \tilde S_\perp\right) \tilde \Delta_\| (1200 +
   c_{12} \delta m^2\nonumber \\
 && + 2550  (2 \tilde S_\|) + c_{15} \tilde \Delta_\|^2) \nonumber \\
&&+ (2 \tilde S_\perp)[c_{16} \delta m + c_{17} \delta m^3 \nonumber \\
&&+ c_{18} \delta m (2 \tilde S_\|) + c_{19} \delta m (2 \tilde S_\|)^2],\\
 V_{\rm c59} &=& (4 \eta)^2 (2 \tilde S_\perp \tilde \Delta_\|) (2160 + c_{12}
\delta m^2 \nonumber \\
 && + 3990  (2 \tilde S_\|) + c_{15} \tilde \Delta_\|^2)\nonumber \\
 &&+ (2 \tilde S_\perp)[c_{16} \delta m + c_{17} \delta m^3 \nonumber \\
 && + c_{18} \delta m (2 \tilde S_\|) + c_{19} \delta m (2 \tilde S_\|)^2],\\
 V_{x0} &=& V_{\rm h} + V_{\rm c0 },\\
 V_{x59} &=& \sqrt{V_{\rm h}^2+V_{\rm c59}^2 + 2 V_{\rm h}
V_{\rm c59} \cos(59\pi/180)}.\label{eq:Vkick_final_final}
\end{eqnarray}
Here $V_{x0}$ indicates a fit assuming the \cross and \hangup
are aligned and $V_{x59}$ assumes they are misaligned by $59^\circ$
(note that $4\eta = 1$ for the equal-mass case and that we have
assumed a leading $\eta^2$ dependence).
Finally $x=4$ indicates a standard fit that includes all terms up
through fourth-order, while $x=4'$ indicates that
again all terms up through fourth-order are used but $S_0$ replaces $S$ in the formula.
We report the fitting parameters in Table~\ref{tab:vkick_fit_param},
and we show the results of fits in Fig.~\ref{fig:Vfits}
and  Fig.~\ref{fig:Vresid}.
The root-mean-square errors in the fits are: $23\ \KMS$ for
$V_{459}$, $25\ \KMS$ for $V_{40}$, $20\ \KMS$ for $V_{4'59}$,
and $19 \KMS$ for $V_{4'0}$.

Examining Fig.~\ref{fig:Vfits}, we can see that quality of the fit changes with
mass ratio. Overall, $V_{4'0}$, $V_{4'59}$, and $V_{p'59}$ (discussed
below) appear to do best at small mass
ratios, at least for the large $\theta$ tail. For $q\leq1/3$, there is a
noticeable oscillation in the predicted recoil from $V_{459}$ and
$V_{40}$ at large $\theta$. On the other hand,
for $q=2$, $V_{40}$ and $V_{459}$ fit the data best with $V_{4'0}$ and
$V_{4'59}$ slightly underestimating the maximum recoil.
As shown in Fig.~\ref{fig:Vresid}, the relative errors in the predicted recoils
for all fitting functions are under $10\%$ for all but one configuration (where the error is
$15-20\%$). For $V_{4'59}$ the relative errors are all less than
10\%, while the absolute errors are
less than $55\ \KMS$ (less than $40\ \KMS$ for all but one
configuration).  Note that at extrapolations down to $q=1/10$,
there is reasonably good agreement between all fitting functions.
Based on the relative and absolute errors, the extrapolation to mass
ratios as small as $q=1/10$, and the fact that $S_{0\|}$ is
approximately conserved in post-Newtonian theory \cite{Racine:2008qv}
and in full numerical simulations
\cite{Lousto:2013vpa,Lousto:2013wta}, we conclude that $V_{4'59}$ has the
best overall performance.

Note that while $S_{0\|}$ is conserved, the other quantities entering
$V_{4'59}$ are not. Thus $V_{4'59}$ is still a function of the
binary's parameters near merger and not at infinite separation.

Motivated by the success of  $V_{4'59}$ in modeling the recoil, we
also reexamined the Pad\'e approximation for the \hangup kick formula
we proposed in~\cite{Lousto:2011kp}. The Pad\'e approximation has the form
\begin{equation}
  V_{\rm hang (pade)} =  \tilde \Delta_\perp 3684.73
 \left(\frac{1 + 0.0705104 (2 \tilde S_\|)}{1 - 0.623831 (2 \tilde
S_\|) }\right),
\label{eq:pade}
\end{equation}
which has pole when $\tilde S_\|\approx0.8015$. This pole can only be
reached for mass ratios smaller than $q=1/8$. However,
by replacing $S_\|$ with $S_{0\|}$ in Eq.~(\ref{eq:pade}), there is no
pole for any physically allowed values for the spins. We were thus
able to fit ($V_1$) the recoil to the form
\begin{equation}
V_{p'59} = \sqrt{V_{h}^2 + V_{k}^2 + 2 V_{h} V_{k} \cos(59\pi/180)},
\label{eq:padeall}
\end{equation}
where
\begin{widetext}
\begin{eqnarray}
V_{h} &=& (4 \eta)^2 \tilde \Delta_\perp \left[3684.73\left(\frac{1 + c_1 \delta m^2 +
        0.0705104 (2 \tilde S_{0\|}) (1 +  c_2 \delta m^2)}{1 -
0.623831 (2 \tilde S_{0\|})}\right) \
    +  c_5 \tilde \Delta_\|^2  + c_7 \Delta_\|^2 (2 \tilde S_{0\|})
\right]\nonumber\\
 &&+ (4 \eta)^2 (c_9 \delta m \tilde \Delta_\perp \tilde \Delta_\|)\\
V_{k}&=&
(4 \eta)^2  \tilde \Delta_\| (2 \tilde S_{0\perp}) (2090 + c_{12} \delta m^2 +
 4150 (2 \tilde S_{0\|}) +
    c_8  (2 \tilde S_{0\|})^2 +  c_{15} \tilde \Delta_\|^2)\nonumber\\
  && +
 (4 \eta)^2 (2 \tilde S_{0\perp})\left(c_{16} \delta m + c_{17} \delta m^3 + c_{18}
\delta m (2 \tilde S_{0\|}) + c_{19} \delta m  (2 \tilde
S_{0\|})^2\right).\label{eq:extendedpade}
\end{eqnarray}
\end{widetext}
The coefficients $2090\pm210$ and $4150\pm690$  in Eq.~(\ref{eq:extendedpade}) were obtained by
fitting to the equal-mass N configurations assuming an angle of
$-59^\circ$ between the cross and hangup components. The remaining
nonzero components are given in Table~\ref{tab:vkick_fit_param}
(we compare the
predictions for the statistical distributions of recoil
velocities for  $V_{p'59}$ to $V_{4'59}$  in
Table~\ref{tab:PlargeKick}).

\begin{table}
\caption{Fitting coefficients in
Eqs.~(\ref{eq:Vkick_final_1})-(\ref{eq:Vkick_final_final}) and
Eqs.~(\ref{eq:padeall})-(\ref{eq:extendedpade}) for the
remnant recoil velocity in.
 All coefficients not given here
were set to zero.}\label{tab:vkick_fit_param}
\begin{ruledtabular}
\begin{tabular}{rD{,}{\pm}{4,4}rD{,}{\,\pm\,}{4,4}rD{,}{\,\pm\,}{4,4}}
  $V_{40}$ \\
  $c_1$ &  -0.747,0.065 & $c_8$ &
-1490,520 & $c_{12}$ & -1670,780 \\
  $c_{16}$ & -480,90 \\
 $c_{19}$ & 2430,250\\
\\
 $V_{459}$\\
  $c_1$ & -0.757,0.069 & $c_{8}$ & -2100,720 & $c_{16}$ &
-880,140\\
   $c_{19}$ & 4200,360\\
\\
$V_{4'0}$\\
  $c_1$ & -0.612,0.044 & $c_2$ & -1.13,0.37 & $c_{16}$ &
-640,80\\
$c_{18}$ & -3430,500\\
\\
$V_{4'59}$ \\
$c_1$ & -0.673,0.051 & $c_{12}$ & -6300,1750 & $c_{16}$ & -1130,160\\
$c_{18}$& -5580,1000
\\
$V_{p'59}$\\
$c_1$ & -0.677,0.046 & $c_9$ & -2540,250 & $c_{16}$ &
-1280,130\\
\end{tabular}
\end{ruledtabular}
\end{table}

In the previous discussion we ignored the in-plane component to the
recoil. The reason is, there is  significant contamination from the
out-of-plane component (e.g., by a small misidentification of the
orientation of the orbital plane) which leads to an in-plane
component that is highly dependent on the procedure used to identify the
plane. We {\it avoid} this issue by
modeling the in-plane recoil using only the nonprecessing 
results of~\cite{Healy:2014yta}. The relative error in doing so can be
large (for the in-plane component). However, as this error is large when
the out-of-plane component is much larger than the in-plane component,
and because the two components
add in quadrature, the net error in the magnitude of the recoil
is less than $10\%$ for all but 3 configurations (where the absolute error is
$< 100\ \KMS$). 
 In Table~\ref{tab:kick_error}, we show the
maximum recoil for a given family and  the RMS  and maximum errors in our prediction of
the total  recoil  and the out-of-plane component of the recoil.
Interestingly, the dominant error
in the total recoil is generally associated with the out-of-plane
component.

\begin{figure}
  \includegraphics[width=.9\columnwidth]{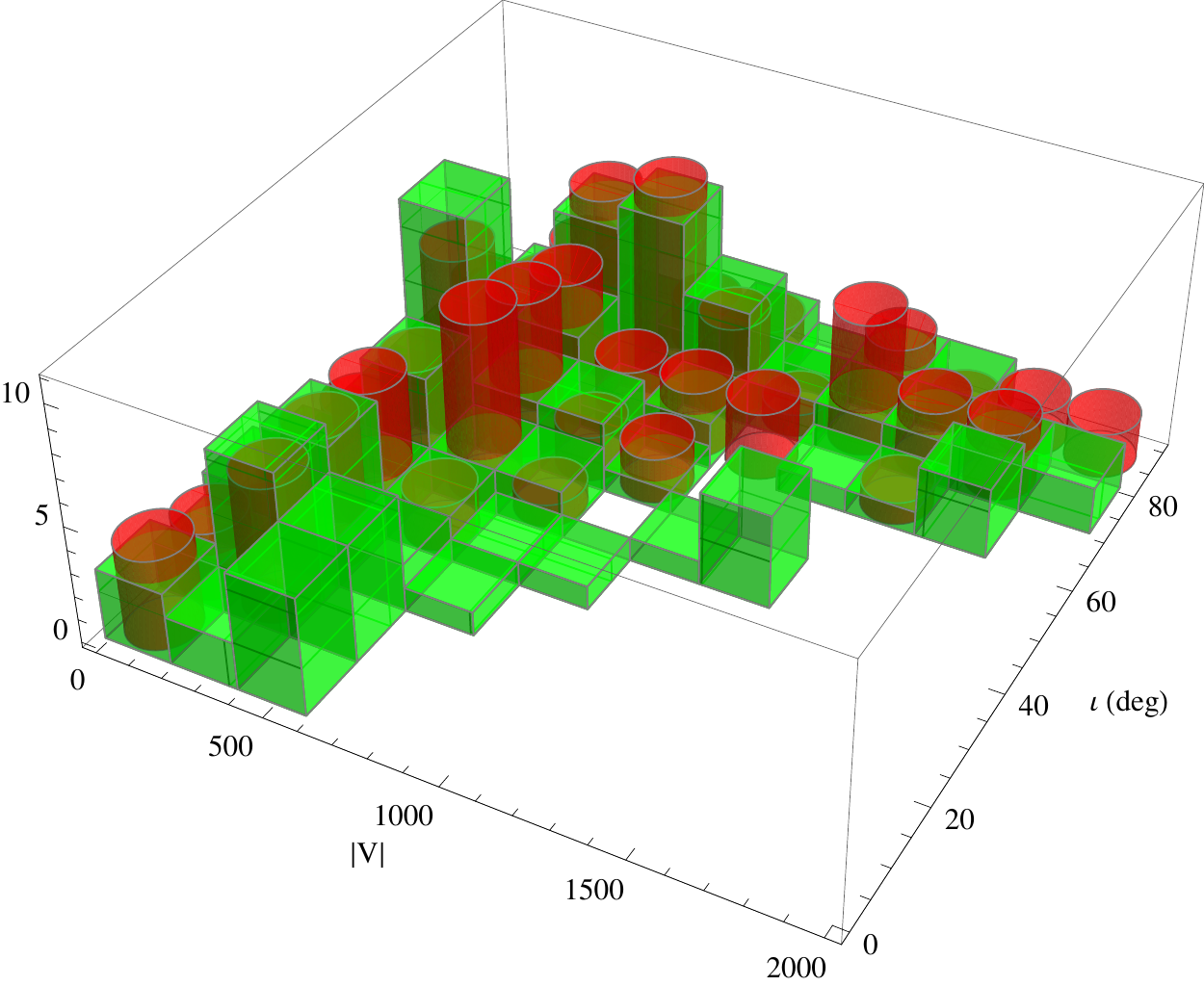}
  \caption{The distribution of recoils for the NQ configurations. The
green boxes indicate the recoil angle as measured with
respect to the initial orbital plane while the red cylinders
indicate the recoil angle is measured with respect to the orbital
plane at merger. Here $\iota$ is the inclination angle of the recoil
with respect to the orbital plane in units of degrees
 and the recoil is measured in units
of $\KMS$.}\label{fig:recoil_ang_dist}
\end{figure}

Finally, we note that while the out-of-plane recoil is the dominant
component, it is important (e.g., for modeling electromagnetic
counterparts to BH mergers) to determine the direction of the recoil
with respect to the orbital plane (more specifically, the orbital
plane when the binary decoupled from any surrounding disk).
 As shown in Fig.~\ref{fig:recoil_ang_dist}, for the
NQ configurations,  the
distribution of recoil angles is quite broad for smaller recoil
velocities ($<700 \KMS$) but is narrow for large recoils ($>1000
\KMS$). There are substantial recoils ($> 1000\ \KMS$) for
inclinations as small as $40^\circ$.

\begin{table}
\caption{ The maximum net recoil measured for each family of NQ
configurations and the RMS and maximum errors
in the predictions of the total recoil (center columns)
 and the out-of-plane component of the recoil (right columns) for each
family.
}\label{tab:kick_error}
  \begin{ruledtabular}
  \begin{tabular}{l|lll|ll}
     Family & $V_{\rm max}$ & RMS & MAX  & RMS 
 & MAX  \\
     \hline
     NQ200TH30 & 434 & 9 & 10 & 4 & 6\\
     NQ200TH60 & 660 & 16 & 21 & 10 & 13\\
     NQ200TH90 & 714 & 15 & 26 & 7 & 11\\
     NQ200TH135 & 467 & 11 & 14 & 5 & 7\\     

     NQ66TH60 & 1920 & 27 & 50 & 32 & 44\\

     NQ50TH30 & 1237 & 70 & 114 & 68 & 109 \\
     NQ50TH60 & 1812 & 87 & 123 & 72 & 94 \\
     NQ50TH90 & 1752 & 81 & 104 & 41 & 52 \\
     NQ50TH135 & 926 & 57 & 66 & 18 & 25 \\

     NQ33TH45 & 1386 & 40 & 53 & 25 & 31 \\ 
     NQ33TH75 & 1424 & 82 & 127 & 58 & 78 \\
     NQ33TH100 & 1288 & 97 & 118 & 27 & 31 \\
     NQ33TH135 & 731 & 56 & 71 & 14 & 18 \\

     NQ25TH30 & 820 & 9 & 14 & 8 & 12 \\
     NQ25TH60 & 1231 & 66 & 91 & 37 & 61 \\
     NQ25TH90 & 1087  & 64 & 93 & 30 & 37 \\
     NQ25TH135 & 569 & 65 & 72 & 7 & 11 \\
     NQ25TH150 & 409 & 38 & 41 & 6 & 9 \\

     NQ16TH45 & 681 & 22 & 32 & 9 & 14 \\
     NQ16TH90 & 551 & 75 & 90 & 18 & 27\\
     NQ16TH115 & 527 & 34 & 69 & 27 & 37 \\
     NQ16TH135 & 361 & 52 & 77 & 8 & 11 \\
     NQ16TH150 & 262& 22 & 26 & 23 & 36\\
   \end{tabular}
  \end{ruledtabular}
\end{table}

\subsection{Fitting the radiated energy and remnant spin}
\label{sec:fit-energy}
The total mass loss of the binary from its complete inspiral (starting
at infinite
separation) is given by
\begin{equation}
  \delta {\cal M} = \frac{M_1^\infty + M_2^\infty - M_{\rm rem}}{M_1^\infty + M_2^\infty},
\end{equation}
where $M_1^\infty$ and $M_2^\infty$ are the initial masses of the two BHs (i.e., at infinite separations) and
$M_{\rm rem}$ is the remnant mass. Since the BH horizon is essentially
constant during the inspiral, we get a very good approximation to
$\delta {\cal M}$ using
\begin{equation}
  \delta {\cal M} \approx \frac{M_1 + M_2 - M_{\rm rem}}{M_1 + M_2},
\end{equation}
where $M_1$ and $M_2$ are the horizon masses of the two BHs in the
binary as measured after the initial burst of radiation.

For each family of NQ configurations  with fixed $q$ and $\theta$, we
fit $\delta {\cal M}$ to the form
\begin{equation}
\delta {\cal M} = E_{c} + E_{\phi}
\cos(2\varphi - 2\phi^m_2),
\label{eq:massrad}
\end{equation}
where $E_c$, $E_\phi$, and $\phi^m_2$ are
fitting constants. We also fit the square of the dimensionless remnant spin
 $\alpha^2$ to the form,
\begin{equation}
\alpha^2 = A_c + A_\phi \cos(2\varphi - 2
\phi_2^a),
\label{eq:asquared}
\end{equation}
where $A_c$, $A_\phi$, and $\phi_2^a$ are fitting constants. The
results are given in Table~\ref{tab:fit_NQ_EJ}.
Note that $E_c$ and $A_c$ dominate the expressions for the mass loss
and remnant spin.
Note also that in
Tables~\ref{tab:fit_NQ_EJ} and
\ref{tab:rem_rad_cmp_part1} there are
missing entries. These missing entries are due to missing remnant
horizon mass and spin data.

\begin{table*}
\caption{Fitting parameters for the NQ families of configurations as a
function of $\varphi$ (see text) to the form
$\delta{\cal M}_\| = E_c + E_\phi \cos(2(\varphi - \phi^m_2)\pi/180)$ and
$\alpha_{\rm rem}^2 = A_c + A_\phi \cos(2(\varphi -
\phi^\alpha_2)\pi/180)$. All angles are measured in degrees.}
\label{tab:fit_NQ_EJ}
\begin{ruledtabular}
\begin{tabular}{l|D{,}{\pm}{-1}D{,}{\,\pm\,}{-1}D{,}{\,\pm\,}{-1}D{,}{\,\pm\,}{-1}D{,}{\,\pm\,}{-1}D{,}{\,\pm\,}{-1}}

Family & 100 E_c & 100 E_\phi & \phi^m_2 &  A_c & 100 A_\phi & \phi^\alpha_2\\
\hline
NQ200TH30 & 4.28514,0.00080& 0.0183,0.0014 & 3.30,2.42 & 0.441972,0.000001 & -0.028,0.001 & -6.8,1.2\\
NQ200TH60 & 4.13687,0.00032& -0.02329,0.00045 & 24.79,0.56 & 0.42292,0.00001 & 0.1154,0.0018  & 12.87,0.46\\
NQ200TH90 & 3.92326,0.00052  & 0.03107,0.00070 & -24.66,0.69 & 0.394768,0.000002 & -0.1160,.0002 & -35.87,0.07\\
NQ200TH135 & \\
NQ66TH60 & 5.5982,0.0025 & 0.12438,0.0035 &  12.62,0.79 & 0.6244,0.0001 & -0.421,0.013 & 11.43,0.91\\
NQ50TH30 & 5.9302,0.0029  &  0.0602,0.0040 & 4.3,2.0 & 0.7215,0.0001 &  -0.1674, 0.0074 & -12.1,1.4\\
NQ50TH60 & 5.1623,0.0065 & 0.1290,0.0092 & 0.0,2.0 & 0.63855,0.00036  & -0.386,0.050 & -5.5,3.7 \\
NQ50TH90 & 4.2528,0.0084 & 0.108,0.011 & 35.9,4.4 & 0.49782,0.00021 & -0.578,0.027  & 38.9,1.5\\
NQ50TH135 & 3.20389,0.00085 & -0.0344,0.0012 & -11.92,0.94 & 0.24812,0.00013 & 0.247,0.018 & -17.0,2.0\\
NQ33TH45 &  4.5458,0.0083  &  0.099,0.011 & -4.7,3.1 & 0.69928,0.00010 & -0.286,0.014 & -5.9,1.4\\
NQ33TH75 &  3.6795,0.0030 & -0.1241,0.0040 & 29.57,0.91 & 0.56914,0.00057  & 0.506,0.071 &  18.5,4.4 \\
NQ33TH100 & 3.0122,0.0038 & 0.0936,0.0056  & 33.5,1.5 & 0.41316,0.00044 & -0.554,0.065  & 32.9,2.9 \\
NQ33TH135 & 2.33489,0.00050 &  -0.03034,0.00069 & 32.55,0.62 & 0.179672,0.000078 & 0.230,0.011 & -35.0,1.3\\
NQ25TH30 & 4.02044,0.00024 &  -0.04259,0.00025 & 33.52,0.24 & 0.735887,0.000061 & 0.1853,0.0089 & -0.3,1.1\\
NQ25TH60 & 3.346,0.013 & -0.1059,0.0048 & 16.2,6.4 & 0.64155,0.00036 & 0.40,0.20 & 12,45\\
NQ25TH90 & 2.5618,0.0021 & -0.0845,0.0019  & 27.97,0.92 & 0.47050,0.00088 & 0.449,0.081 & 23.4,7.3\\
NQ25TH135 & 1.7785,0.00052 & 0.02772,0.00070 & 38.53,0.76 & 0.157916,0.000096 & -0.178,0.012 & 34.0,2.2\\
NQ25TH150 &1.64828,0.00057 & -0.01193,0.00078 & 135.02,0.19 &  0.074246,0.000057 & -0.0702,0.0078 & 38.3,3.3\\
NQ16TH45 & 2.5836,0.0057 & 0.065,0.011 &  -36.9,3.3 & 0.6870,0.0011 & -0.11,0.24 & 33,16\\
NQ16TH90 & \\
NQ16TH115 & 1.36377,0.00094  & 0.0411,0.0010  & 40.60,0.90 & 0.30713,0.00033 & -0.208,0.023 & 30.3,7.1\\
NQ16TH135 & \\
NQ16TH150 &  1.06412,0.00037 & 0.00893,0.00052 & -11.3,1.6 & 0.093108,0.000065  & -0.0483,0.0091  & -27.8,5.2\\
\end{tabular}
\end{ruledtabular}
\end{table*}


To fit $E_c$ as a function of $q$ and $\theta$, we start by
refitting the configurations of  Healy \etal~\cite{Healy:2014yta}
(all these configurations were nonprecessing). We
need to refit the results there because our fitting formulas are
different. There the fits were to the remnant mass and
 here we are fitting to the mass loss.

As in~\cite{Healy:2014yta}, we keep terms up through fourth-order in
the spins and $\delta m$, and enforce the particle limit. Our fitting
function for $E_c$ is given by
\begin{eqnarray}
E_c^{\|}&&= \left( 4\eta\right)^2\left(E_{\rm HU}+ k_{2 a} \delta m \tilde\Delta_\|
+ (0.000743) \tilde\Delta_\|^2 + k_{2d} \delta m^2\right.\nonumber \\
  && \left.+ k_{3 a} \delta m \tilde\Delta_\| \tilde S_\|
+k_{3 b} \tilde S_\| \tilde\Delta_\|^2 + k_{3 d} \delta m^2 \tilde S_\|\right.\nonumber \\
 &&\left. + k_{4a} \delta m
\tilde\Delta_\| \tilde S_\|^2 + k_{4 b} \delta m \tilde\Delta_\|^3+ (0.000124)
\tilde\Delta_\|^4 +
k_{4 e} \tilde\Delta_\|^2 \tilde S_\|^2 \right.\nonumber \\
&&\left.+ k_{4 f} \delta m^4+ k_{4 g} \delta m^3 \tilde\Delta_\|
\right) + \delta m^6 \eta (1-E_{\rm isco}),\label{eq:Ec_aligned}
\end{eqnarray}
where $E_{\rm HU}$ is given by~\cite{Hemberger:2013hsa}
\begin{equation}\label{eq:EHU}
   E_{\rm HU} = 0.0025829 - \frac{0.0773079}{2 \tilde {S}_{0\|} -
1.693959},
\end{equation}
$E_c^{\|}$ denotes that spins are aligned or antialigned with the
orbital angular momentum,
and $E_{\rm isco}$ is the energy of the innermost stable circular
orbit (ISCO). For the fits here and below we approximate $E_{\rm
isco}$ by the ISCO energy of a particle on an equatorial geodesic on a
Kerr background with spin parameter $\alpha = \tilde S_\|$.
Note that we define $ E_{\rm HU}$ using the $\vec S_0$ variable. This
is due to the fact that $ E_{\rm HU}$ would have a pole at small mass
ratios if we defined it using $\vec S$. In the equal-mass limit, both
definitions are equivalent.

As in our fits to the recoil, we successively remove the most uncertain
of the fitting coefficients. Our final fitting parameters are
summarized in Table~\ref{tab:Ec_all}. We then fit $E_c$ from each family of the
NQ configurations to
\begin{eqnarray}
  E_{c} &=& E_c^{\|} + \left(4 \eta\right)^2\left(|\tilde S_\perp|^2
(e_1 + e_2\tilde S_\|  +
e_3\tilde S_\|^2 )\right.\nonumber\\
&&\left. + |\tilde \Delta_\perp|^2 (\epsilon_1  +
\epsilon_2 \tilde S_\| + \epsilon_3 \tilde S_\|^2) + \delta m^2
|\tilde S_\perp|^2 (E_{A} + \tilde S_\| E_{B})
\right.\nonumber \\
&& \left.+
 \delta m^2 |\tilde \Delta_\perp|^2 (E_{D} + \tilde S_\| E_{E}) +
E_{F}
\delta m  |\tilde \Delta_\perp| |\tilde S_\perp|\right.\nonumber \\
&&\left. + E_{G} \tilde \Delta_\|^2
|\tilde \Delta_\perp|^2 + E_{H} \tilde \Delta_\|^2 |\tilde S_\perp|^2\right),
\label{eq:Ec_all_fit}
\end{eqnarray}
where $E_c^{\|}$ is given by Eq.~(\ref{eq:Ec_aligned}),
and the constants $e_1$, $e_2$, $e_3$, $\epsilon_1$,
$\epsilon_2$, and $\epsilon_3$ were determined
in~\cite{Lousto:2013wta} (note that the constants $e_1$, $\cdots$ here
are denoted by $e'_1$, $\cdots$ in ~\cite{Lousto:2013wta}).
For the
convenience of the reader, those constants are also
given in Table~\ref{tab:Ec_all}.
The remaining terms in Eq.~(\ref{eq:Ec_all_fit}) were chosen by adding
even powers in $\delta m$ to terms present in the equal-mass case. In
addition, we found that term odd in $\delta m$ ($E_F$) was needed in
order to fit the $q=2$ family.

Unlike in the equations for the recoil in the preceding section
[e.g., 
Eqs.~(\ref{eq:Vkick_final_1})-(\ref{eq:Vkick_final_final}) and
Eqs.~(\ref{eq:padeall})-(\ref{eq:extendedpade})],
here $S_\perp$ and $\Delta_\perp$ arise from
the magnitudes of the in-plane components of these two vectors rather 
than dot products with unit
vectors in the plane. We therefore use the notation $|S_\perp|$, etc.,
to distinguish between these types of  terms and those in for the
recoil. This distinction will be important when generalizing to
arbitrary binaries.

When fitting the remaining constants in Eq.~(\ref{eq:Ec_all_fit}) we
take all previously fitted constants as exact (i.e., we do not include
the uncertainties in these constants in subsequent fits).
Once again, we successively remove the least certain of the new fitting
constants. The final fitting parameters are given in
Table~\ref{tab:Ec_all}. In the table, we report on fits using our
standard choice of variables \{$ \vec S, \vec \Delta, \delta m$\} and
the alternative choice  \{$ \vec S_0, \vec \Delta, \delta m$\}. In
both
cases, $E_{\rm isco}$ is calculated using $\vec S$, and
$E_{\rm HU}$  is calculated using $\vec S_0$.

\begin{table}
\caption{Fitting coefficients in Eqs.~(\ref{eq:Ec_aligned}) and
(\ref{eq:Ec_all_fit}). A prime (') indicates that the variable
$S$ was replaced by $S_0$ in the fitting formula (except in $E_{\rm
isco}$, which always takes $\tilde S_\|$ as its arguments).
 All coefficients not
given here were set to zero identically.
Note that the equal-mass terms $e_1$, $e_2$, $e_3$, $\epsilon_1$,
$\epsilon_2$, and $\epsilon_3$ are unaffected by the change from
$S$ to $S_0$.}\label{tab:Ec_all}
\begin{ruledtabular}
\begin{tabular}{rD{,}{\pm}{4,4}rD{,}{\,\pm\,}{4,4}rD{,}{\,\pm\,}{4,4}}

$k_{2a}$& -0.024,0.003 & $ k_{3a} $& -0.055,0.003 & $ k_{3d} $& -0.019,0.009 \\
$ k_{4a} $& -0.119,0.029 & $ k_{4b} $& 0.005,0.004 & $ k_{4f} $& 0.035,0.005 \\
$ k_{4g} $& 0.022,0.016 & $ E_{B} $& 0.59,0.31 &
$ E_{E} $& -0.51,0.20 \\
$ E_{F} $& 0.056,0.004 &
$ E_{G} $& -0.073,0.016 &
 $e_1$ & 0.0356,0.0025  \\
 $e_2$ & 0.096,0.012 &
  $e_3$ & 0.122,0.067 &
 $\epsilon_1$ &  0.0043,0.0012  \\
 $\epsilon_2$ & 0.0050,0.0021 & $\epsilon_3$ & -0.009,0.026\\
$ k'_{2a} $& -0.017,0.003 &
$ k'_{3a} $& -0.091,0.008 & $
k'_{4a} $& -0.146,0.022 \\
$ k'_{4b} $& -0.01,0.005 &
$ k'_{4f} $& 0.037,0.007 &
$ E'_{A} $& -0.075,0.001 \\
$ E'_{B} $& -0.29,0.14 &
$ E'_{D} $& -0.019,0.006 &
$ E'_{H} $& -0.244,0.063 \\
 $e'_1$ & 0.0356,0.0025 &  $e'_2$ & 0.096,0.012 & $e'_3$ &
0.122,0.067 \\
 $\epsilon'_1$ &  0.0043,0.0012  & $\epsilon'_2$ & 0.0050,0.0021 & $\epsilon'_3$ & -0.009,0.026\\
\end{tabular}
\end{ruledtabular}
\end{table}

{\it The following section corrects an error in the published version
of the manuscript.}
We use a very similar procedure for fitting $A_c$. 
Here, we have several choices for how we incorporate previous results.
First, we can start with the
form of Healy \etal~\cite{Healy:2014yta} for the remnant spin for the
aligned case and add non-aligned corrections. Alternatively, we may
start with alternate forms of the aligned spin remnant spin function.

The form of the  Healy \etal spin is
\begin{eqnarray}\label{eq:p4spin}
\alpha_{\rm align} = 
 (4\eta)^2\Big\{L_0 + L_{1}\,\Spar+\nonumber\\ 
                     L_{2a}\,\Dpar\dmt+
                     L_{2b}\,\Spar^2+
                     L_{2c}\,\Dpar^2+
                     L_{2d}\,\dmt^2+\nonumber\\
                     L_{3a}\,\Dpar\Spar\dmt+
                     L_{3b}\,\Spar\Dpar^2+
                     L_{3c}\,\Spar^3+\nonumber\\
                     L_{3d}\,\Spar\dmt^2+
                     L_{4a}\,\Dpar\Spar^2\dmt+
                     L_{4b}\,\Dpar^3\dmt+\nonumber\\
                     L_{4c}\,\Dpar^4+
                     L_{4d}\,\Spar^4+
                     L_{4e}\,\Dpar^2\Spar^2+\nonumber\\
                     L_{4f}\,\dmt^4+
                     L_{4g}\,\Dpar\dmt^3+\nonumber\\
                     L_{4h}\,\Dpar^2\dmt^2+
                     L_{4i}\,\Spar^2\dmt^2\Big\}+\nonumber\\
                     \Spar(1+8\eta)\dmt^4+\eta{L}_{\rm isco}\dmt^6,
\end{eqnarray}
where the values of these coeefficients are reproduced in
Table~\ref{tab:Ac1}, and the last line in Eq.~(\ref{eq:p4spin}) is due to the particle limit
and
\begin{eqnarray*}
  {L}_{\rm isco} = \frac{2}{\sqrt{3 r_{\rm isco}}}\left(3
        \sqrt{r_{\rm isco}} - 2 \Spar\right),\\
        r_{\rm isco} = 3 + Z_2 - {\rm SIGN}(\Spar) \sqrt{(3-Z_1)(3+Z_1
        + 2 Z_2)},\\
        Z_1 = 1 + (1 - \Spar^2)^{1/3} ((1 + \Spar)^{1/3} + (1 - \Spar)^{1/3}),\\
        Z_2 = \sqrt{3 \Spar^2 + Z_1^2}.
      \end{eqnarray*}
We will also consider replacing $\vec S$ with $\vec S_{0} = \vec S +
1/2 \dmt \vec \Delta$ and using
Hemberger \etal's \cite{Hemberger:2013hsa} coefficients for some of the terms in
Eq.~(\ref{eq:p4spin}).

\begin{widetext}
The fitting formula for the generic case is
\begin{eqnarray}
  A_{c} &=& A_c^{\|} +
      \left(4 \eta\right)^2\left(|\tilde S_\perp|^2
(a_1 + a_2\tilde S_\|  +
a_3\tilde S_\|^2 )
+ |\tilde \Delta_\perp|^2 (\zeta_1 +
\zeta_2 \tilde S_\| + \zeta_3  \tilde S_\|^2) +
 \delta m^2 |\tilde S_\perp|^2 (A_{A} + \tilde S_\| A_{B})
\right.\nonumber \\
&& \left.+
 \delta m^2 |\tilde \Delta_\perp|^2 (A_{D} + \tilde S_\| A_{E}) +
A_{F}
\delta m  |\tilde \Delta_\perp| |\tilde S_\perp|
 + A_{G} \tilde \Delta_\|^2 |\tilde \Delta_\perp|^2 + A_{H} \tilde \Delta_\|^2
|\tilde S_\perp|^2\right)
+\nonumber\\
        && \dmt^6 (1+ 12 \eta) S_\perp^2 
,
\label{eq:Ac_all_fit}
\end{eqnarray}
where $A_c^{\|}\equiv \alpha_{\rm align}^2$ and the last line in
Eq.~(\ref{eq:Ac_all_fit}) is due to the particle limit.
\end{widetext}

Here we consider three alternatives denoted by $A_c$, $A'_c$, and
$A_{\rm alt}$. The $A_c$ fit is obtained by using the Healy {\it et al.}
parameters as given there (reproduced here in Table~\ref{tab:Ac1}) for
Eq.~(\ref{eq:p4spin}) and then fitting the remaining coefficients in
Eq.~(\ref{eq:Ac_all_fit}). For the $A'_{c}$ fit, we replace $\vec S$ by
$\vec S_0$ for all terms in both Eq.~(\ref{eq:p4spin}) and
Eq.~(\ref{eq:Ac_all_fit}), except
for terms associated with the particle limit (last line in
Eq.~(\ref{eq:p4spin}) and 
Eq.~(\ref{eq:Ac_all_fit})). To do this, we need to refit the
Healy {\it et al.} data to the new form of Eq.~(\ref{eq:p4spin}). In addition,
we also used the Hemberger {\it et al.} choices for some of the
parameters (see Table~\ref{tab:Ac2}). 
$A_{\rm  alt}$ is like $A'_c$ except we do not replace
$S$ by $S_0$ (but do refit the Healy {\it et al.} data). The fitting
coefficients are given in Tables~\ref{tab:Ac2} and~\ref{tab:Ac3}.
Relative RMS errors are 0.020, 0.024, 0.005. Absolute RMS
errors are 0.0029, 0.0027, 0.0009. 

\begin{table}
  \caption{Coefficients in Eq.~(\ref{eq:p4spin}) for the $A_c$ fit.
    These are from Healy \etal and
   reproduced here for reference.}\label{tab:Ac1}
  \begin{ruledtabular}
    \begin{tabular}{llll}
      L0 & 0.686710 &
      L1 & 0.613247 \\
      L2a & -0.145427 &
      L2b & -0.115689 \\
      L2c & -0.005254 &
      L2d & 0.801838 \\
      L3a & -0.073839 &
      L3b & 0.004759 \\
      L3c & -0.078377 &
      L3d & 1.585809 \\
      L4a & -0.003050 &
      L4b & -0.002968 \\
      L4c & 0.004364 &
      L4d & -0.047204 \\
      L4e & -0.053099 &
      L4f & 0.953458 \\
      L4g & -0.067998 &
      L4h & 0.001629 \\
      L4i & -0.066693
    \end{tabular}
  \end{ruledtabular}
 \end{table}

\begin{table}
  \caption{Coefficients in Eq.~(\ref{eq:p4spin}). Left values are
    for $A'_c$  and right for $A_{\rm alt}$. The coefficients below the
  line are common to both and are derived from Hemberger \etal. All
unspecified coefficients are zero.}\label{tab:Ac2}
  \begin{ruledtabular}
    \begin{tabular}{llll}
      $ L2a $& $ -0.517\pm0.01 $ & $ L2a $& $ -0.148\pm0.007 $ \\
      $ L2d $& $ 0.81\pm0.017 $ &  $ L2c $& $ -0.003\pm0.001 $ \\
      $ L3d $& $ 1.587\pm0.061 $ & $ L2d $& $ 0.814\pm0.008 $  \\
      $ L4f $& $ 0.692\pm0.054 $ &  $ L3a $& $ -0.099\pm0.043 $ \\
      $ L4g $& $ 1.003\pm0.081 $ & $ L3d $& $ 1.615\pm0.01 $ \\
       & & $ L4b $& $ -0.013\pm0.011 $ \\
       & & $ L4e $& $ -0.077\pm0.058 $ \\
       & & $ L4f $& $ 0.832\pm0.05 $ \\
       & & $ L4i $& $ -0.093\pm0.067 $ \\
      \hline
      $L0$ & 0.686403 \\
      $L1$ & 0.613203 \\
      $L2b$ & -0.107373 \\
      $L3c$ & -0.0784152 \\
      $L4d$ & -0.079896

    \end{tabular}
  \end{ruledtabular}
\end{table}

\begin{table}
  \caption{Coefficients in Eq.~(\ref{eq:Ac_all_fit}) for $A_c$ (top), $A'_c$
  (middle), $A_{\rm at}$ (bottom). Coefficients common to all are
listed below the double line.}\label{tab:Ac3}
  \begin{ruledtabular}
    \begin{tabular}{llll}
      $ A_A $& $ 3.008\pm0.862 $ &
      $ A_B $& $ -7.334\pm4.39 $ \\
      $ A_D $& $ -2.011\pm0.552 $ &
      $ A_E $& $ 5.1\pm2.816 $ \\
      $ A_G $& $ -2.804\pm0.864 $ &
      $ A_H $& $ 5.143\pm1.539 $ \\
      \\
\hline
$ A_A $& $ 2.189\pm0.025 $ &
$ A_D $& $ 0.547\pm0.099 $  \\
\\
\hline
$ A_A $& $ 3.402\pm0.859 $ &
$ A_B $& $ -4.571\pm4.373 $  \\
$ A_D $& $ -2.216\pm0.55 $ &
$ A_E $& $ 3.713\pm2.805 $  \\
$ A_G $& $ -3.099\pm0.86 $ &
$ A_H $& $ 5.999\pm1.534 $   \\
\\
\hline
\hline
 $a_1$ & $0.8401 \pm 0.0061$ &
 $a_2$ & $-0.328 \pm 0.029$ \\
$a_3$ & $-0.61 \pm 0.16$ &
 $\zeta_1$ & $-0.0209 \pm 0.0070$ \\
$\zeta_2$ & $-0.038 \pm 0.012$ &
 $\zeta_3$ & $0.04 \pm 0.16$\\

    \end{tabular}
 \end{ruledtabular}
\end{table}

 \begin{figure*}
   \includegraphics[width=.48\textwidth]{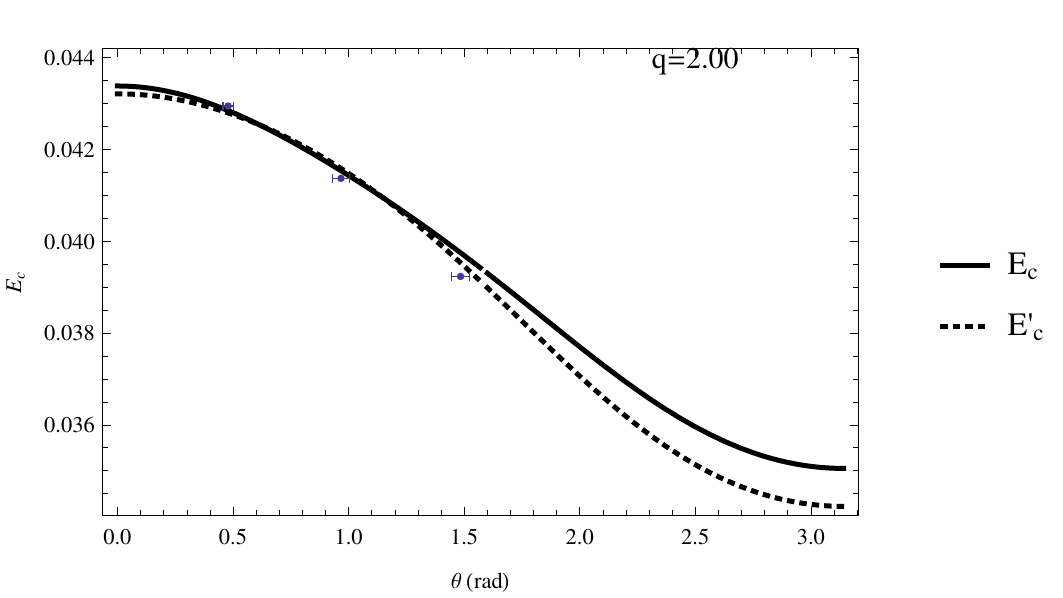}
   \includegraphics[width=.48\textwidth]{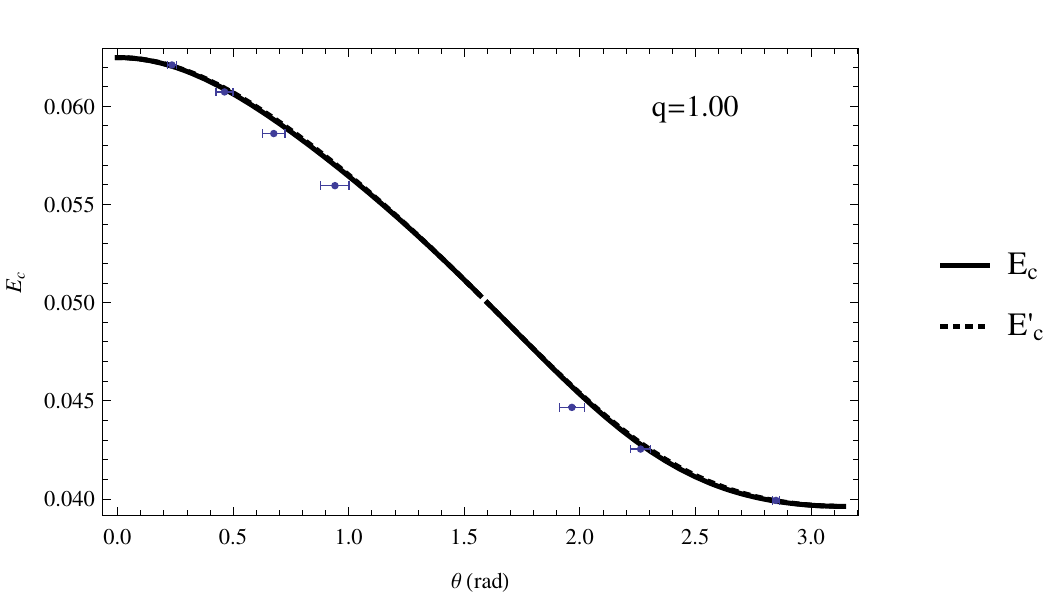}
   \includegraphics[width=.48\textwidth]{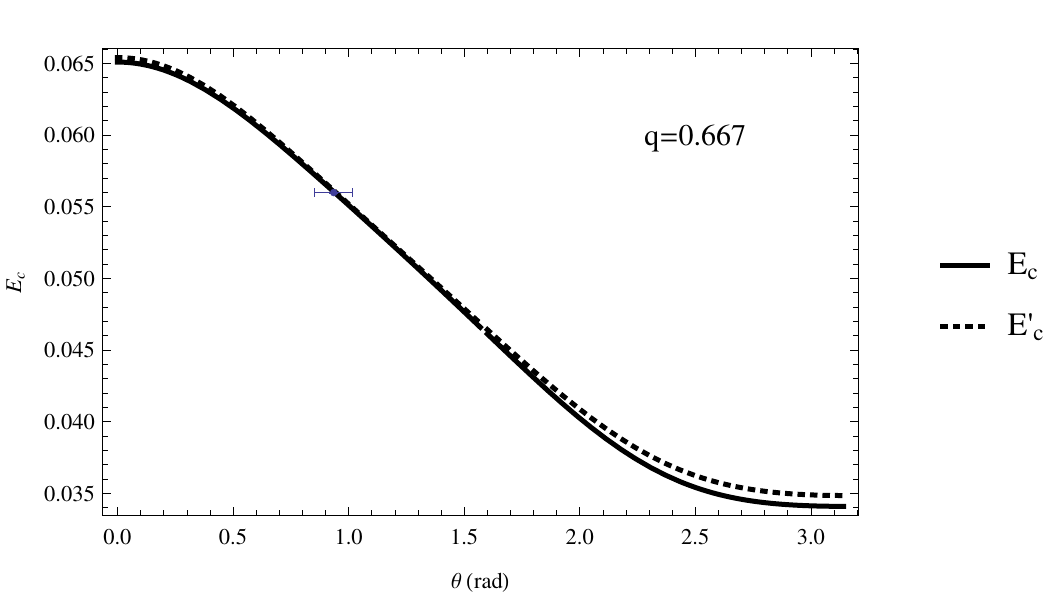}
   \includegraphics[width=.48\textwidth]{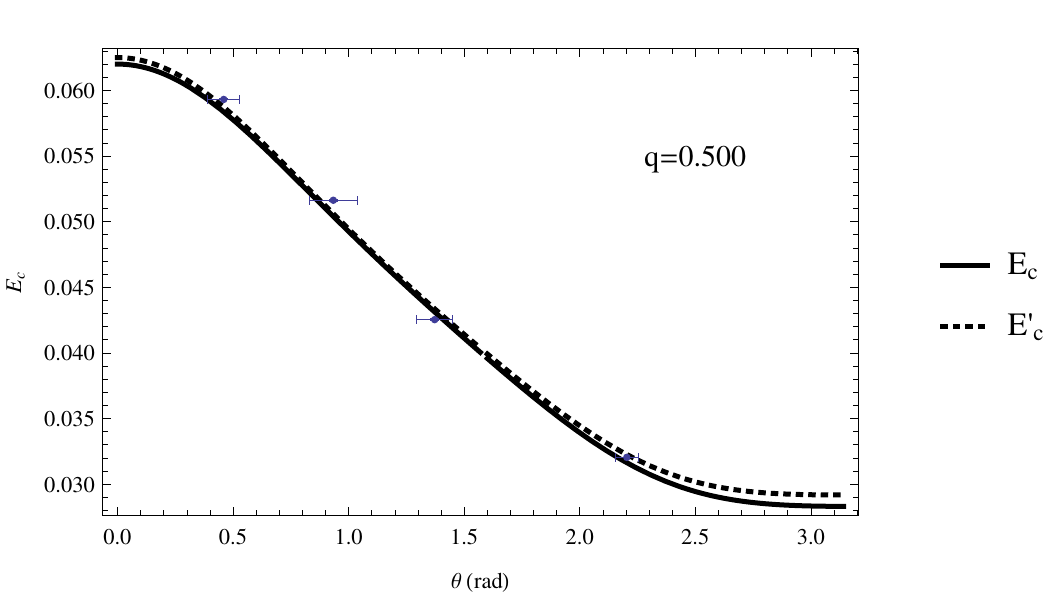}
   \includegraphics[width=.48\textwidth]{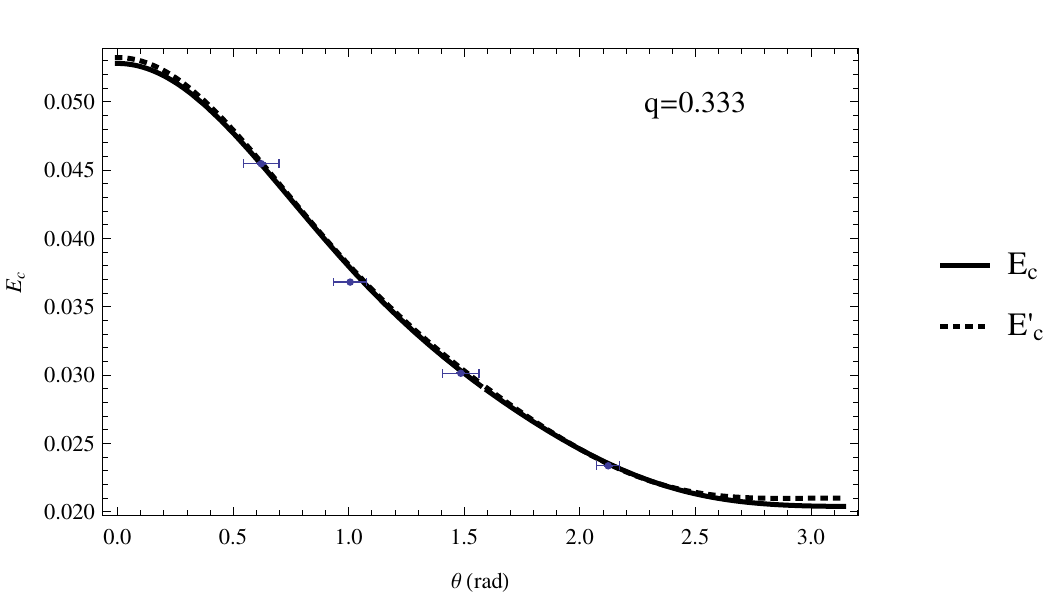}
   \includegraphics[width=.48\textwidth]{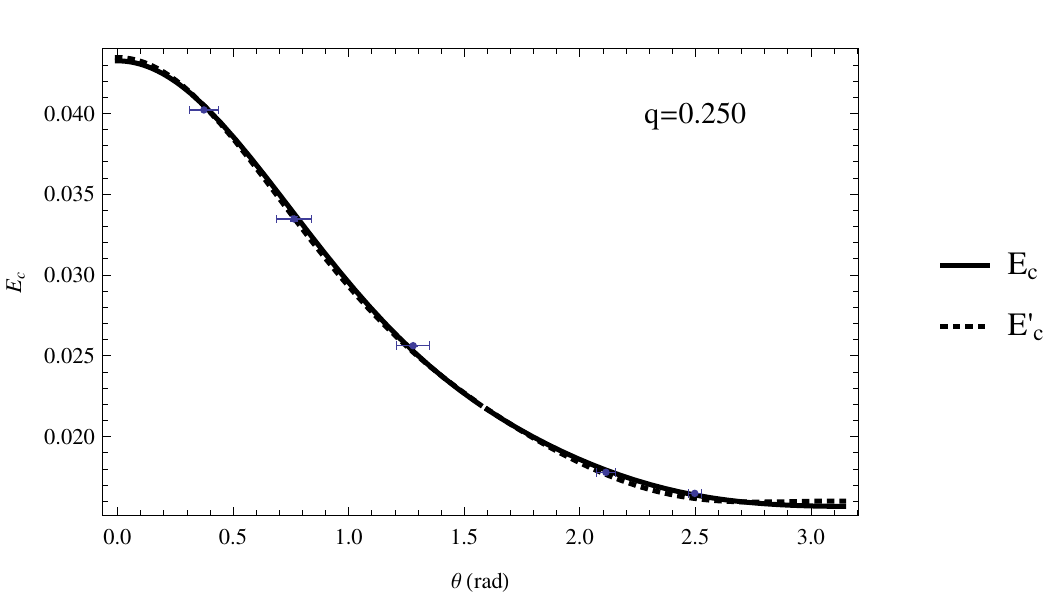}
   \includegraphics[width=.48\textwidth]{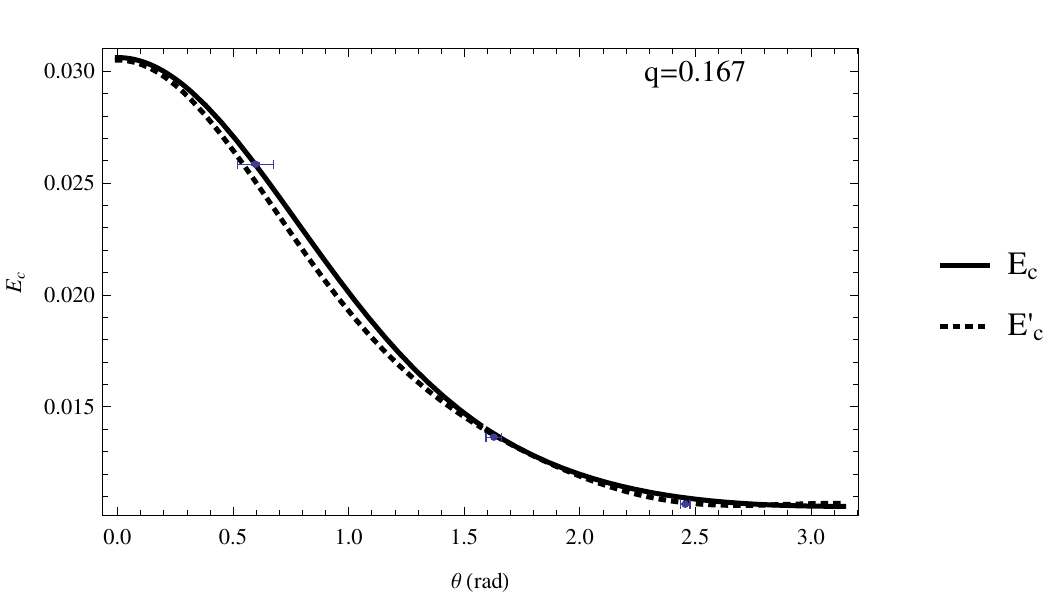}
  \caption{ Plots of the fitted $E_c$ versus inclination angle
$\theta$
and $q$ for the NQ
configurations. Each data point represents the value  of $E_c$ for a
family of azimuthal configurations with the same inclination angle and
mass ratio. Note $\delta M\approx E_c$, and that a prime denotes that
$\vec S_0$ was used in the fits, rather than $\vec S$.}\label{fig:Ec_fits}
 \end{figure*}

\begin{figure}
   \includegraphics[width=.9\columnwidth]{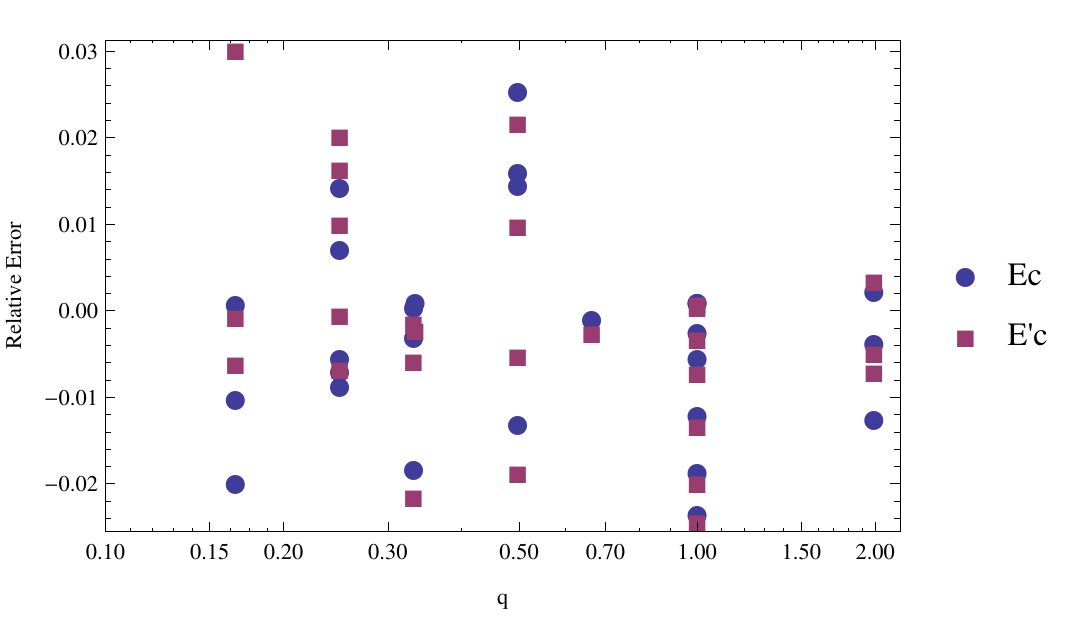}
   \includegraphics[width=.9\columnwidth]{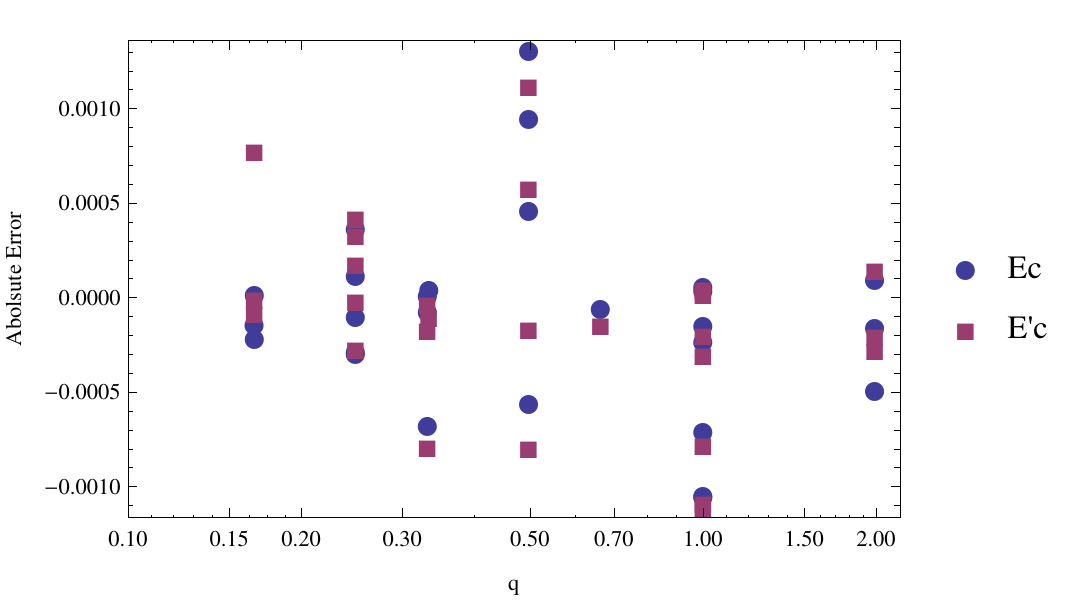}
  \caption{The relative errors (residuals) in the fit  of $E_c$ as a
function of the binary's parameters.
Note that there are multiple data points for each $q$,
 and that a prime denotes that
$\vec S_0$ was used in the fits, rather than $\vec S$.}\label{fig:Ec_resid}
\end{figure}

\begin{figure*}
   \includegraphics[width=.45\textwidth]{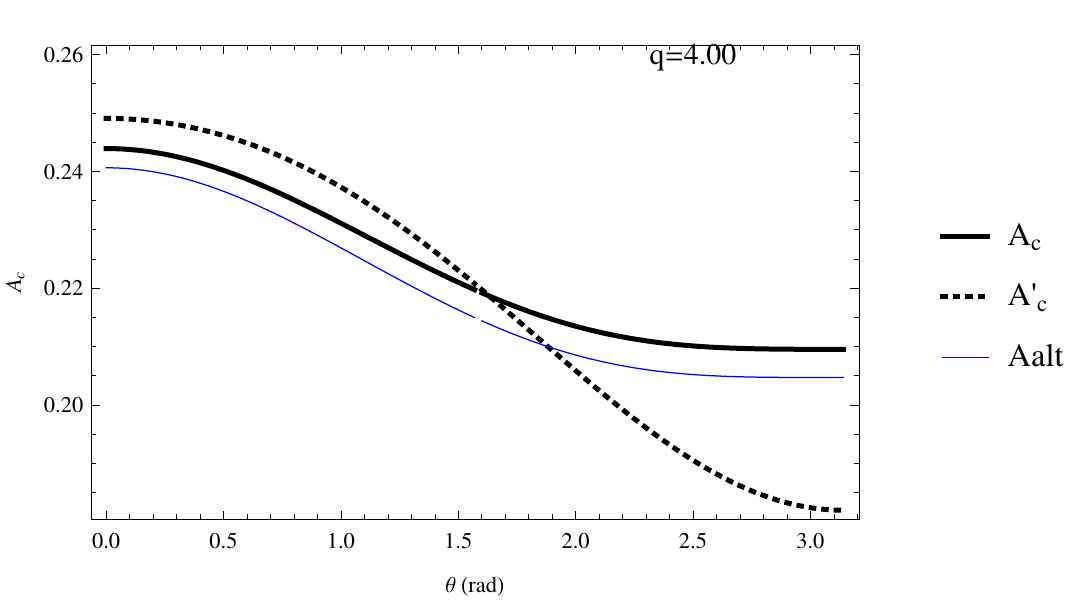}
   \includegraphics[width=.45\textwidth]{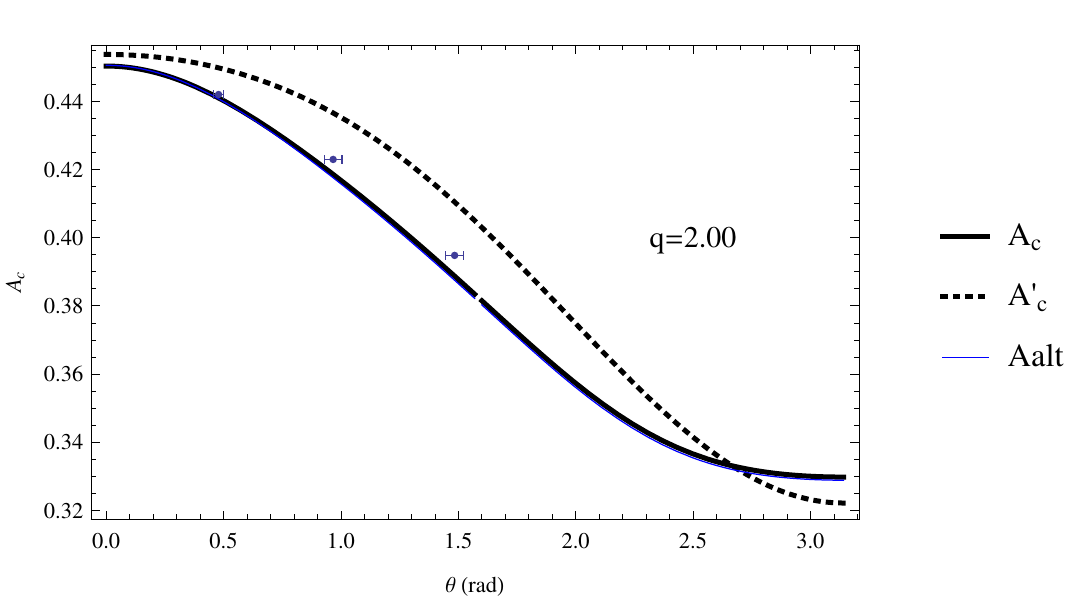}
   \includegraphics[width=.45\textwidth]{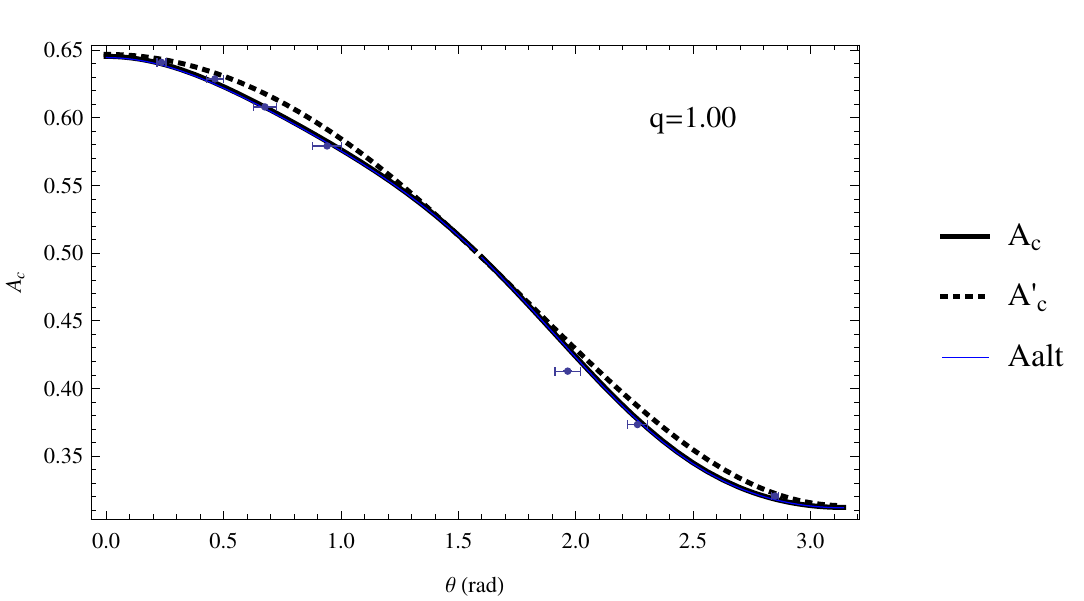}
   \includegraphics[width=.45\textwidth]{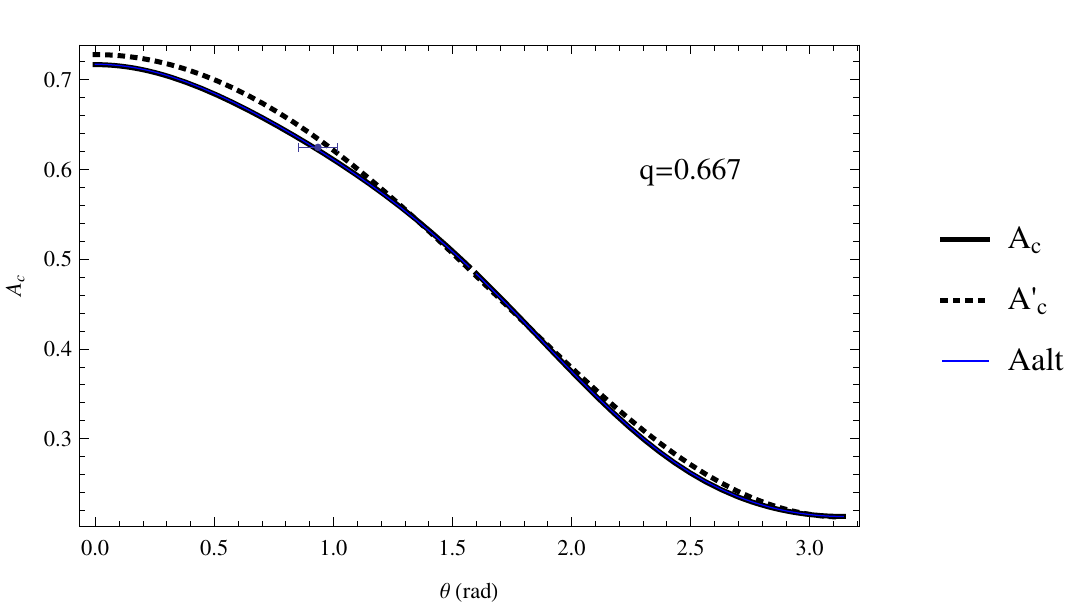}
   \includegraphics[width=.45\textwidth]{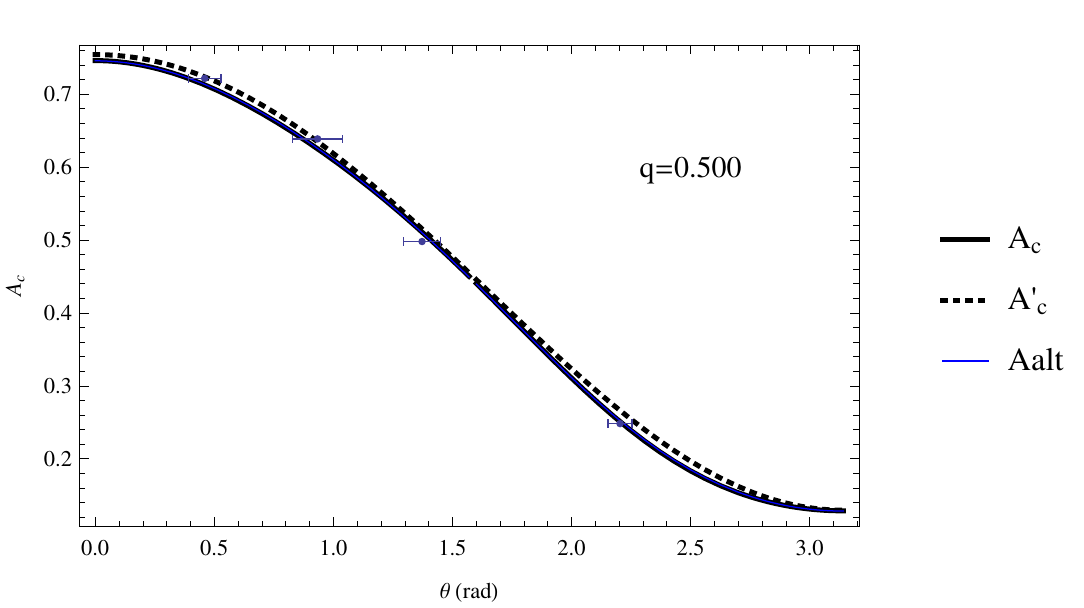}
   \includegraphics[width=.45\textwidth]{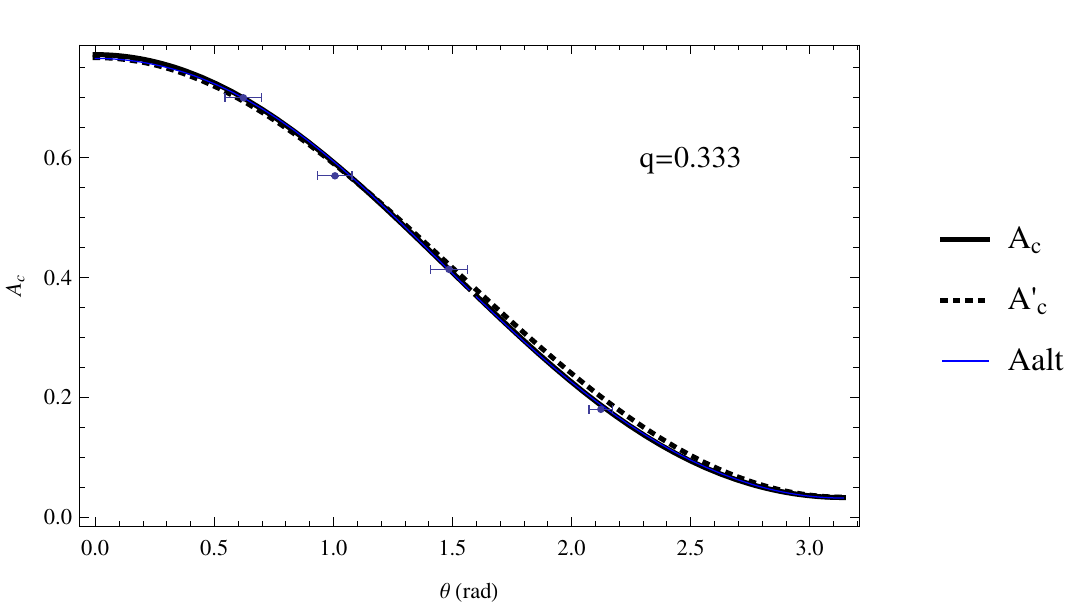}
   \includegraphics[width=.45\textwidth]{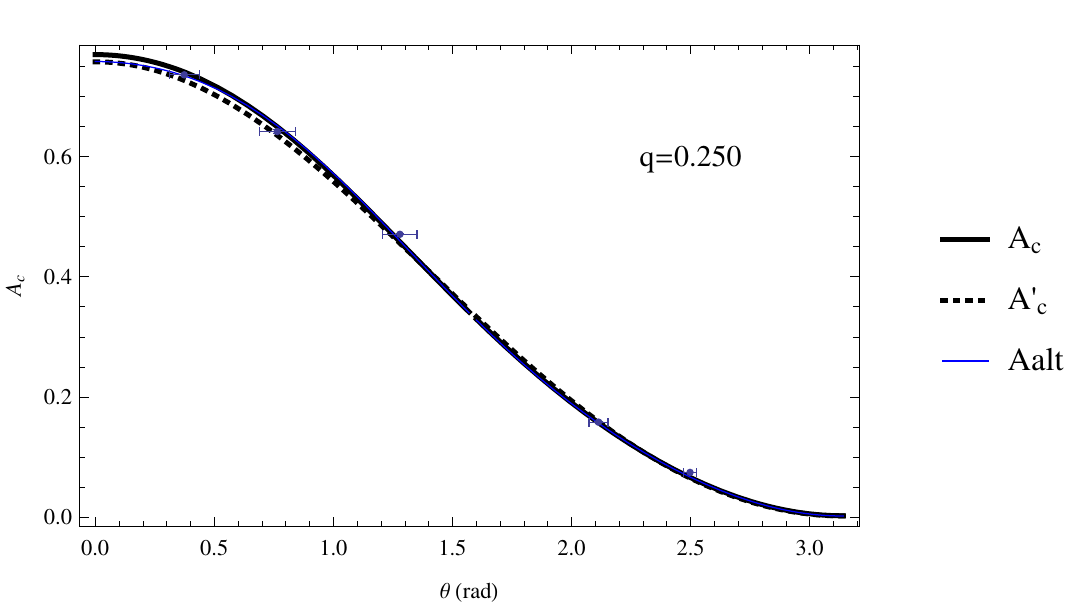}
   \includegraphics[width=.45\textwidth]{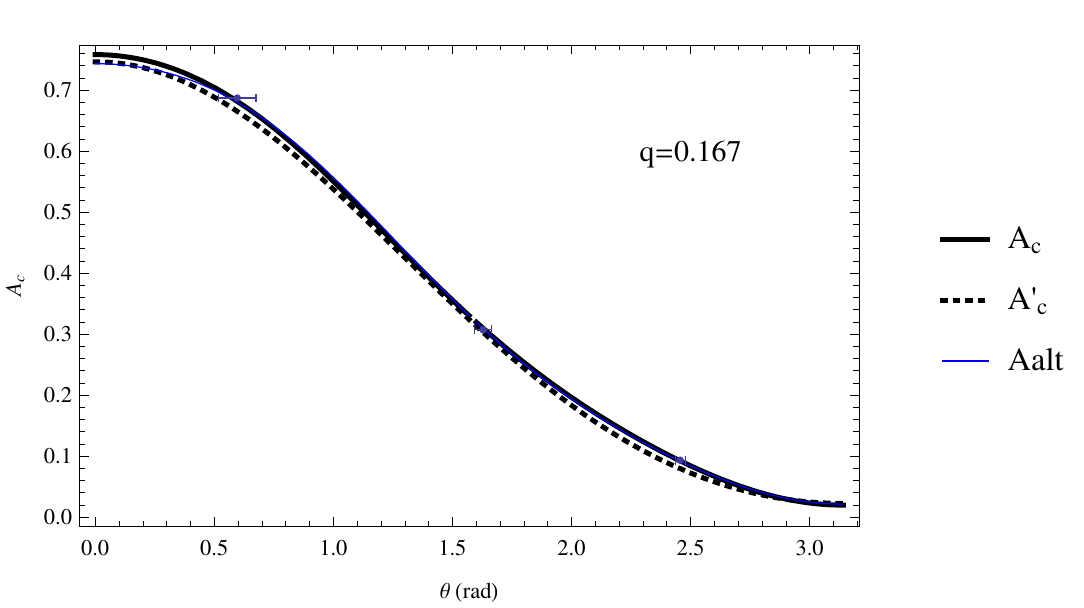}
   \includegraphics[width=.45\textwidth]{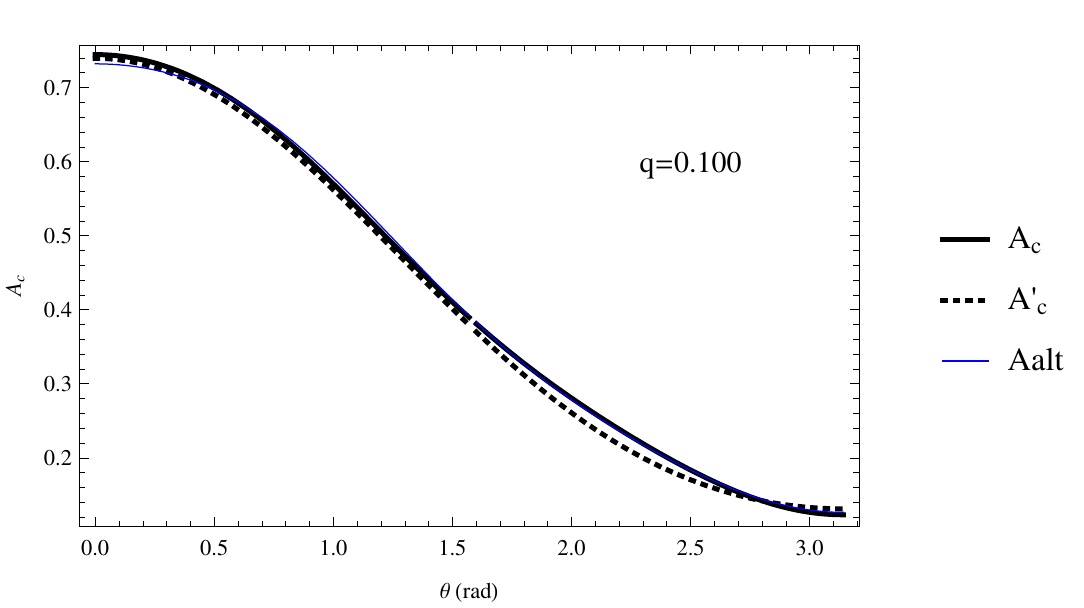}
   \includegraphics[width=.45\textwidth]{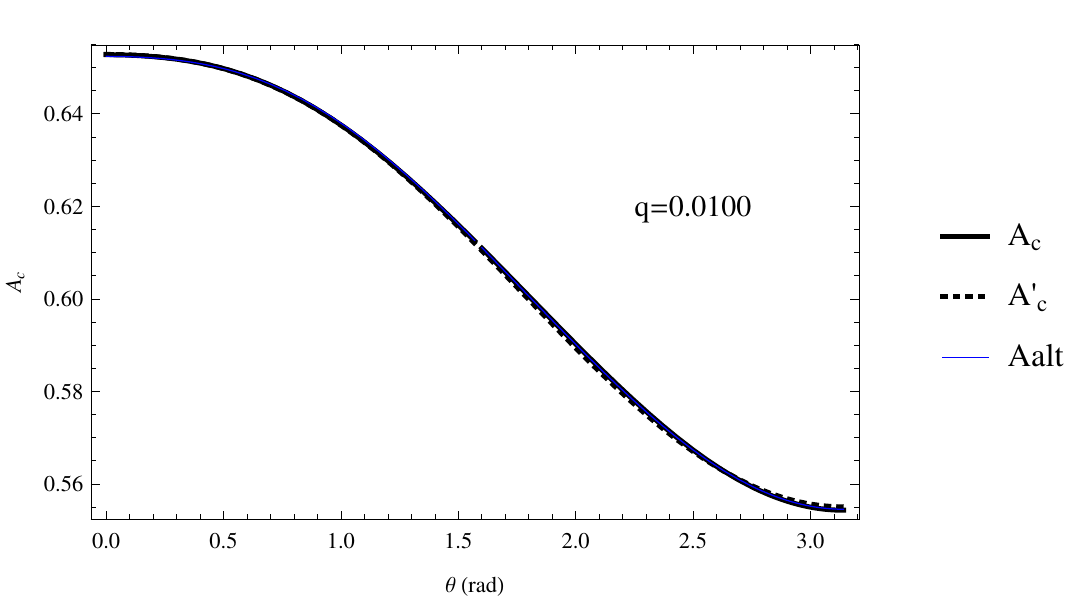}
  \caption{Plots of the fitted $A_c$ versus inclination angle $\theta$
and $q$ for the NQ
configurations. Each data point represents the value  of $A_c$ for a
family of azimuthal configurations with the same inclination angle and
mass ratio. Note that $\alpha \approx \sqrt{A_c}$ and that a prime
denotes that
$\vec S_0$ was used in the fits, rather than $\vec S$.
}\label{fig:Ac_fits}
 \end{figure*}

\begin{figure}
   \includegraphics[width=.9\columnwidth]{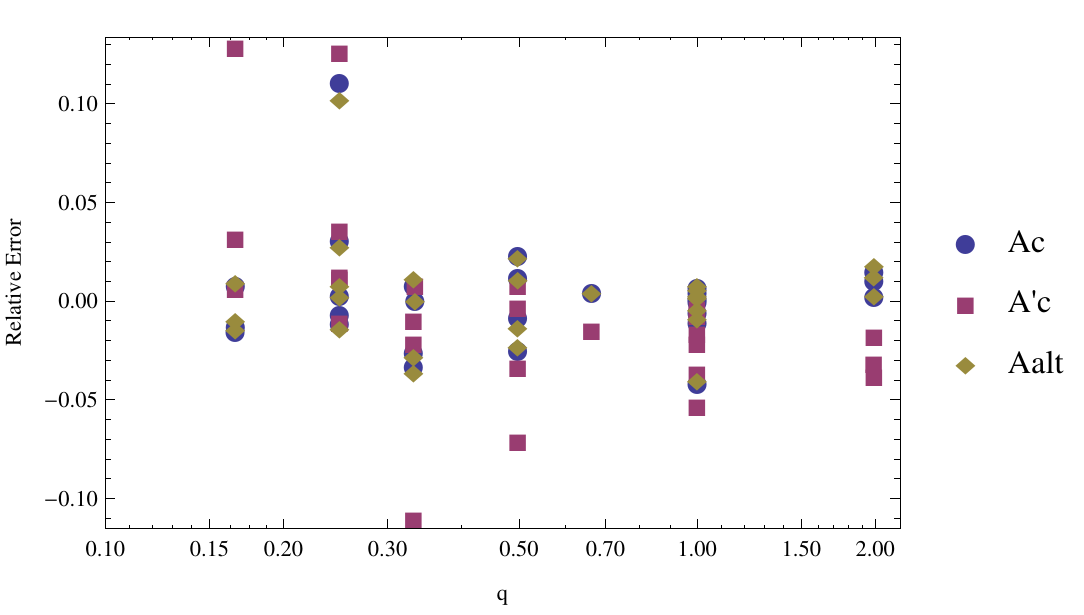}
   \includegraphics[width=.9\columnwidth]{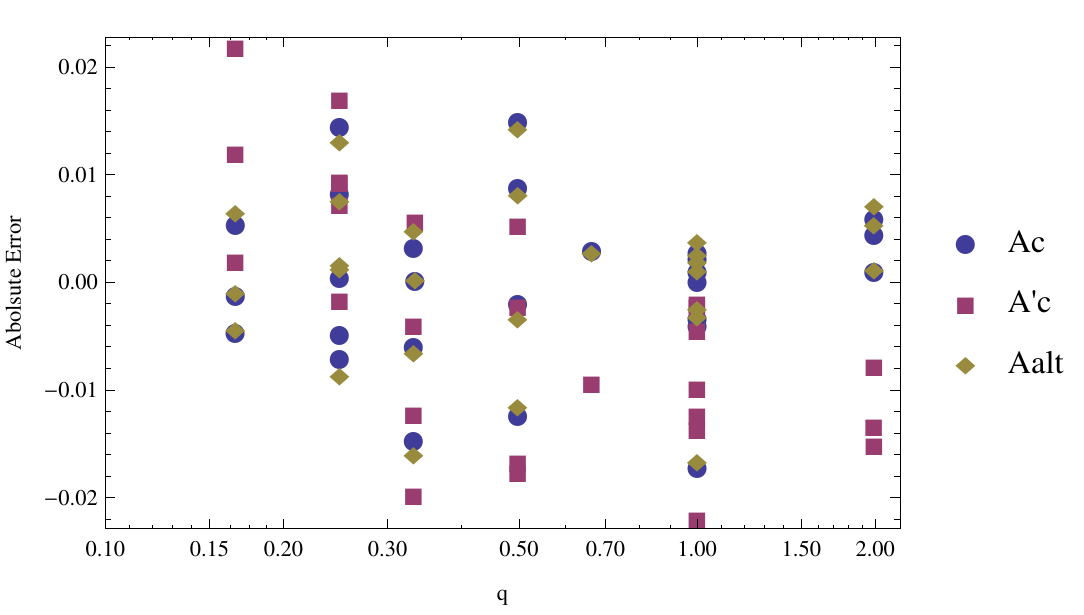}
  \caption{The relative errors (residuals) in the fit of $A_c$ as a
function of the binary's parameters.
Note that there are multiple data points for each $q$
and a prime
denotes that
$\vec S_0$ was used in the fits, rather than $\vec S$.}\label{fig:Ac_resid}
\end{figure}

Overall the fits for the radiated energy $\delta {\cal M}$ are
accurate to within $3\%$ (that is a 3\% error in the radiated mass,
not a 3\% error in the remnant mass) and the fits for the square of
remnant spin are accurate to within $10\%$. There is no clear
advantage here for using $\vec S_0$ or $\vec S$ as the expansion
variable.

\section{Statistical Analysis}\label{sec:analysis}

Now that we have empirical formulas for the recoil velocity of the
remnant BH that are shown to be accurate at least for moderate
spins and mass ratios, we can begin to model the distributions of
astrophysical recoils. Our formulas are based on the spin-magnitudes
and direction measured during the final plunge. We will thus be using
several assumptions to tie the statistical distributions of spins for
distant binaries to the statistical distributions of spins near
merger. Our primary assumption will be that the distribution of
inclination angles at merger is the same as the distribution of
inclination angles for distant binaries (note, this is an assumption
on distributions, we are not assuming that a given binary's
spin-inclination angle will not change). We expect that for distant
binaries the azimuthal orientations of the spins are uniformly
distributed. However, post-Newtonian spin resonances can align or
antialign the two spins in the binary azimuthally
\cite{Schnittman:2004vq, Kesden:2010ji, Berti:2012zp}. To account for
this, we will consider three azimuthal distributions, spins
aligned azimuthally, spins antialigned azimuthally, and random
azimuthal alignments.

To this end, we consider binaries with spin magnitudes
 $\alpha_1$ and $\alpha_2$
given  by the  hot and cold accretion models (i.e., wet accretion)
described
in Ref.~\cite{Lousto:2012su}, and the
``dry'' merger model described in Ref.~\cite{Lousto:2009ka}.
The distributions are given by
$P(\alpha) \propto (1-\alpha)^{(b-1)}  \alpha^{(a-1)}$, where
$a=3.212$, $b=1.563$; $a=5.935$,
$b=1.856$; and $a=10.5868$,  $b=4.66884$, for
hot, cold, and dry mergers, respectively.

For the directions of the spins $\hat{S}_1$ and $\hat{S}_2$, we
use the distributions
$P(\theta) \propto (1-\theta)^{(b-1)}  \theta^{(a-1)}$,
there $\theta$ is measured in radians
and
$a=2.018$, $b=5.244$ and $a=2.544$,
$b=19.527$ for hot and cold accretion, respectively.
For dry mergers, we choose a distribution uniform in $\cos\theta$.
Note that the distributions for $\theta$ assume $0\leq\theta\leq1$.
The probabilities for $\theta>1$ are taken to be identically
zero.

In addition, we use a  mass ratio distribution motivated by
cosmological simulations  $P(q) \propto q^{-0.3} (1-q)$,
as given in Ref.~\cite{Yu:2011vp, Stewart:2008ep, Hopkins:2009yy}.
Note that our formulas are constructed so that the same recoil /
remnant properties are given when the labels of the two BHs are
interchanged ($1\leftrightarrow2$). Hence we need to only consider
$0\leq q\leq1$.

We performed the statistical analysis itself by analyzing the recoil,
radiated mass, and remnant spin from $10^9$ randomly chosen
configurations consistent with the above distributions for the
parameters
 of the binary. Note that we did not make any assumptions about
correlations between these parameters (with the exception of the above
mentioned azimuthal distributions).

The total radiated energy and final remnant spin for a generic BHB is
given by 
Eqs.~(\ref{eq:Ec_aligned})--(\ref{eq:Ac_all_fit}) directly.  The total
recoil, however, is given by $V_{\rm rec}^2 = V_\|^2 + V_\perp^2$,
where $V_\|$ is given by one of
(for the sake of brevity, we only give the explicit formulas for
$V_{4'59}$ and $V_{p'59}$)
\begin{widetext}
\begin{eqnarray}
V_{\|4'59} &=& \left(4 \eta^2\right)\Bigg[
\vec{\tilde \Delta}\cdot\hat n_0 (3678. - 2475
\delta m^2 + 4962. \tilde S_{0\|} + 
    7170. \tilde S_{0\|}^2 + 12050. \tilde S_{0\|}^3) \nonumber \\
  && + 
 \vec{\tilde S}_{0}\cdot\hat m_{59} \left( {\tilde \Delta}_\| \left[4315.
-1262 \delta m^2  + 15970 \tilde S_{0\|}\right] - 2256 \delta m -2231
\delta m \tilde S_{0\|}\right)\Bigg],
\label{eq:gen_v4'59}\\
V_{\|p'59} &=&\left(4 \eta^2\right)\Bigg[\vec {\tilde \Delta}\cdot\hat
n_0 \left( \frac{3685 (1 -
0.6766 \delta m^2 + 0.1410 \tilde S_{0\|})}{1 - 1.248 \tilde S_{0\|}}
-2537 {\tilde \Delta}_\| \delta m\right) \nonumber\\
&&
    + \vec{\tilde S}_{0}\cdot \hat m_{59}
\left(\Delta_\|\left[4180+1660
\tilde S_{0\|}\right] - 2565  \delta m
\right)\Bigg{]},\label{eq:gen_vp'59}
\end{eqnarray}
and $V_\perp^2$ is given by
\begin{eqnarray}
V_{\perp}^2 &=& (4\eta)^4 \Big(2.106\times10^5 \tilde \Delta_\|^2 + 4.967\times10^5 \tilde
\Delta_\| \delta m -
     2.116\times10^5 \tilde \Delta_\|^3 \delta m - 5.037\times10^5 \tilde
\Delta_\|^2 \delta m^2 \nonumber \\
   && -
    1.269\times10^5 \tilde \Delta_\| \delta m^3 
   - 3.384\times10^5 \tilde
\Delta_\|^2 \tilde S_{0\|} -
    6.440\times10^5 \delta m^2 \tilde S_{0\|} + 2.138\times10^6 \tilde
\Delta_\|^2 \tilde S_{0\|}^2 \nonumber \\
   && -
    4.905\times10^6 \tilde \Delta_\| \delta m \tilde S_{0\|}^2 -
1.100\times10^6 \delta m^2 \tilde S_{0\|}^2 
   - 1.024\times10^7 \tilde \Delta_\|^2 \tilde S_{0\|}^3\Big) +\Big[
1.2\times10^{4} \eta^2 \delta m (1-0.93
\eta)\Big]^2.\label{eq:gen_vperp}
\end{eqnarray}
\end{widetext}
 In Eqs.(\ref{eq:gen_v4'59}) and
(\ref{eq:gen_vp'59}) there are two unspecified unit vectors $\hat n_0$ and
$\hat m_{59}$. As explained in Sec.~\ref{sec:fit-recoil}, $\hat n_0$
is a unit vector in the final orbital plane and $\hat
m_{59}$ is another unit vector in this plane rotated by $-59^\circ$
with respect to $\hat n_0$.
The direction of $\hat n_0$ is unknown
(we only know that it must lie in the final orbital plane).
From a practical point of view, this means that if we choose azimuthal
distributions that are uniform with respect to some external reference
frame, then the choice of $\hat n_0$ will not affect the resulting
distributions of recoils. In practice we take
$\hat n_0 = (1,0,0)$ and $\hat m_{59} = (\cos 59^\circ, -\sin
59^\circ,0)$.  Finally, Eq.~(\ref{eq:gen_vperp}) was
obtained by fitting the square of the recoil for the nonprecessing runs
in~\cite{Healy:2014yta}. This formula has the advantage that there is
no explicit dependence on the angle $\xi$, but does have the drawback
that it can predict negative values for $V_\perp^2$. This
can only happen when $V_\perp^2$ is small and we therefore take
$V_\perp^2=0$ in these cases.

\begin{figure}
  \includegraphics[width=.9\columnwidth]{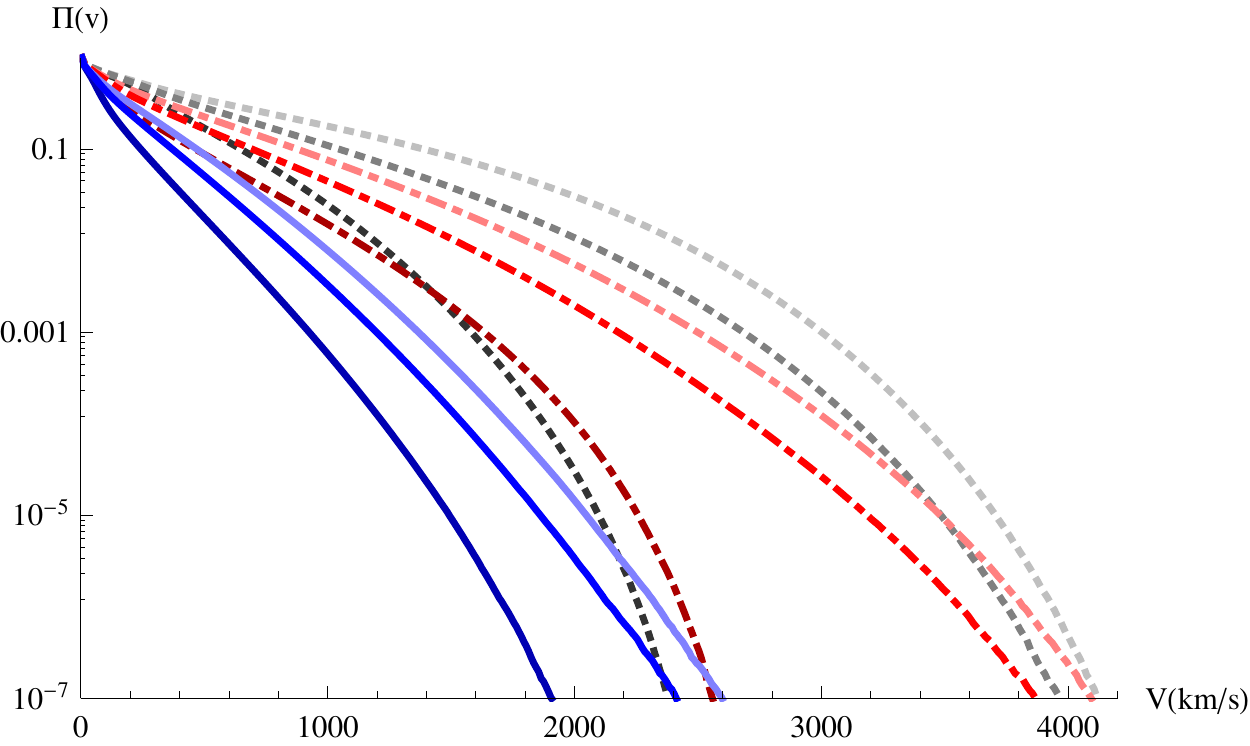}
  \includegraphics[width=.9\columnwidth]{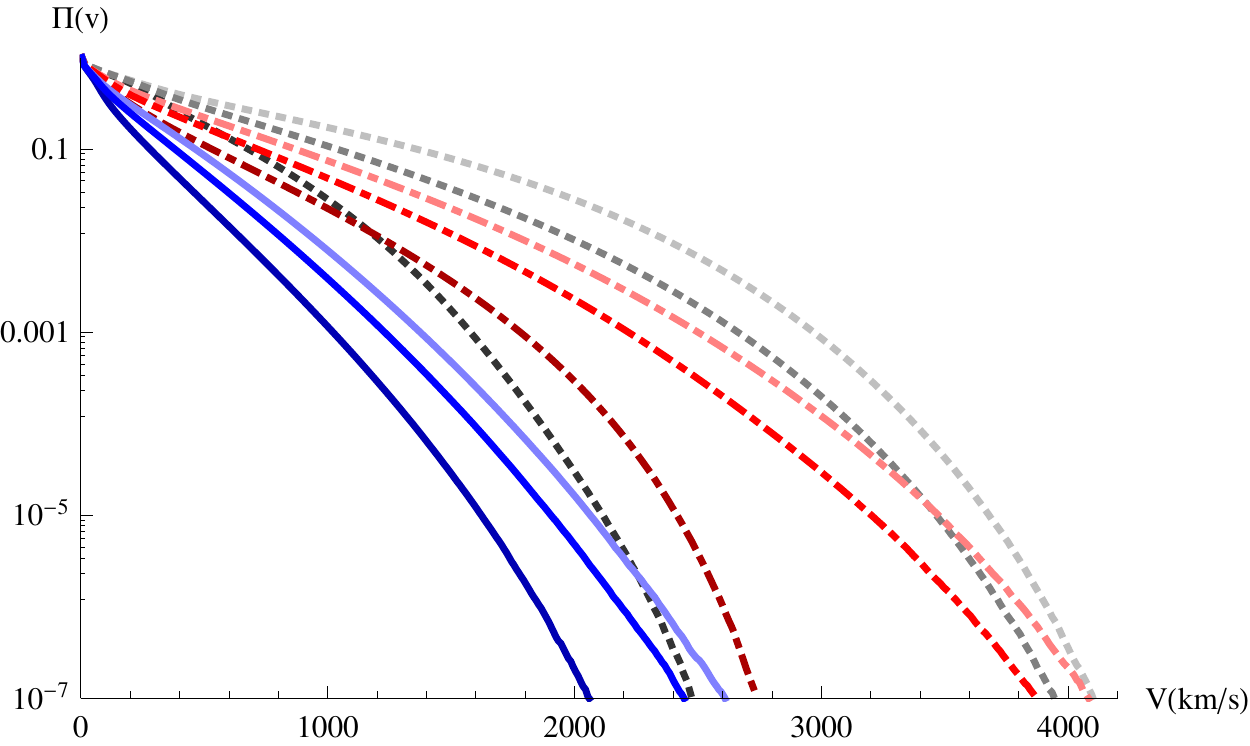}
\caption{The integrated probability $\Pi(v)$ for the remnant to recoil
at speed $v$ or higher (in $\KMS$). The blue (solid) curves are for cold
accretion models, the red (dot-dashed) curves are for hot accretion
models, and gray (dotted)
curves are for dry mergers. Within a given color/line style (blue, red, gray),
the dark shade indicates that the spins were aligned azimuthally,
the light shade indicates that the spins were antialigned
azimuthally, and the intermediate shade indicates random azimuthal
alignment.
The top plot shows the probabilities when modeling the recoil with $V_{459}$
and the lower panel shows the probabilities when modeling the recoil
with $V_{4'59}$.
\label{fig:vstats}
}
\end{figure}

\begin{figure}
  \includegraphics[width=.9\columnwidth]{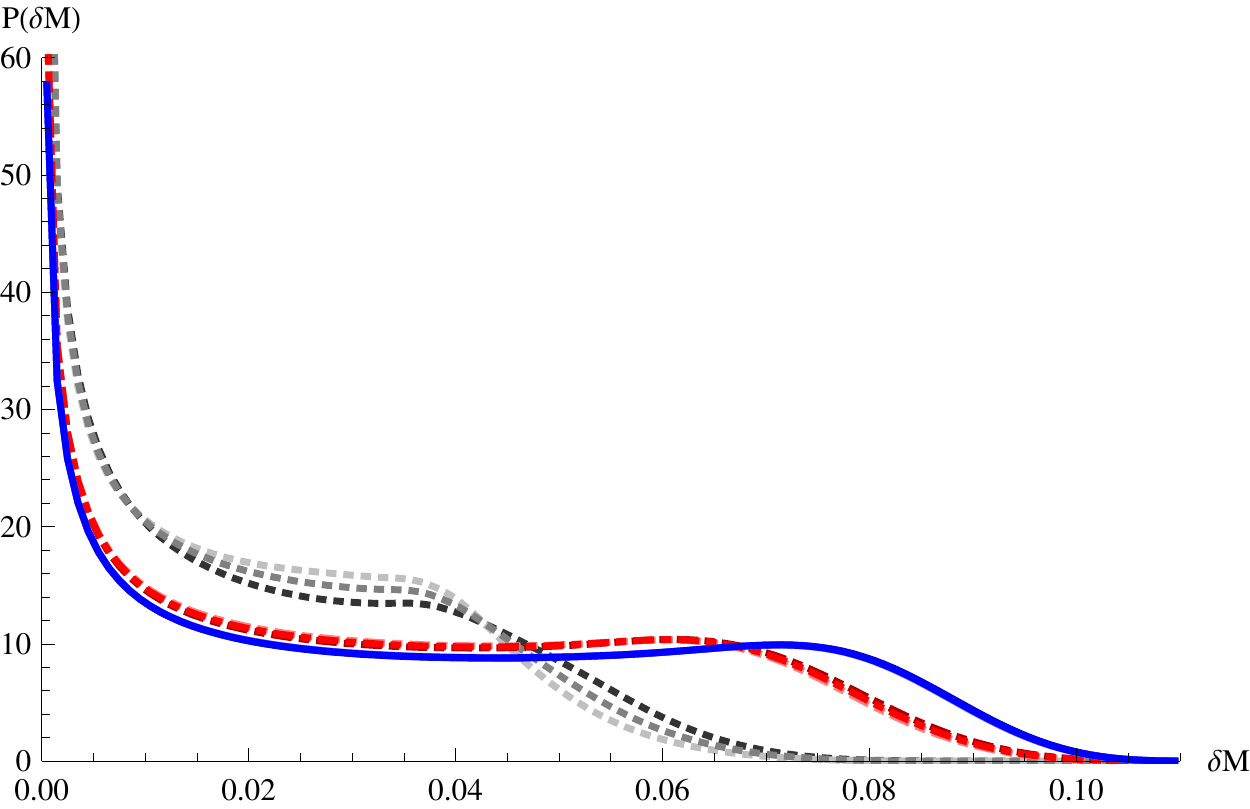}
  \includegraphics[width=.9\columnwidth]{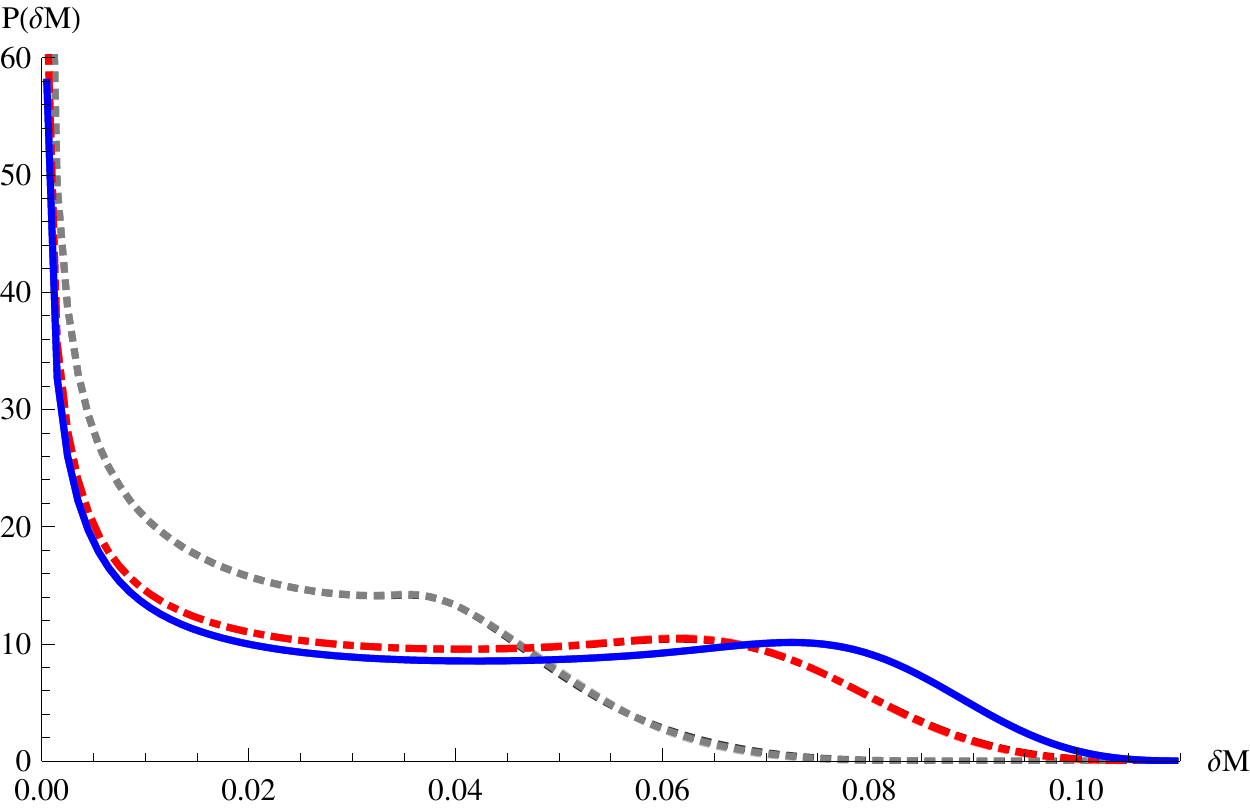}
\caption{The probability (nonintegrated)
for a mass loss of $\delta {\cal M}$.
The blue (solid) curves are for cold
accretion models, the red (dot-dashed) curves are for hot accretion
models, and gray (dotted)
curves are for dry mergers. Within a given color/line style (blue, red, gray),
the dark shade indicates that the spins were aligned azimuthally,
the light shade indicates that the spins were antialigned
azimuthally, and the intermediate shade indicates random azimuthal
alignment. The upper panel displays probabilities when the radiated
energy is modeled in terms of the spin variable $\tilde{S}$ and the
lower panel when the variable is chosen to be $\tilde{S}_0$.
}
\label{fig:mstats}
\end{figure}

\begin{figure}
  \includegraphics[width=.9\columnwidth]{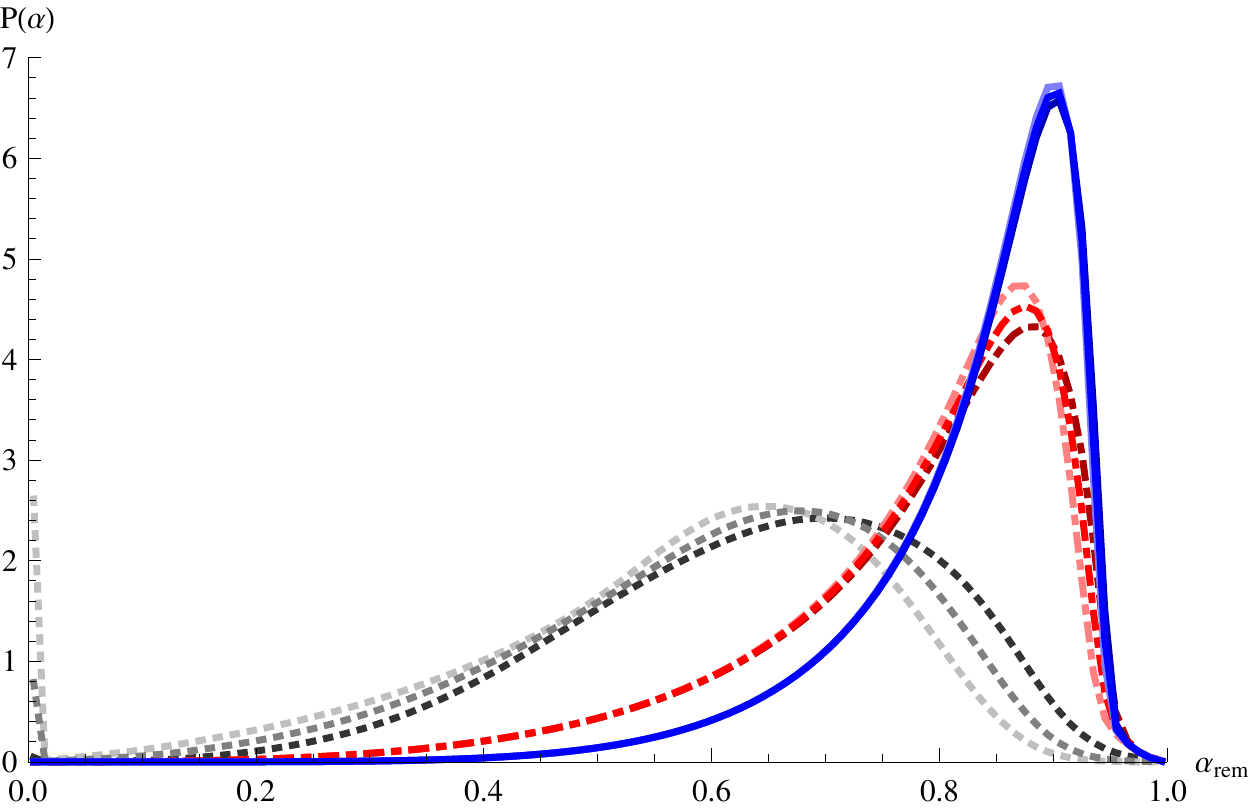}
  \includegraphics[width=.9\columnwidth]{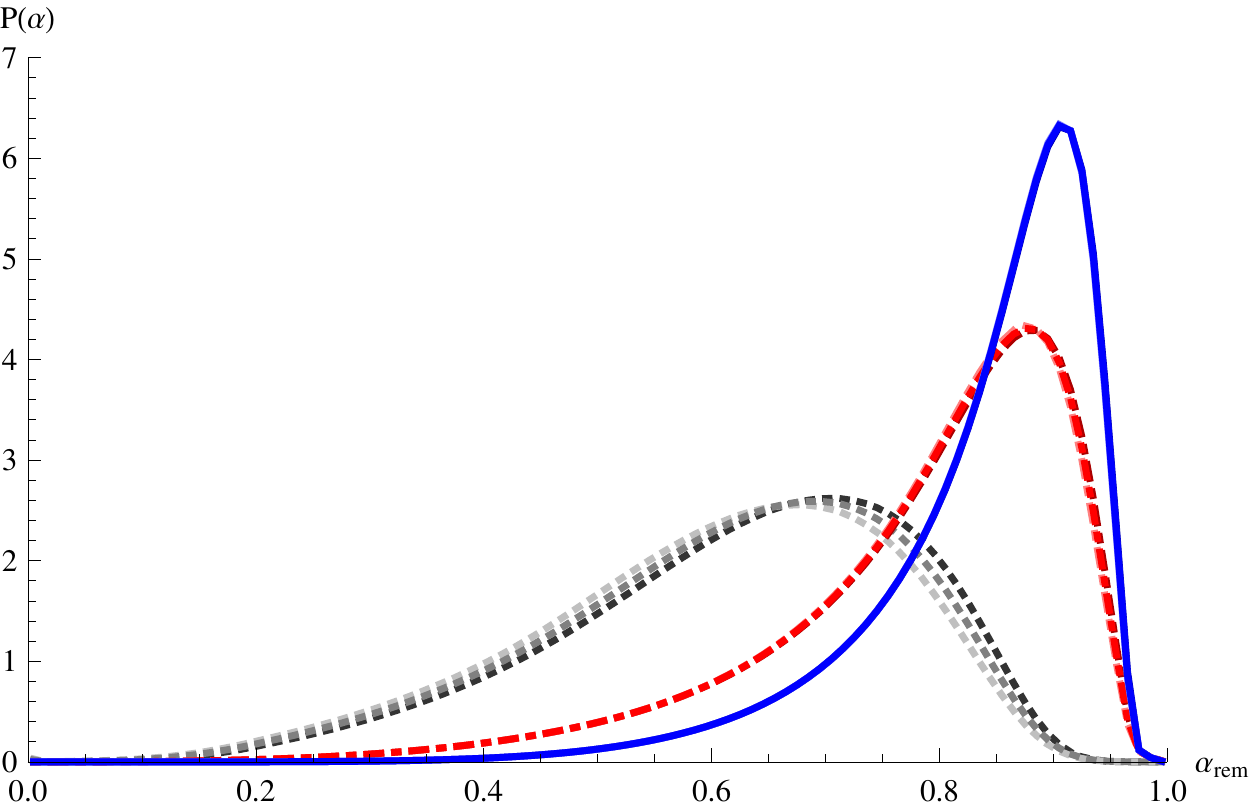}
\caption{The probability (nonintegrated)
for a remnant spin $\alpha$.
The blue (solid) curves are for cold
accretion models, the red (dot-dashed) curves are for hot accretion
models, and gray (dotted)
curves are for dry mergers. Within a given color/line style (blue, red, gray),
the dark shade indicates that the spins were aligned azimuthally,
the light shade indicates that the spins were antialigned
azimuthally, and the intermediate shade indicates random azimuthal
alignment. The upper panel displays probabilities when the final remnant spin
magnitude is modeled in terms of the total spin variable $\tilde{S}$ and the
lower panel when the variable is chosen to be $\tilde{S}_0$.
}
\label{fig:astats}
\end{figure}

Figure \ref{fig:vstats} shows the resulting probabilities for
a given recoil $v$ or larger (i.e., an integrated probability).
Perhaps not too surprisingly, the {\it dry} distribution with
antialigned spins (azimuthally) give the largest probabilities for
high recoils. We summarize the probabilities for very high
recoils in Table~\ref{tab:PlargeKick}. Assuming the most
pessimistic distribution (cold accretion, azimuthal alignment),
there is a 2 in $10^7$ chance of a supermassive BH recoiling
at $2000\ \KMS$. For dry mergers with azimuthal alignment, on the other hand, the probability
would be 142 times larger (for dry mergers with random alignment the
probability would be 47 000 times larger).

\begin{figure*}
   \includegraphics[width=.3\textwidth]{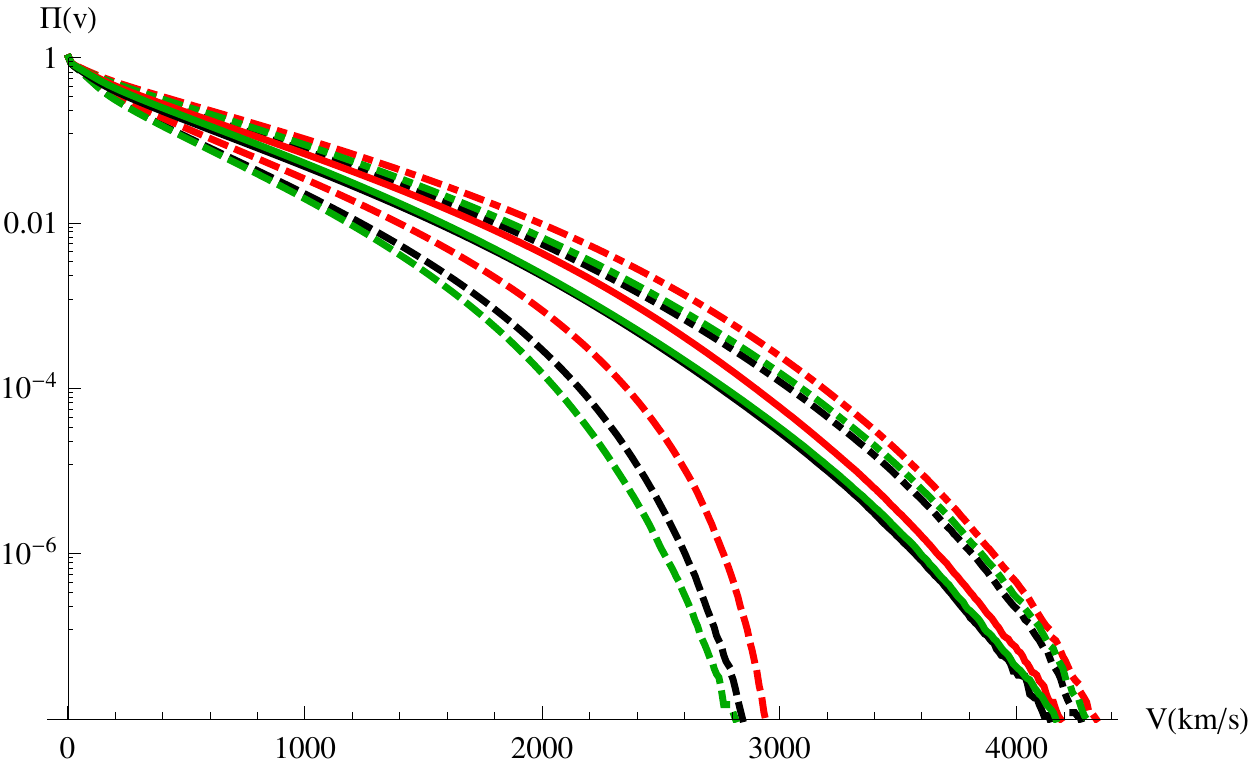}
   \includegraphics[width=.3\textwidth]{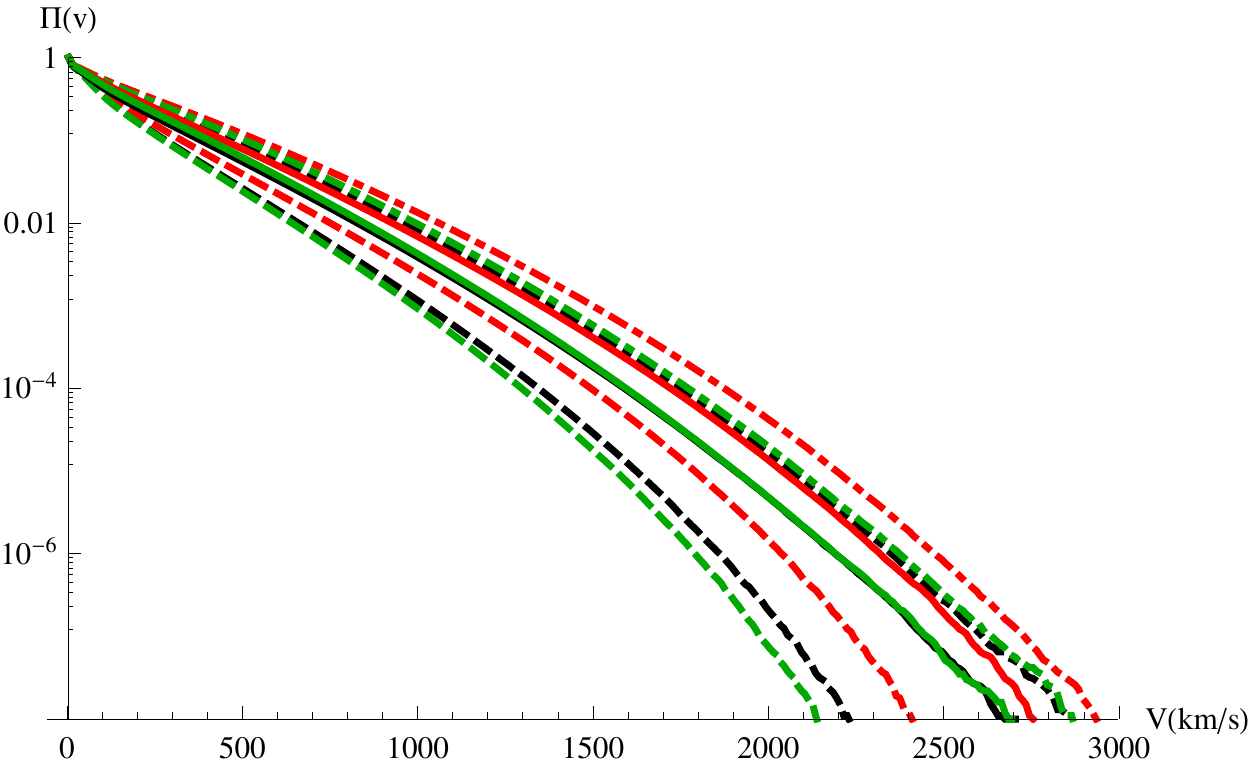}
   \includegraphics[width=.3\textwidth]{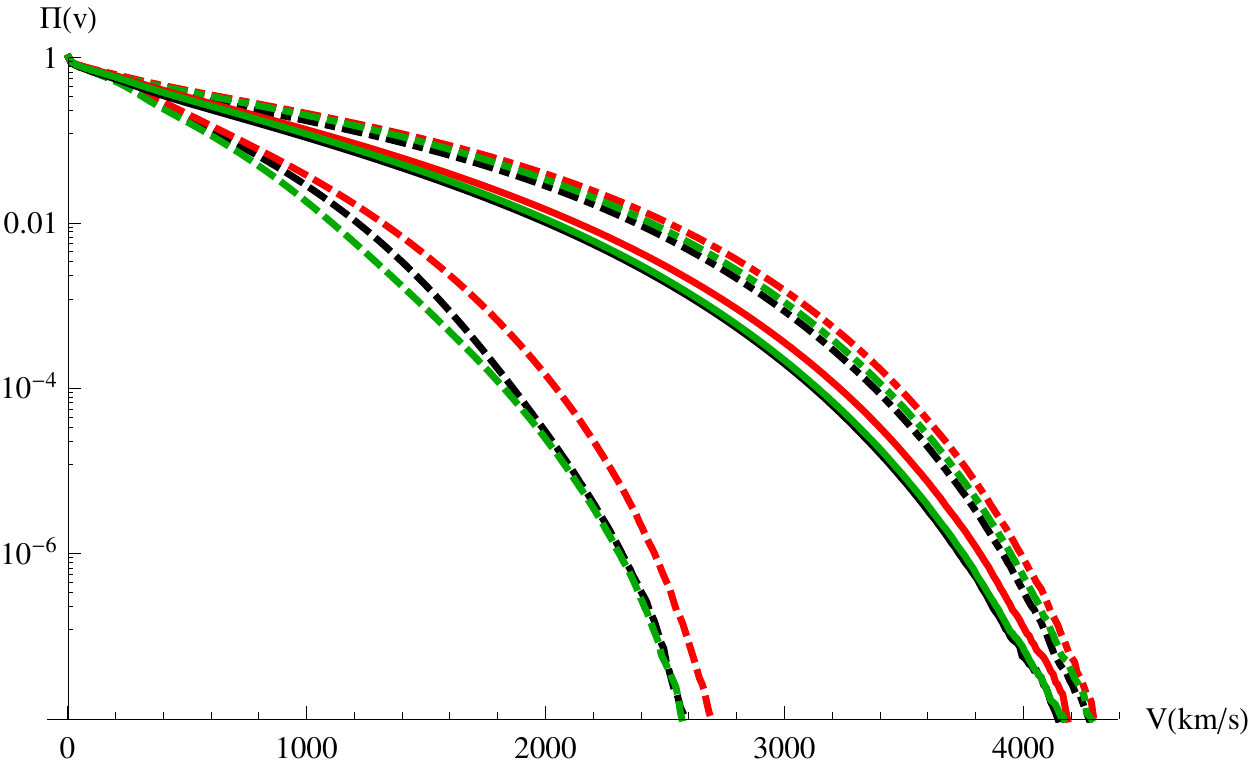}
  \caption{The integrated probability $\Pi(v)$ for the remnant to
recoil
at speed $v$ or higher (in $\KMS$) as predicted using $V_{p'59}$
(black curves), the older \cross formula (red curves), and the
original \hangup formula (green curves). The plot on the left shows the
results  for hot
accretion, the plot in the center shows the results for cold accretion, and the plot
on the right shows the results for dry mergers. In each plot, solid curves are for
random azimuthal alignment, dashed curves are for azimuthal alignment,
and dot-dashed curves are for antiazimuthal
alignment.}\label{fig:new_old_kick_comp}
\end{figure*}

 In
Figs.~\ref{fig:new_old_kick_comp}, we compare
the new predicted distributions with the \hangup and \cross
predictions.  Note that if we assume random azimuthal alignment, the
predictions of $V_{4'59}$ match very closely to the \hangup
predictions. Interestingly, while the \cross and \hangup predictions were
based on simple ans\"atze for how the equal-mass contributions to the recoil
generalize, the predictions are not too different (within a factor of 2)
from the results obtained by our new fitting to unequal-mass 
configurations. This gives some assurance that further modifications to the empirical
formula for the recoil will give incremental improvements in accuracy.

\begin{table}
  \caption{The probability in units of percent of a recoil $v$ or larger assuming
    the Hot, Cold, and Dry merger models and assuming
    the spins are antialigned (AA) azimuthally,
    aligned (A) azimuthally, or randomly distributed (R) azimuthally.
In all cases the recoil was calculated using
$V_{4'59}$ for the number outside the parenthesis and
using $V_{p'59}$ for the number inside the parenthesis.}\label{tab:PlargeKick}
  \begin{ruledtabular}
  \begin{tabular}{lrrr}
     Model & $\Pi(1000\KMS)$ &  $\Pi(2000\KMS)$ & $\Pi(3000\KMS)$\\
     \hline
     Hot A  & 2.292 (2.280 ) & 0.30 (0.029 ) & 0(0)\\
     Hot R  & 4.884 (4.721 ) & 0.233 (0.229 ) & 0.003 (0.003 )\\
     Hot AA & 7.568 (7.220 ) & 0.563 (0.550 ) & 0.012 (0.014 )\\
     Cold A  & 0.120 (0.126 ) & $2\cdot10^{-5}$ ( $3\cdot10^{-5}$) & 0(0)\\
     Cold R & 0.398 (0.418 ) & $5\cdot10^{-4}(7\cdot10^{-4})$ &
$0(2\cdot 10^{-7})$\\
     Cold AA & 0.814 (0.846 ) & 0.002 (0.003 ) &
$1\cdot 10^{-7}(6\cdot 10^{-7})$ \\
     Dry A  & 2.900 (3.216 ) & 0.003 (0.003 ) & 0(0)\\
     Dry R & 10.932 (11.006 )& 1.033 (1.009 ) & 0.020 (0.019 )\\
     Dry AA & 17.404 (17.327 ) & 2.849 (2.759 ) & 0.088 (0.082 )\\
  \end{tabular}
  \end{ruledtabular}
\end{table}

In Figs.~\ref{fig:mstats} and~\ref{fig:astats},
we show the probability distributions for a binary losing $\delta{\cal
M}$ of its total mass to gravitational radiation [i.e., $P(\delta{\cal
M})$] and the probability that the remnant will have a spin $\alpha$
[i.e.,
$P(\alpha)$].  Unlike in the previous figures, here we show the raw
probabilities rather than the integrated ones.
The probability distribution $P(\delta{\cal M})$ has three distinct
regions: a large peak centered at $\delta{\cal M} = 0$, which is
produced by the small mass ratio binaries, a plateau where the
distribution is almost constant, and decaying tail at {\it high}
energies. The plateau ends at $\delta{\cal M}\approx 4\%$ for
dry mergers, $\delta{\cal M}\approx 7\%$ for hot accretion, and
$\delta{\cal M}\approx 8\%$ for cold accretion.
The plateau extends to the highest energies for the cold
accretion model, indicating that such binaries will, on average, be
the loudest gravitational wave sources.

This long plateau in the radiated energy distribution is related to the
probability that merger remnant will have high spin, which in turn is
related to the probability that the binary will have a large net spin
in the direction of the orbital angular momentum. As shown in
Fig.~\ref{fig:astats}, the probability  distribution for the remnant
spin magnitude for dry mergers is very broad and peaks at
$\alpha\approx0.7$, with very low probabilities for high spins.
The hot and cold accretion models lead to much narrower peaks centered
at higher spins (near  $\alpha\approx0.9$ for cold accretion).
Both these models have the spins of the two BHs strongly aligned with
the orbital angular momentum. This leads to both large radiated
energies and large remnant spins~\cite{Lousto:2012su}.

The fact that black holes merging in an accretion dominated environment
have a non-negligible probability of radiating up to $8 - 9\%$ of their
total mass make them more {\it visible} for gravitational wave detectors
than binary black holes merging in a relatively dry scenario. In
particular, according to Fig.~\ref{fig:mstats} wet mergers
produce nearly double the radiation of dry mergers (and hence
roughly 1.4 times the gravitational wave strain),
which means that merging BHs from accretion dominated systems
are detectable in a volume 2.8 times larger than for
dry mergers~\cite{Aylott:2009ya, Aylott:2009tn, Ajith:2012az, Aasi:2014tra}.

\section{Discussion}\label{sec:discussion}

In this paper we  revisited the question of generating empirical
formulas describing the remnant mass, spin, and recoil from the
mergers of black-hole binaries.
We extended the formulas of
Refs.~\cite{Lousto:2012gt,Lousto:2013wta}
to include explicit mass difference dependence.

Our final formula for the recoil along the orbital angular
momentum at merger is given by
Eqs.~(\ref{eq:Vkick_final_1})-(\ref{eq:Vkick_final_final})
with fitting coefficients provided in
Table~\ref{tab:vkick_fit_param}.
While we provide several alternatives of fitting to
study the robustness of the empirical formula, our results
favor formulas $V_{4'59}$ and $V_{p'59}$.

While Fig.~\ref{fig:Vfits} provides an overview of the quality of
the fittings for the recoil velocities from our simulations
in the range of $1/6\leq q\leq2$, Fig.~\ref{fig:Vresid}
gives a more quantitative measure of the absolute and
relative errors of the fits. We observe that all but one
point lies within a relative error of 12\%
(which translates to an absolute error bound of
within  $60\KMS$).
These errors should be acceptable for most astrophysical
applications, and in particular to estimate the probability
of observing recoiling black holes near active galactic
nuclei with peculiar features such as displaced narrow and
wide spectral lines, displaced luminosity centers, etc.
(see \cite{Komossa:2012cy, Bogdanovic:2014cua} for a review).
An important factor to consider is also the lifetime
of accretion disks carried by
recoiling black holes \cite{Blecha:2010dq, Blecha:2008mg},
as they can only be observed for a few million years and
at distances from the center of the colliding galaxies
that depend strongly on the angle of the recoil with respect to the final
orbital plane. Importantly, large recoil velocities are strongly beamed
along the orbital angular momentum (see Figs. 11-14 of
Ref.~\cite{Lousto:2012su} and Fig.~\ref{fig:recoil_ang_dist} here).

We also provide fitting formulas adapted to include the
unequal mass parameter $\delta{m}$ for the total radiated
gravitational energy of the binary. The leading terms of
the radiated energy are given by expression (\ref{eq:Ec_all_fit}),
with fitting coefficients
 given in Table~\ref{tab:Ec_all}.
Figure~\ref{fig:Ec_fits} shows the actual fitting curves
for the alternative variables based on $\vec{S}$ or
$\vec{S}_0$, which provides a  measure of the errors in
truncating the fitting formula.
Figure~\ref{fig:Ec_resid}
shows the residuals of such fits (the
error is within $3\%$ of the total radiated
energy).

The final spin magnitude of the remnant black hole can also be
fitted with our new approach. The leading term of
the final spin (actually $\alpha^2$) is given by expression
 (\ref{eq:Ac_all_fit}), with fitting coefficients
 given in Tables~\ref{tab:Ac1}-\ref{tab:Ac3}.
Figure~\ref{fig:Ac_fits} shows the actual fitting of the curves
for the alternative variables based on $\vec{S}$ or
$\vec{S}_0$, which  provides a measure of the errors in
truncating the fitting formula.
Figure~\ref{fig:Ac_resid}
shows the residuals of such fittings. The
relative errors (except for one point)
are within $10\%$ for the square of the spin.
Regarding the final spin direction, as reported in the last two
columns of Table~\ref{tab:rem_rad_cmp_part1},
we observe that the net deviation of the total angular momentum
with respect to its initial orientation is
always small (within a few degrees) for comparable mass ratios
(within 1:2). This deviation  increases,
reaching up to 20 degrees,
with smaller mass ratios and spins pointing in the opposite
direction to the orbital angular momentum, in agreement with
the studies in Refs.~\cite{Lousto:2013vpa, Lousto:2013wta}.

We note that our modeling is based on configurations with one
BH spinning and the other nonspinning extrapolated to both BHs
spinning. In the small-mass-ratio regime, where the spin of the
lighter component will have a relatively small effect, this
extrapolation should be accurate.  However, when the spins of both BHs
are dynamically important, we can foresee two main sources of error. First
$\vec S$ and $\vec \Delta$ will not be aligned, which means that
our formulas should depend on the azimuthal orientations of $\vec
S_\perp$ and $\vec \Delta_\perp$ independently. Second, the magnitude
of $\vec S$ and $\vec \Delta$ can be effectively double the magnitudes
achievable with the NQ configurations  (but only in the similar mass
regime). We partially addressed the first source of error when we refit the K
configurations. Our new model is based on the terms
$\Delta_\perp = \vec \Delta \cdot \hat n_0$ and $S_\perp = \vec S
\cdot \hat m_0$ where $\hat n_0$ and $\hat m_0$ are rotated with
respect to each other by $59^\circ$, which we found to be the correct
azimuthal dependence of the recoil for K configurations. And while
this modification allowed for an accurate modeling of the K
configurations, it had a negligible effect on the statistical
distribution of recoils. We thus have good evidence that this first
source of error is acceptable for statistical studies. The second
source of error is potentially  more problematic because we use terms
up to fourth-order in the spin (and hence errors can increase by a
factor of 16).  Fortunately, these terms tend to be largest for the
equal-mass configurations that we previously studied and used to
construct our empirical formulas.

Pad\'e approximants give an alternative to the Taylor-like 
expansions to fit the remnant recoil. We first used a Pad\'e
approximation when modeling 
the ``hangup-recoil'' configurations of 
Ref.~\cite{Lousto:2011kp,Lousto:2012su}. There our goal
was to resum the $S_\|$ dependence (which proved to be a very
slowly converging series).
Here,
in the unequal mass context, we use the more  appropriate variable 
$\vec{S}_0\cdot\hat{L}$. This variable has the
advantage of being ``essentially'' conserved \cite{Racine:2008qv}
during evolution thus allowing us to relate the parameters of the
binary at  large separations
 with the parameters around merger.
The Pad\'e expansion has also been used in Ref.~\cite{Hemberger:2013hsa}
to better fit the energy radiated (or final mass) of equal-mass,
(anti)aligned spins of merging binaries.

One clear avenue for improvement of our modeling concerns the fact that
we base our formulas on the spin orientations near merger, rather than
at large separations. Such a program would implicitly entail modeling
the precession of the spins and orbital plane from the distant PN
regime down to merger. The use of $S_{0\|}$ in our modeling is a step
in this direction and the work of Ref~\cite{Galley:2010rc} to find
other constants of the motion and of Ref~\cite{Kesden:2014sla} to
model the precision may prove to be useful.

One can foresee a further decomposition of the modeling of the recoil
into three distinct characteristic regimes: the inspiral (where most
of the recoil is representing by an almost self-canceling wobbling
of the center of mass with the orbital period), the merger, where most
of the anisotropic radiation of linear momentum takes place, and
the ringdown of the final, highly distorted, black holes which
gives rise
to the antikick phenomenon~\cite{Rezzolla:2010df}.

While we have included in the modeling the particle limit through
the ISCO energy and angular momentum and the $\eta^2$ leading dependence
of the recoil, a set of simulations in the region around mass ratios
$q=1/10$ would be beneficial to improve the accuracy of the interpolation
fits. Another area of improvement would be to use  near maximally spinning
black holes, i.e. intrinsic spins above $\alpha=0.99$. This is particularly
interesting for the modeling of the recoil since it has been pointed out
in \cite{Hirata:2010xn, vandeMeent:2014raa} that resonance effects in the
for small-mass ratio inspirals around a highly-spinning primary
can lead to mass ratio dependences in the recoil that scale
as $\sim\eta^{1.5}$ rather than $\eta^2$.

Finally, we note that as seen in Fig.~\ref{fig:new_old_kick_comp}, our
new formula for the recoil is consistent (within a factor of 2) with
our older formulas, which were based on ans\"atze on how the
equal-mass contributions to the recoil generalize.  This gives some
assurance that further modifications to the empirical formula for the
recoil will give incremental improvements in accuracy.

In conclusion, we provided a set of formulas that
 describe the final state of the mergers of black hole binaries
 within reasonable errors for astrophysical applications
and tested in the comparable
mass ratio regime of $1/6\leq q\leq6$ and spins $S_i/M_i^2\leq0.8$
with reasonable extrapolation properties.

\acknowledgments

We thank the referee for their careful review and for the many
helpful suggestions on how we could improve the paper.
The authors gratefully acknowledge the NSF for financial support from
Grants
PHY-1305730, PHY-1212426, PHY-1229173,
AST-1028087, PHY-0969855, OCI-0832606, and
DRL-1136221. Computational resources were provided by XSEDE allocation
TG-PHY060027N, and by NewHorizons and BlueSky Clusters
at Rochester Institute of Technology, which were supported
by NSF grant No. PHY-0722703, DMS-0820923, AST-1028087, and
PHY-1229173.


\appendix
\section{Data from the full numerical evolution}\label{app:data}

In this appendix we provide detailed data for the 126 new BHB
configurations studied here. Our configurations have one BH spinning
(generally the larger one, except for the $q=2$ configurations)
 and the other nonspinning. The initial data parameters are given in
Table~\ref{tab:ID_part1}. The radiated angular
momentum, mass, and recoil (all in the original frame) are
given in Table~\ref{tab:rad_part1}. In
Table~\ref{tab:rem_rad_cmp_part1} we
compare the radiated mass and angular momentum as measured by the isolated horizons
formalism to the radiated mass and angular momentum as measured
directly from $\psi_4$. The difference between the two measures
provides an error estimate for the $\psi_4$-based measure of these
quantities. The missing entries in these tables are due to missing
horizon data for certain configurations. Finally, in
Table~\ref{tab:rot_part1}, we give the BH spins
and remnant recoil in the rotated frame of the final plunge.

\begin{longtable*}{r|lllllllllll}
\caption{Initial data parameters. In all cases the puncture masses
were
chosen such that the total ADM mass of the binary was
$1.0\pm10^{-6}M$. Here
the punctures are located at $(x_{1,2},0,0)$
with momenta $\pm(0, p,0)$ and spins
$\vec S_2 = (S_x, S_y, S_z)$ [note that $\vec S_1=0$ for all
configurations]. For the NQ200 configurations $\vec P_1 =
-(0,p,0)$, for all other configurations
$\vec P_1 = (0,p,0)$. In all cases, $\vec P_2 = - \vec P_1$.
The approximate initial
eccentricities $e_i$,
eccentricities measured over the last orbit $e_f$, and the number of
orbits $N$,
are also given.
}
\label{tab:ID_part1}\\

\hline
\hline
CONF & $m_{p1}/M$ & $m_{p2}/M$ & $x_1/M$ & $x_2/M$  & $p/M$ &
$S_x/M^2$ & $S_y/M^2$ & $S_z/M^2$ & $M_{H1}$ & $M_{H2}$ &
$N_{e_{\rm f}}^{e_{\rm i}}$\\
\hline
\endfirsthead
\multicolumn{12}{c}
{{\tablename\ \thetable{} -- continued from previous
page}}\\
\hline
CONF & $m_{p1}/M$ & $m_{p2}/M$ & $x_1/M$ & $x_2/M$  & $p/M$ &
$S_x/M^2$ & $S_y/M^2$ & $S_z/M^2$ & $M_{H1}$ & $M_{H2}$ &
$N_{e_{\rm f}}^{e_{\rm i}}$\\
\hline
\endhead
\hline \multicolumn{12}{r}{{Continued on next page}} \\
\hline
\endfoot
\hline\hline
\endlastfoot
NQ200TH30PH0  &  $0.657661$  &  $0.202748$  &  $-2.771460$  &
$5.527396$  &  $0.097847$  &  $0.000000$  &  $0.045426$  &  $0.078681$
&  $ 0.672522 $   &  $ 0.338027 $   &  $5_{0.005}^{0.02}$ \\
 PH30  &  $0.657672$  &  $0.202779$  &  $-2.771460$  &
$5.527396$  &  $0.097847$  &  $-0.022713$  &  $0.039340$  &
$0.078681$  &  $ 0.672533 $  &  $ 0.338026 $   & \\
 PH60  &  $0.657694$  &  $0.202838$  &  $-2.771460$  &
$5.527396$  &  $0.097847$  &  $-0.039340$  &  $0.022713$  &
$0.078681$  &  $ 0.672554 $ &  $ 0.338021 $   & \\
 PH90  &  $0.657706$  &  $0.202865$  &  $-2.771460$  &
$5.527396$  &  $0.097847$  &  $-0.045426$  &  $0.000000$  &
$0.078681$  &  $ 0.672564 $  &  $ 0.338016 $   & \\
 PH120  &  $0.657695$  &  $0.202837$  &  $-2.771460$  &
$5.527396$  &  $0.097847$  &  $-0.039340$  &  $-0.022713$  &
$0.078681$  &  $ 0.672552 $   &  $ 0.338018 $   & \\
 PH150  &  $0.657672$  &  $0.202780$  &  $-2.771460$  &
$5.527396$  &  $0.097847$  &  $-0.022713$  &  $-0.039340$  &
$0.078681$  &  $ 0.672531 $   &  $ 0.338024 $   & \\
\\
NQ200TH60PH0  &  $0.657769$  &  $0.202707$  &  $-2.891856$  &
$5.572900$  &  $0.096793$  &  $0.000000$  &  $0.078662$  &  $0.045416$
&  $ 0.672365 $   &  $ 0.338011 $   &  $4^{0.02}_{0.005}$ \\
\end{longtable*}

\begin{longtable*}{r|llllllllll}

\caption{Radiated mass, angular momentum, and the
remnant recoil (in original frame) as calculated from $\psi_4$. Errors
quoted are from differences between two extrapolations to $r=\infty$.
See
Table~\ref{tab:rem_rad_cmp_part1} for
more accurate measurement of the error. }
\label{tab:rad_part1}\\
\hline
\hline
CONF & $\delta M_{\rm rad}$ & $\delta J_{x}$ & $\delta J_{y}$ &
$\delta J_{z}$ & $V_{x}$ & $V_{y}$ & $V_{z}$ \\
\hline
\endfirsthead

\multicolumn{10}{c}
{{\tablename\ \thetable{} -- continued from previous
page}}\\
\hline
CONF & $\delta M_{\rm rad}$ & $\delta J_{x}$ & $\delta J_{y}$ &
$\delta J_{z}$ & $V_{x}$ & $V_{y}$ & $V_{z}$ \\
\hline
\endhead
\hline \multicolumn{11}{r}{{Continued on next page}} \\
\hline
\endfoot
\hline\hline
\endlastfoot
NQ200TH30PH0 & $ 0.0323\pm0.0001 $  & $ -0.0011\pm0.0002 $  & $
0.0129\pm0.0002 $  & $ 0.2709\pm0.0027 $  & $ -63\pm2 $  & $ -178\pm2
$  & $ 390\pm2 $  \\      
PH30 & $ 0.0323\pm0.0001 $  & $ -0.0073\pm0.0001 $  & $
0.0105\pm0.0002 $  & $ 0.2707\pm0.0027 $  & $ -64\pm2 $  & $ -184\pm2
$  & $ 309\pm2 $  \\     
PH60 & $ 0.0322\pm0.0001 $  & $ -0.0116\pm0.0001 $  & $
0.0053\pm0.0001 $  & $ 0.2704\pm0.0027 $  & $ -55\pm2 $  & $ -190\pm2
$  & $ 152\pm2 $  \\     
PH90 & $ 0.0322\pm0.0001 $  & $ -0.0128\pm0.0002 $  & $
-0.0013\pm0.0000 $  & $ 0.2703\pm0.0027 $  & $ -47\pm2 $  & $ -189\pm2
$  & $ -46\pm2 $  \\    
PH120 & $ 0.0322\pm0.0001 $  & $ -0.0105\pm0.0002 $  & $
-0.0076\pm0.0000 $  & $ 0.2704\pm0.0027 $  & $ -47\pm2 $  & $ -183\pm2
$  & $ -238\pm2 $  \\  
PH150 & $ 0.0323\pm0.0001 $  & $ -0.0055\pm0.0002 $  & $
-0.0118\pm0.0002 $  & $ 0.2707\pm0.0027 $  & $ -54\pm2 $  & $ -177\pm2
$  & $ -365\pm2 $  \\  
\\
NQ200TH60PH0 & $ 0.0303\pm0.0001 $  & $ -0.0046\pm0.0021 $  & $
0.0250\pm0.0028 $  & $ 0.2516\pm0.0017 $  & $ -145\pm26 $  & $ 83\pm8
$  & $ 131\pm4 $  \\
\end{longtable*}
\begin{longtable*}{r|clclclclcl}
  \caption{{\it This table corrects errors in the published version}. Comparison between remnant horizon properties and radiated
quantities. Differences between the two are a much better measurement
of the true error than either the (very small) variations in the
isolated horizon quantities with time or the extrapolation error in
the radiative quantities. Here $\delta\Theta$ is the angle, in
degrees, that the
total angular momentum $\vec J$ precesses from the start of the
simulation to the final ringdown. Rows with a bold header differ from
the original manuscript.}
\label{tab:rem_rad_cmp_part1}\\
\hline\hline
CONF & $M_{\rm rem}$(IH) & $M_{\rm rem}$(r) &  $S_{x{\rm rem}}$(IH) &
$S_{x{\rm rem}}$(r) & $S_{y{\rm rem}}$(IH) &
$S_{y{\rm rem}}$(r) & $S_{z{\rm rem}}$(IH) & $S_{z{\rm rem}}$(r) &
$\delta \Theta$(IH) & $\delta \Theta$(r) \\
\hline
\endfirsthead
\multicolumn{11}{c}
{{\tablename\ \thetable{} -- continued from previous
page}}\\
\hline

CONF & $M_{\rm rem}$(IH) & $M_{\rm rem}$(r) &  $S_{x{\rm rem}}$(IH) &
$S_{x{\rm rem}}$(r) & $S_{y{\rm rem}}$(IH) &
$S_{y{\rm rem}}$(r) & $S_{z{\rm rem}}$(IH) & $S_{z{\rm rem}}$(r) &
$\delta \Theta$(IH) & $\delta \Theta$(r) \\
\hline
\endhead

\hline \multicolumn{11}{r}{{Continued on next page}} \\
\hline
\endfoot
\hline\hline
\endlastfoot
NQ200TH30PH0 & $ 0.9670 $& $ 0.9677 $& $ -0.0063 $& $ 0.0011 $& $ 0.0241 $& $ 0.0325 $& $ 0.6210 $& $ 0.6198 $& $ 0.9 $& $ 0.1 $\\
\hspace{20pt}PH30 & $  $& $ 0.9677 $& $  $& $ -0.0154 $& $  $& $ 0.0288 $& $  $& $ 0.6200 $& $  $& $ 0.1 $\\
\hspace{20pt}PH60 & $ 0.9672 $& $ 0.9678 $& $ -0.0253 $& $ -0.0277 $& $ 0.0260 $& $ 0.0174 $& $ 0.6210 $& $ 0.6203 $& $ 1.0 $& $ 0.1 $\\
\hspace{20pt}PH90 & $ 0.9673 $& $ 0.9678 $& $ -0.0282 $& $ -0.0326 $& $ 0.0060 $& $ 0.0013 $& $ 0.6215 $& $ 0.6204 $& $ 0.6 $& $ 0.1 $\\
\hspace{20pt}PH120 & $ 0.9672 $& $ 0.9678 $& $ -0.0252 $& $ -0.0288 $& $ -0.0077 $& $ -0.0151 $& $ 0.6215 $& $ 0.6203 $& $ 0.8 $& $ 0.1 $\\
\hspace{20pt}PH150 & $ 0.9671 $& $ 0.9677 $& $ -0.0280 $& $ -0.0172 $& $ -0.0367 $& $ -0.0275 $& $ 0.6200 $& $ 0.6200 $& $ 1.4 $& $ 0.1 $\\
\\
NQ200TH60PH0 & $ 0.9687 $& $ 0.9697 $& $ 0.0060 $& $ 0.0046 $& $ 0.0488 $& $ 0.0537 $& $ 0.6091 $& $ 0.6131 $& $ 0.8 $& $ 0.5 $\\
\end{longtable*}
\begin{longtable*}{r|llllllllll}
\caption{BH spins during final plunge, recoil velocity,
and the angle between $\vec\Delta_\perp$ for PHYYY and
$\vec\Delta_\perp$ of the corresponding PH0 configuration; all
calculated
in
a rotated frame where the infall occurs in the $xy$ plane. Note that
in this frame, $\vec V_\perp = (V_x, V_y)$ and $V_\| = V_z$.}
\label{tab:rot_part1}\\
\hline\hline
CONF & $S_{x1}$ & $S_{y1}$ & $S_{z1}$ & $S_{x2}$ & $S_{y2}$ & $S_{z2}$
& $V_{x}$ & $V_{y}$ & $V_{z}$ & $\varphi$ \\
\hline
\endfirsthead
\multicolumn{11}{c}
{{\tablename\ \thetable{} -- continued from previous
page}}\\
\hline
CONF & $S_{x1}$ & $S_{y1}$ & $S_{z1}$ & $S_{x2}$ & $S_{y2}$ & $S_{z2}$
& $V_{x}$ & $V_{y}$ & $V_{z}$ & $\varphi$ (deg.) \\
\hline
\endhead
\hline \multicolumn{11}{r}{{Continued on next page}} \\
\hline
\endfoot
\hline\hline
\endlastfoot
NQ200TH30PH0 & $ 0 $  & $ 0 $  & $ 0 $  & $ 0.0044 $  & $ 0.0426 $  &
$ 0.0817 $  & $ -75 $  & $ -184 $  & $ 386 $  & $ 0.$  \\
PH30 & $ 0 $  & $ 0 $  & $ 0 $  & $ -0.0168 $  & $ 0.0412 $
& $ 0.0806 $  & $ -76 $  & $ -182 $  & $ 308 $  & $ 28.1$  \\
PH60 & $ 0 $  & $ 0 $  & $ 0 $  & $ -0.0329 $  & $ 0.0296 $
& $ 0.0807 $  & $ -73 $  & $ -182 $  & $ 154 $  & $ 53.8$  \\
PH90 & $ 0 $  & $ 0 $  & $ 0 $  & $ -0.0410 $  & $ 0.0103 $
& $ 0.0818 $  & $ -70 $  & $ -183 $  & $ -41 $  & $ 81.7$  \\
PH120 & $ 0 $  & $ 0 $  & $ 0 $  & $ -0.0384 $  & $ -0.0126 $
& $ 0.0830 $  & $ -68 $  & $ -185 $  & $ -231 $  & $ 114.$  \\
PH150 & $ 0 $  & $ 0 $  & $ 0 $  & $ -0.0247 $  & $ -0.0322 $
& $ 0.0829 $  & $ -71 $  & $ -185 $  & $ -358 $  & $ 148.3$  \\
\\
NQ200TH60PH0 & $ 0 $  & $ 0 $  & $ 0 $  & $ -0.0680 $  & $ 0.0346 $  &
$ 0.0500 $  & $ -72 $  & $ -140 $  & $ 142 $  & $ 0.$  \\
\end{longtable*}

\clearpage
\bibliographystyle{apsrev4-1}
\bibliography{../../../../Bibtex/references}

\end{document}